\documentclass[twocolumn,showpacs,amsmath,amssymb,prd,floatfix,preprintnumbers]{revtex4}

\usepackage[dvipdfmx]{graphicx}
\usepackage{bm}
\usepackage{amsmath}
\usepackage{float}
\usepackage{color}
\usepackage{xcolor}
\usepackage{appendix}
\usepackage{hyperref}
\usepackage{comment}
\usepackage{multirow}  
\usepackage[varg]{txfonts}

\hypersetup{
  bookmarksnumbered=true,
  linkcolor=red,
  citecolor=green,
  urlcolor=cyan
}

\newcommand{\simgt}{\lower.5ex\hbox{$\; \buildrel > \over \sim \;$}}
\newcommand{\simlt}{\lower.5ex\hbox{$\; \buildrel < \over \sim \;$}}

\begin{document}

\title[]{The multipole expansion of the local expansion rate}

\author{ Basheer Kalbouneh}
\author{Christian Marinoni}
\author{Julien Bel}
\email{basheer.kalbouneh@cpt.univ-mrs.fr, \\
christian.marinoni@cpt.univ-mrs.fr, julien.bel@cpt.univ-mrs.fr}
\affiliation{Aix Marseille Univ, Universit\'e de Toulon, CNRS, CPT, Marseille, France}

\date{\today}

\begin{abstract}

We design a new observable, the expansion rate fluctuation $\eta$, to characterize deviations from the linear relation between redshift and dUniverseistance in the local Universe.  We also show how to compress the resulting signal into spherical harmonic coefficients in order to better decipher the structure and symmetries of the anisotropies in the local expansion rate.  We apply this analysis scheme to several public catalogs of redshift-independent distances, the Cosmicflows-3 and Pantheon data sets, covering the redshift range $0.01<z<0.05$. 

The leading anisotropic signal is stored in the dipole. Within the standard cosmological model, it is interpreted as a bulk motion ($307 \pm 23$ km/s) of the entire local volume in a direction aligned at better than $4$ degrees  with the bulk component of the Local Group velocity  with respect to the CMB. This term alone, however, provides an  overly simplistic and inaccurate description of the angular anisotropies of the expansion rate. We find that the quadrupole contribution  is non-negligible ($\sim 50\%$ of the anisotropic signal), in fact, statistically significant, and signaling a  substantial  shearing of gravity  in the volume covered by the data. In addition, the 3D structure of the quadrupole is axisymmetric, with the expansion axis aligned along the axis of the dipole.  

Implications for the determination of the $H_0$ parameter are discussed. We find that Hubble constant estimates may show variation as high as  $\Delta H_0 =
(4.1 \pm  1.1)$ km/s/Mpc between  antipodal directions along the dipole axis. In the case of the Pantheon sample,  this systematic difference  is reduced 
to $\Delta H_0 = (2.4 \pm  1.1)$ km/s/Mpc  once model-dependent correction for peculiar velocity flows are implemented. Notwithstanding,  the axial anisotropy in the general direction of the CMB dipole is still detected. We thus show  how to optimally subtract redshift anisotropies from Pantheon data in a fully model-independent way  by exploiting the  $\eta$ observable. As a result, the value of the best fitting $H_0$ is systematically revised upwards by nearly $0.7$ km/s/Mpc (about $2 \sigma$) compared to the value deduced from the Hubble diagram using the uncorrected observed redshift. The goodness of fit is also improved. 
\end{abstract}

\maketitle

\section{Introduction}
\label{intro}
In accordance with the cosmological principle (CP), the spatial sections of the Universe are maximally symmetric, that is, rotationally and translationally invariant (e.g. \cite{Pee1980}).  
This statement about the symmetries of the Universe, sealed in the Robertson \& Walker line element,  can only be interpreted in a statistical sense, after convolving the spatial distribution of matter with large smoothing kernels. This  makes its empirical confirmation difficult and subject to non-trivial systematicity. 

Despite observational hurdles (e.g. \cite{Clarkson:2010uz}), convincing proofs of isotropy are provided by the 
angular distribution of the temperature fluctuations of the cosmic microwave background (CMB) \cite{Bennet2013,Planck:2015igc}. Also, 3D supporting evidence continues to grow as spectroscopic studies reveal the  structure of ever larger and deeper regions of the Universe  \cite{Sarkar2009, MBB2012, Appleby2014, Bengaly2019, Sarkar2019, Ntelis2019,Andrade:2019kvl, Payne2020, Pandey2022, Goncalves:2020erb}.
Analysis of the spatial distribution of supernovae, i.e. objects whose distances are estimated using redshift independent techniques, also provides tentative confirmation 
\cite{Cai2013, Chang2018, DengWei2018, SunWang2018, Andrade:2018eta, Zhao2019, Hu:2020mzd, Rahman:2021mti}.
The nature of the confirmations remains preliminary, however, and there is no shortage of evidence to the contrary \cite{Schwarz:2007wf, Kashlinsky:2008ut, Antoniou:2010gw, Kalus:2012zu, Yoon:2014daa, Tiwari:2015tba, Javanmardi2015, Colin:2017juj, Migkas:2020fza,Migkas:2021zdo, Secrest:2020has,Siewert:2020krp, Luongo:2021nqh, Krishnan:2021jmh,Dainotti:2021pqg}. If some of these signals are not statistically significant enough to reject outright the CP, others appear as improbable in the framework of the standard cosmological model.

However, it has long been known that in the local outskirts of the Milky Way, at scales $r<150 h^{-1}$Mpc, the CP is violated (e.g. \cite{Hoffman:2017ako, Shaya}). 
This region represents about half of the volume used to determine the Hubble parameter $H_0$, a fundamental constant of the standard model and a consequence of the cosmological principle hypothesis.  Traditionally, deviations from CP predictions are treated perturbatively by expanding the cosmological quantities into a smooth background component and a fluctuating part. Among the latter, a central role is occupied by peculiar velocities. 
These super-Hubble motions contain a lot of interesting cosmological information (e.g.\cite{Koda:2013eya}) and indeed their amplitude confirms that the deviations from the CP 
are in general agreement with the limits imposed by the perturbation theory of the standard cosmological model (e.g. \cite{Ma,Hellwing,Huterer,Larena}). 
However, it was soon realised that many subtle systematic errors, if not properly subtracted, can compromise their use  as efficient cosmological probes; non-Gaussian issues, homogeneous and in-homogeneous Malmquist biases, incompleteness of mass catalogs used to predict the amplitude of peculiar velocities are among the pitfalls that most hamper the analysis \cite{Strauss:1995fz}. 

In order to free the investigation of local inhomogeneities from certain statistical and observational complications, we explore in this paper another direction. We develop a completely non-perturbative approach to inhomogeneities that focuses directly on the scale factor of the Universe as a relevant variable to quantify deviations from uniformity (see also \cite{Macpherson, Dhawan}  for a similar approach). 
This is precisely the parameter that is kept invariant in perturbative analyses, defining the reference background against which deformations in the spatial sector of the metric are compared.  

In this spirit, we design  an observable, the expansion rate fluctuation $\eta$,   that provides information about fluctuations in the local expansion rate and is, at the same time, easily comparable with theoretical predictions. Indeed we will show that  it provides a model-independent means of analysing inhomogeneities, not even requiring the CP assumption as a prerequisite. This cosmographic approach (e.g. \cite{Kristian, ellis85}) 
allows  the results to be  directly interpretable in alternative spacetimes and can ideally guide the search for unconventional line elements that capture the essential features of the local inhomogeneities. 

From an observational point of view, the goal is to investigate the existence and significance of anisotropies in the local Universe through new methods of investigation. In this specific case, by decomposing the angular fluctuations of the expansion rate into spherical harmonics and compressing  information about anisotropies into a set of independent Fourier coefficients.  

Multipolar expansion in spherical harmonics provides an orthogonal  insight into the nature of the local redshift-distance  relation and 
allows to go beyond the simple dipole model with which anisotropies are traditionally described in the nearby Universe.  
At the same time, it allows  to deepen  and extend studies, such as those of \cite{Lilje_1986, Courtois, Parnovsky,Hoffman:2017ako, Dhawan} which attempt to constrain the tidal field component by analysing the shear of the velocity field generated by local gravitational fluctuations. In this respect, we focus on the study of the symmetries and geometric structure of the harmonic multipoles, showing how their analysis gives a simple and inexpensive description of the structure of the anisotropies in the Hubble flow. We demonstrate that the three-parameter formula encoding such information has predictive power comparable to that of much more complex numerical studies of peculiar motions.

The paper is organized as follows:
in Sec.~\ref{sec:obs} we  introduce the observable that optimally extract information about the fluctuation in the angular expansion rate,  while in \ref{sec_shd} we present the method implemented to estimate the signal  from discrete datasets and to compress it into spherical harmonic  coefficients. We also discuss how we  estimate reconstruction errors, both statistical and systematic. In Sec.~\ref{sec:data}, we describe the data analyzed.   Results are presented and interpreted  in \ref{sec_results}.  
Sec.~\ref{sec:conclusions} provides summary and conclusion.

In the following,  we present results  in natural units  ($c=1$) and we refer  to the standard  $\Lambda$CDM model, as the  flat Friedmann-Robertson-Walker (FRW)  spacetime which best fits the Planck18 data \cite{Planck:2018nkj}.
Redshift is expressed with respect to the CMB rest frame. 

\section{The expansion rate fluctuations \boldmath$\eta$\unboldmath}\label{sec:obs}

We model the angular anisotropies in the  redshift-distance relation  by directly exploiting the local expansion rate  as a target observable. In a perfectly uniform FRW universe, the ratio $z/d$ between the  redshift and the proper distance of comoving particles  is predicted to be constant, independent from the  particular line-of-sight along which it is estimated. 

In any generic metric model describing the structure of local space-time, i.e. the inhomogeneous distribution of mass at the periphery of the Local Group of galaxies, it is possible, at least in the limit of small separations, to relate the  redshift $z$ and the proper distance as follows
\begin{equation}
    z=\tilde{H}_0(l,b) d.
\label{ansatz}
\end{equation}
In this expression, $\tilde{H}_0$ is a continuous function that depends only on the angular coordinates ($l,b$)  and can be constrained experimentally.  It is clear that the angular dependence is in principle theoretically determinable as soon as a line element is provided. Note that if the observer is only at rest relative to the CMB, but not comoving with respect  to the surrounding matter, then we expect  a dependence of $\tilde{H}_0$ on the radial distance even in very local regions of the Universe. As a matter of fact, 
in a generic spacetime, the characteristic distance scale at which the linear limit of the equation (\ref{ansatz}) is reached is not known apriori. 

We actually characterize deviations from isotropy in the local expansion rate via the (decimal) logarithmic relation 
\begin{equation}
     \eta \equiv \log\left[ \frac{\tilde{H}_0}{H_0}\right]. 
     \label{eta3D}
\end{equation} 
 Here $H_0$ is a normalizing factor that,  in the standard model of cosmology, coincides with the  value of the  Hubble constant. We fix its amplitude by requiring that the average value of $\eta$ over the volume covered by data vanishes.

The justification for the choice of this observable is statistical in nature.
 Errors are Gaussian only in the distance modulus $\mu$ and not in the redshift-independent distance $d$ if these latter are   estimated as
\begin{equation}
\hat{d}=10^{\frac{\mu-25}{5}}.
\label{ldi}
\end{equation}
Therefore, given a sample of objects at spatial position ${\bf r}$, the discrete estimator of the continuous field  (\ref{eta3D}) 
\begin{equation}
     \hat{\eta}({\bf r}) = \log\left[ \frac{z}{H_0}\right]+5-\frac{\mu}{5}, 
     \label{eta3D2}
\end{equation} 
 is a random variable that follows a Gaussian distribution.
 Indeed, we assume that the uncertainty $\delta$ on $\hat{\eta}$ is induced only by the imprecision with which the redshift-independent  distances are estimated ($\delta=\sigma_{\mu}/5$), i.e. we consider that any error in the redshift estimate is negligible.   
 As a consequence, $\hat{\eta}$  provides an unbiased estimate of $\eta$, as can be easily verified. As an added bonus, Eq. (\ref{eta3D}) also makes it possible to quantify anisotropies in the expansion rate, regardless of the value of the Hubble constant parameter used to normalize the distance modules $\mu$. 
 A subtlety must be pointed out.  It is implicitly assumed, in the above   argument, that in the limit $z<<1$, the range we are concerned in this paper,  $d$ is a fair proxy for both  the luminosity and  angular diameter distance, {\it i.e. $d \approx d_L \approx d_A$}.

 The  expansion rate fluctuations $\eta$ 
  is not specifically tailored  to have only nice statistical properties. It also has a physical content. Linear perturbation theory  of the standard cosmological model provides a framework for interpreting this observable. According to it, the redshift observed in the CMB rest frame is given  by
\begin{equation}
    z=z_c+v(1+z_c),
\label{zdefi}
\end{equation}
where $z_c$ is the cosmological redshift and $v$ is the line of sight component of the peculiar velocity of the source (assumed to be non-relativistic) with respect to the CMB rest frame. 
By inserting this last relation into \ref{eta3D} we get
\begin{equation}
\eta=\log \left(1+\frac{v(1+z_c)}{z_c} \right)  \approx  \frac{v(1+z)}{z\ln 10},
\label{lpte}
\end{equation}
where we have assumed  $v<<z_c \approx z$. The fluctuations in the expansion rate are excited by radial peculiar  velocity and suppressed in inverse proportion to the object's distance.

\section{Expansion rate fluctuations: the  spherical harmonic decomposition}\label{sec_shd}

We can  compress the information contained in the $\eta$ observable into a few coefficients. To this end, we  expand the  expansion rate field $\eta$ in spherical harmonic (SH) components.  We orthogonally decompose  $\eta$  on  a sphere as follows
 \begin{equation}
 \eta=\sum_{\ell=0}^{\infty} \sum_{m=-\ell}^{\ell} a_{\ell m}Y_{\ell m}(\theta, \phi) =\sum_{\ell=0}^{\infty} \eta_\ell,
 \label{etamodo}
 \end{equation}
 where 
\begin{equation}
    Y_{\ell m}(\theta,\phi)=\sqrt{\frac{(2 \ell+1)(\ell -m)!}{4\pi (\ell +m)!}} P^{m}_{\ell }(\cos{\theta})e^{i m \phi},
\end{equation}
and $P^{m}_{\ell}$ are associated Legendre polynomials. Thus, the Fourier coefficients $a_{\ell m}$ can be expressed as 
\begin{equation}\label{fouco}
a_{\ell m} \equiv \int_{0}^{2\pi}\int_{0}^{\pi}\eta(\theta,\phi)Y^{*}_{\ell m}(\theta,\phi)\sin{\theta}\; d\theta d\phi.
\end{equation}
Note that, due to  the definition, the monopole ($\ell = 0$) of $\eta$ vanishes.  

In addition, one can define the  angular  power  spectrum  of  the $\eta$  anisotropy  as
\begin{equation}
     C_{\ell}=\langle |a_{\ell m}|^{2}\rangle_e,
     \label{cl}
\end{equation}
where the expectation is intended to be over a statistical ensemble of universes. 

\begin{equation}
    \hat{C}_{\ell}=\frac{1}{2\ell +1}\sum^{\ell}_{m = -\ell}|a_{\ell m}|^{2},
    \label{clhat}
\end{equation}
is an unbiased estimator for $C_\ell$

The following section (\S \ref{sec_method}) describes in detail the procedure  adopted to reconstruct the field $\eta(\Omega)$  from discrete 3D data with non-uniform sampling rates on the sky. In section (\ref{sec_alm}) we describe how the SH coefficients  $a_{\ell m}$ are estimated.  We present the analytical formulas and numerical recipes for evaluating measurement errors, both statistical and systematic, in the sections \ref{sec_errors} and \ref{sec_system}, respectively.

\subsection{Estimation of the angular \boldmath $\eta$ \unboldmath field}    \label{sec_method}

 The  expansion rate fluctuation estimator  $\hat{\eta}({\bf r})$  is a  discrete  random variable.  The analysis of this observable can be simplified, and the underlying  theoretical model (\ref{eta3D})  can be better traced if we  convert it  into a stochastic field.  We thus average $\hat{\eta}({\bf r})$  over all the objects at position ${\bf r}$  within a given volume $V(\Omega, R)$, where $\Omega$ is a solid angle centered  on the observer and $R$ the depth of the catalog (i.e. its upper edge). The angular anisotropies seen by the observer are thus piece-wise defined as
 \begin{equation}
 \eta (\Omega) =\frac{\sum_i^N \hat{\eta}({\bf r}_i) w({\bf r}_i) W({\bf r}_i|V(\Omega,R))} {\sum_i^N w({\bf r}_i) W({\bf r}_i|V(\Omega, R))}  
\label{eta2D}
 \end{equation}  
 where $N$ is the number of objects in the catalog, 
 $w({\bf r}_i)=1/\delta^2_i$ is a weight that takes into account the precision in the measurement of the distance of the $i$-$th$ object in the catalog.  $W({\bf r}|V(\Omega,R))$ is a window function which evaluates to unity if ${\bf r}_i \in V$ and is null otherwise. 
 
 It is clear that averaging has the advantage of reducing noise at the cost of a lower angular resolution.  The latter is essentially controlled by the aperture of the solid angle $\Omega$, although it also depends, in principle, on the depth $R$ of the sample on which the spatial averaging is performed.  

\begin{figure}
\includegraphics[scale=0.52]{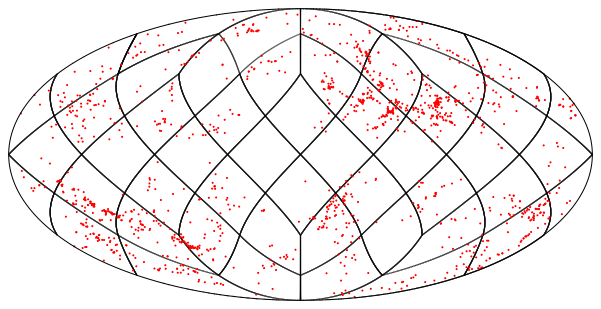}
\includegraphics[scale=0.52]{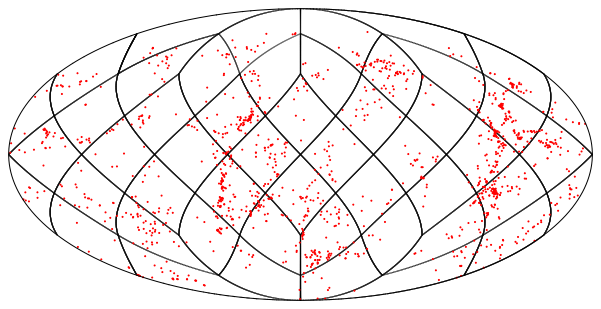}
\includegraphics[scale=0.52]{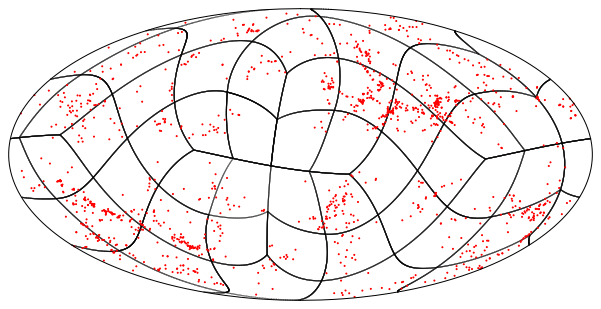}
\centering
\caption{Illustration of the rotation strategy to improve the estimation of SH coefficients. {\it Upper:} the standard HEALPix pixels ($N_{side}=2$ and $N_{pix}=48$) tessellating the distribution of  galaxies in the galactic coordinates. Note the presence of an empty pixel. {\it Center :} rigid rotation applied to the sample so that in each pixel falls at least one galaxy (the minimum number is 5 in this example).  {\it Lower :} the inverse rotation is applied to both galaxies and pixels.}
\label{FIG_rot}
\end{figure}

In practice, we construct the  $\eta$ 2D field out of a discrete point process    $\hat{\eta}({\bf r})$, by first partitioning the sky  in $N_{pix}$  identical pixels (each subtending a solid angle $\Omega_i$)  using HEALPix \cite{Gorski:2004by} and  then by  applying eq. (\ref{eta2D})  to objects within the volume $V$ subtended by each pixel $\Omega_i$.  
HEALPix is an algorithm  which tessellates a spherical surface into curvilinear quadrilaterals, each covering the same area as every other. Although characterized by a different shape, the resulting pixels are located on lines of constant latitude. This property is essential for  speeding up computation but is less than optimal for pixelating the discrete  $\eta$ field. In fact, the counts can show large variations from pixel to pixel. In addition, some of the pixels may end up containing no data at all.  To tackle this issue,
we first rigidly rotate the galaxy field randomly, by looking for configurations in which all the HEALPix pixels are filled with objects {\it and} the least populated cell contains a maximum number of objects. 
 If as a result of different rotations, 
 the maximum number of galaxies in the least populated cells stays the same,  we pick up  the configuration for which the  distribution of the number of the galaxies in the pixels has the minimum variance. This allows to minimize pixel-to-pixel fluctuations in the reconstructed value of $\eta$  and increase the signal-to-noise ratio in the determination of the Fourier coefficients. Note that the rotation trick does not affect the estimation of the angular power spectrum $\hat{C}_\ell$, which is, by definition, invariant under rotation. However, once the Fourier coefficients have been estimated, we apply an inverse rotation to the pixels and $\eta$ maps so that, for the sake of clarity, the results are presented in standard galactic $(l,b)$ coordinates.  The whole strategy is graphically illustrated in Fig. \ref{FIG_rot}.

The resolution of the HEALPix grid is calculated as $N_{pix}=12 N_{side}^2$ where $N_{side}=2^t$, and $t\in \mathbb{N}$. The baseline grid,  corresponding to $t=0$,   has 12 pixels.  Our choice of the  resolution in the reconstruction of the angular $\eta$ map, as explained in \ref{sec_results}, is dictated by two criteria: the SH decomposition must result in multipoles that have a sufficiently high signal-to-noise ratio and a sufficiently low  probability $p$ to   occur by chance in a randomly fluctuating $\eta$ field.

\subsection{Estimation of the SH coefficients}\label{sec_alm}
We estimate the Fourier coefficients of the spherical harmonic decomposition by slightly modifying the reconstruction scheme provided by the HEALPix algorithm. HEALPix accomplishes that by means of an iteration scheme, the so-called {\it Jacobi iteration}.  The zeroth order estimator of the coefficients of the expansion field is
\begin{equation}
    \hat{a}_{\ell m}^{(0)}= \frac{4\pi}{N_{pix}}\sum^{N_{pix}}_{p=1}\eta(\Omega_p) Y_{\ell m}^{*} (\theta_p,\phi_p),
    \label{alm0estimate}
\end{equation}
where ($\theta_p$,$\phi_p$) are the angular coordinates  of the center of each pixel $p$
and  $\eta(\Omega_p)$ is calculated by eq. (\ref{eta2D}). The Fourier coefficients  $a_{\ell m}$ are then calculated up to the order $\ell_{max}=3N_{side}-1$, and the higher orders are

\begin{equation}
\hat{a}_{\ell m}^{(k+1)}=\hat{a}_{\ell m}^{(k)}+ \frac{4\pi}{N_{pix}}\sum^{N_{pix}}_{p=1}\bigg(\eta(\Omega_p)-\eta^{(k)}(\theta_p,\phi_p)\bigg) Y_{\ell m}^{*} (\theta_p,\phi_p),
    \label{almkestimate}
\end{equation}
where
\begin{equation}
    \eta^{(k)}(\theta_p,\phi_p)=\sum^{\ell_{max}}_{\ell=0}\sum^{\ell}_{m = -\ell}\hat{a}^{(k)}_{\ell m}Y_{\ell m}(\theta_p,\phi_p).
\end{equation}
In  matrix notation,
\begin{equation}
    \boldsymbol{a}^{(0)} = \boldsymbol{A}\boldsymbol{\eta}
    \label{matrixA}
\end{equation}
\begin{equation}
    \boldsymbol{a}^{(k+1)} =\boldsymbol{a}^{(k)}+ \boldsymbol{A}(\boldsymbol{\eta}-\boldsymbol{\eta}^{(k)})
    \label{akiter}
\end{equation}
\begin{equation}
    \boldsymbol{\eta}^{(k)} = \boldsymbol{S}\boldsymbol{a}^{(k)},
\end{equation}
where $\boldsymbol{a}$ is the vector of the spherical harmonic coefficients (containing $(\ell_{max}+1)^2$ elements), while $\boldsymbol{\eta}$ and $\boldsymbol{\eta}^{(k)}$ are vectors representing the measured and estimated values of $\eta$ in each pixel (and thus contain $N_{pix}$ elements). Moreover,  $\boldsymbol{A}= \frac{4\pi}{N_{pix}} Y_{\ell m}^{*}(\theta_p,\phi_p)$ and $\boldsymbol{S}= \frac{N_{pix}}{4\pi}\boldsymbol{A}^{* T}$. The calculation of $\boldsymbol{a}^{(k)}$ is repeated until convergence i.e. until the residual has zero Fourier coefficients up to $\ell_{max}$. 

Instead of going through the iteration scheme, we proceed in a different way. We estimate analytically the asymptotic limit  $\boldsymbol{a}^{(\infty)}$ that should be ideally obtained in the limit of an 
 infinite number of iterations. We first  write equation (\ref{akiter}) as

\begin{equation}
    \boldsymbol{a}^{(k+1)} =\boldsymbol{a}^{(k)}+ \boldsymbol{A}\boldsymbol{\eta}-\boldsymbol{A}\boldsymbol{S}\boldsymbol{a}^{(k)},
\end{equation}
so
\begin{equation}
    \boldsymbol{a}^{(k+1)} = \boldsymbol{A}\boldsymbol{\eta}+(\boldsymbol{I}-\boldsymbol{A}\boldsymbol{S})\boldsymbol{a}^{(k)},
\end{equation}
where $\boldsymbol{I}$ is the identity matrix with size $(\ell_{max}+1)^2\times(\ell_{max}+1)^2 $. By using $\boldsymbol{a}^{(0)} = \boldsymbol{A}\boldsymbol{\eta}$ we obtain 

\begin{equation}
    \boldsymbol{a}^{(k+1)} = \boldsymbol{a}^{(0)}+(\boldsymbol{I}-\boldsymbol{A}\boldsymbol{S})\boldsymbol{a}^{(k)}.
\end{equation}
Under the assumption that this is convergent, for $k\rightarrow \infty$, $ \boldsymbol{a}^{(k+1)}\rightarrow  \boldsymbol{a}^{(k)}$ we get 
\begin{equation}
    \boldsymbol{a}^{(\infty)} = \boldsymbol{a}^{(0)}+(\boldsymbol{I}-\boldsymbol{A}\boldsymbol{S})\boldsymbol{a}^{(\infty)},
\end{equation}
which results in  
\begin{equation}
    \boldsymbol{a}^{(\infty)} = \boldsymbol{M} \boldsymbol{a}^{(0)}  ,  \label{closedex}
\end{equation}
where $\boldsymbol{M}=(\boldsymbol{A}\boldsymbol{S})^{-1}$. Note that we cannot take the inverse of $\boldsymbol{A}$ or $\boldsymbol{S}$ individually because they are not square matrices.
By this trick, we achieve two goals. First, we minimize the computing time, moreover, and more importantly,   we obtain a closed form  expression which simplifies the estimation of the error on the SH coefficients, as we detail  in the next section and in Appendix \ref{app_varalm}. 
The elements of the vector $\boldsymbol{a}^{(\infty)}$ represents our best  
estimate  ($\hat{a}_{\ell m}$)  of the coefficients of the spherical harmonic decomposition $a_{\ell m}$.

\subsection{Statistical measurement errors}
\label{sec_errors}

 In the following, we  consider the SH coefficients $a_{\ell m}$ as a  deterministic variable whose estimate 
\begin{equation}
    \hat a_{\ell m} = a_{\ell m} + \epsilon_{\ell m},
    \label{almhat}
\end{equation}
fluctuates due to measurement errors $\epsilon_{\ell m}$ induced by uncertainties in the reconstruction of $\eta$. The expectation over different observational measurements made on the same sample is thus $E[\hat a_{\ell m}] \equiv \langle  \hat{a}_{\ell m} \rangle =a_{\ell m}. $
i.e. we assume that the  estimator  provides an unbiased estimate of the coefficients of the spherical harmonic expansion. 
 
The variance of the estimator is  defined as 
\begin{equation}
    V[\hat{a}_{\ell m}]=V[\epsilon_{\ell m}]\equiv \sigma_{\ell m}^{2}, 
    \label{sigmalm}
\end{equation}
and, in general,  depends on both modes $\ell$ and $m$. An exact analytical expression for $\sigma_{\ell m}$ is far from trivial and unenlightening. It can be evaluated from the knowledge of the uncertainties in the distance modulus measurements (see Appendix \ref{app_varalm} and cf. eq. \ref{varalmex} and \ref{closedex}).

An estimator of the power locked in each harmonic moment $\ell$ is the angular power spectrum (cf. eq. \ref{clhat}) estimator
\begin{equation}
    \hat{\hat{C}}_\ell=\frac{1}{2\ell +1}\sum_{m = -\ell}^{\ell}|a_{\ell m}+\epsilon_{\ell m}|^2.
    \label{defestCl1}
\end{equation}
which can be expressed as
\begin{equation}
    \hat{\hat{C}}_\ell=\frac{1}{2\ell +1}\sum_{n=0}^{2l}w_{n}^{(\ell)},
\end{equation}
where 
\begin{equation}
w_{n}^{(\ell)} = \left\{
        \begin{array}{ll}
            \hat{a}_{\ell0}^{2} & \quad n = 0 \\
            2\Re[\hat{a}_{\ell n}]^{2} & \quad \ell \geq n > 0  \\
            2\Im[\hat{a}_{\ell(n-\ell)}]^{2} & \quad 2\ell \geq n > \ell  \\
        \end{array}
    \right.
\end{equation}
This decomposition is conveniently chosen so to take into account that $a_{l-m}$ and $a_{\ell m}$ are conjugate variables ($a_{\ell -m} =(-1)^m a_{\ell m}^*$). 
Thus, the variance of $\hat{\hat{C}}_{\ell}$ reads
\begin{equation}
    V\left [\hat{\hat{C}}_\ell\right ]=\left(\frac{1}{2\ell +1}\right)^2\sum_{n=0}^{2\ell}V[w_{n}^{(\ell)}],
    \label{varCl1}
\end{equation}
where

\begin{equation}
V[w_{n}^{(\ell)}] \! = \! \left\{
        \begin{array}{ll}
            2\sigma_{\ell0}^{4}+4\sigma_{\ell0}^{2} a_{\ell0}^{2}  &  n = 0 \\
            8\sigma_{\ell n}^{(R)4}+16\sigma_{\ell n}^{(R)2}\Re[a_{\ell n}]^{2} & \ell \geq n > 0  \\
            8\sigma_{\ell(n-\ell)}^{(I)4}+16\sigma_{\ell(n-\ell)}^{(I)2} \Im[a_{\ell(n-\ell)}]^{2} &  2l \geq n > \ell  \\
        \end{array}
    \right. ,
    \label{varw2}
\end{equation}
and where we have defined
$\sigma_{\ell m}^{(R)2}=V[\Re[\hat{a}_{\ell m}]]$ and $ \sigma_{\ell m}^{(I)2}=V[\Im[\hat{a}_{\ell m}]]$.

In appendix \ref{app_MC} we show that the analytical estimates obtained using eq. (\ref{varCl1}) provides a fairly good approximation to those obtained via a numerical Monte Carlo analysis (see the comparison in TABLE \ref{Tab_4}). Expression (\ref{varCl1}) can be further simplified by making assumptions about the nature of the errors. If the distribution of the galaxies is isotropic and all of them have the same error in $\eta$ ($\delta_i=\delta$), then, the real and imaginary parts of $\hat{a}_{\ell m}$ are characterised by the same variance, and it will not depend on the considered harmonics ( $\sigma_{\ell m}^{2}= \sigma^{2} \approx \frac{4\pi}{N}\delta^{2} $), so,
\begin{equation}
    V[\hat{\hat{C}}_l]=\frac{2}{2\ell +1} \left[ (\sigma^{2}+\hat{C}_{\ell})^2-\hat{C}_l^2 \right].
    \label{varCla}
\end{equation}
The above formula neatly  isolates the two quantities contributing to the observed variance: the error in the estimation of the distance modulus and the measured amplitude of the angular power spectrum $\hat C_\ell$. 

Note, incidentally, that eq. (\ref{varCla}) differs  from  the expression  that would be obtained by averaging over a statistical ensemble. Indeed, in this latter case  $E_e[\hat{a}_{\ell m}] = 0$  (so  $\hat{C}_\ell$ will be replaced by zero) and also  $V_e[\hat{a}_{\ell m}]=C_\ell+\sigma^{2}$. It thus follows  that  the theoretical expression factoring in  the contributions of cosmic variance is  
$V_e[\hat{\hat{C}}_\ell]= 2/(2\ell +1) (\sigma^{2}+C_{\ell})^2$, if, again, it is assumed that $\sigma_{\ell m}$ is isotropically distributed.

\subsection{Systematic measurement errors}\label{sec_system}

 Eq.  (\ref{defestCl1})
 provides a biased estimate of the local value of the angular power spectrum $\hat C_\ell$ since its 
 expectation over measurements is 
\begin{equation}
    E\left [\hat{\hat{C}}_{\ell}\right ] = \hat{C}_\ell+\frac{1}{2\ell +1} \sum_{m = -\ell}^{\ell}\sigma_{\ell m}^{2}.
    \label{meancl}
\end{equation}
Measurements errors on the distance modulus lead to a systematic overestimation of the angular power spectrum $\hat C_\ell$, and  the statistical bias term calculated in eq. (\ref{meancl}) might not fully remove the systematic shift.  
Indeed, expression  (\ref{meancl}) is strictly valid if the estimator $\hat{a}_{\ell m}$ (cf. eq. (\ref{almhat})) is, as we assumed, affected only by statistical errors.
It is true, however, that incompleteness and anisotropies in the sky distribution, as well as the pixelization and resolution strategy adopted to transform the discrete $\eta$ observable  into a field, could bias the $\hat{a}_{\ell m}$ estimator.
Although any constant systematic term added in \ref{almhat} does not affect the variance of the coefficients of $\hat a_{\ell m}$, it will result in an additional (and analytically nontrivial) term in eq. (\ref{meancl}). 

The total systematic bias $\Delta C_{\ell}$ affecting  the estimator $\hat{\hat{C}}_{\ell}$  is thus more conveniently quantified using  Monte Carlo simulations. The way Monte Carlo simulations are constructed and analysed is discussed in Appendix \ref{app_MC}. There we also report the values of the systematic bias induced on the $\hat{C}_\ell$ estimates (see TABLE \ref{Tab_3}).

  A different systematic error  results from the SH decomposition of the logarithm of $H_0$ and not of $H_0$ itself. 
This choice of the observable, motivated by statistical reasons (cf $\sec$ \ref{sec:obs}), induces spurious higher order multipoles even if they are absent in $H_0$. This bias is however negligible. Indeed, in the ideal case in which 
$\tilde{H}_0$ is purely dipolar in nature 
$\tilde{H}_0/H_0=(1+\epsilon \cos{(\alpha)})$ (here $\epsilon$ is a small expansion parameter), 
one gets $\hat{C}_0=4\pi$, $\hat{C}_1=\frac{4\pi}{9}\epsilon^2$ and $\hat{C}_\ell=0$ for $l\geq 2$.
The logarithmic transformation gives a new dipole power spectrum $\hat{C}_1=\frac{4 \pi}{9 (\ln{(10)})^2}\epsilon^2+o(\epsilon^4)$ and induces a parasitic quadrupole  $\hat{C}_2=\frac{4\pi}{225(\ln{10})^2}\epsilon^4+o(\epsilon^6)$. Its amplitude is however negligible $(\frac{\hat{C}_2}{\hat{C}_1}\approx\frac{1}{25}\epsilon^2)$, 
as it is that of the high order spurious multipole. This effect is thus largely subdominant (a factor roughly $\sim 30$) compared to the typical errors that plague our analysis.

An additional  effect that should in principle be considered is that, although we are not interested in the monopole of $\log \tilde H_0$ field, since we are not interested in calibrating the absolute scale of the Hubble constant, a redshift-dependent value of the parameter $H_0$ in eq. \ref{eta3D}  could lead to a spurious dipole component. This effect, which occurs if   anisotropically distributed data are projected onto the sky and analyzed in two dimensions, does not affect   the sample we analyze,  for which the monopole amplitude appears to be fairly independent of redshift. Indeed, we have verified that $\langle \log \tilde H_0\rangle $ calculated
at  $z=0.05$ and $z=0.01$ changes by roughly $0.0006$ ($1.8628\pm0.0010$ and  $1.8622\pm0.0014$ respectively for the CF3 sample) which  corresponds to $\Delta H_0\approx\Delta\eta H_0 \ln{(10)}=0.1$ for $H_0=70 \; \rm km/s/Mpc$)  a negligible amount if compared to the errors affecting the analysis.

\begin{figure}
\begin{center}
	\includegraphics[scale=0.21]{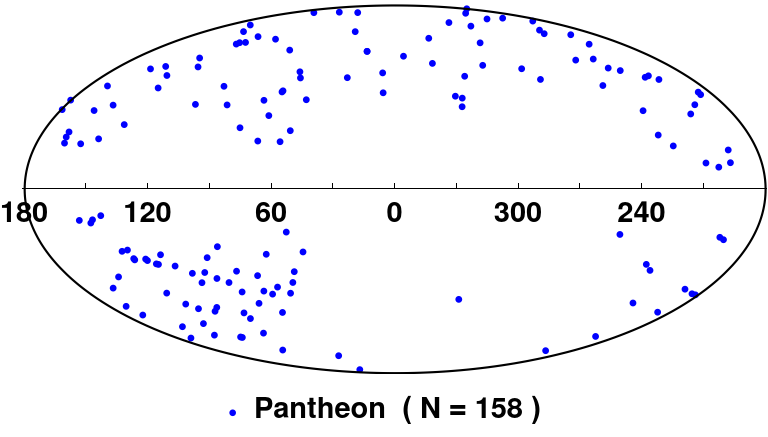}
	\includegraphics[scale=0.33]{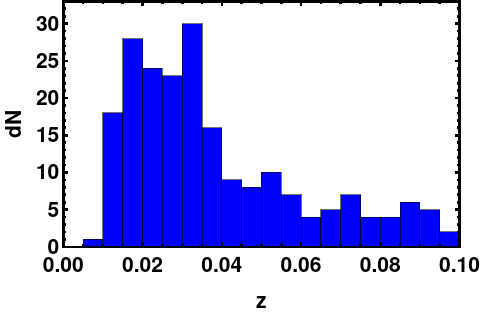}
        \includegraphics[scale=0.21]{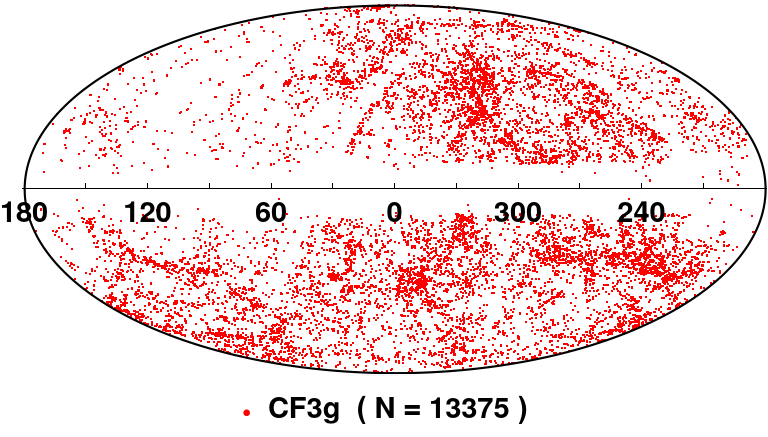}
        \includegraphics[scale=0.33]{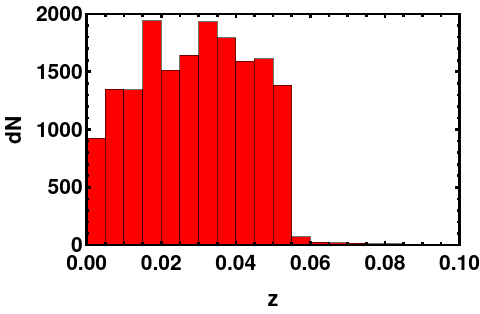} 
        \includegraphics[scale=0.21]{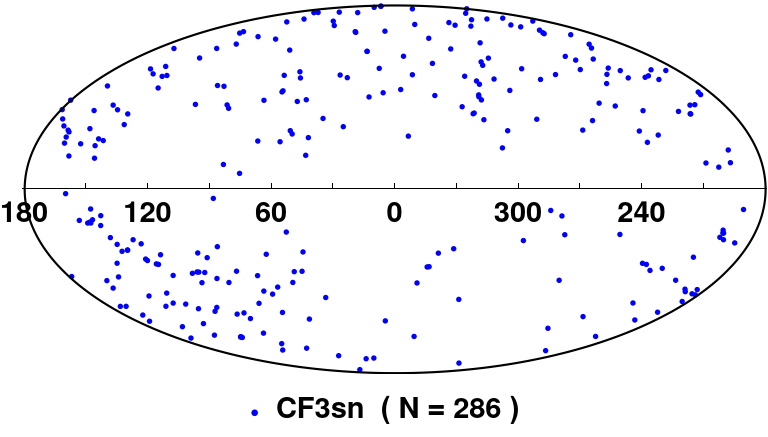}
        \includegraphics[scale=0.33]{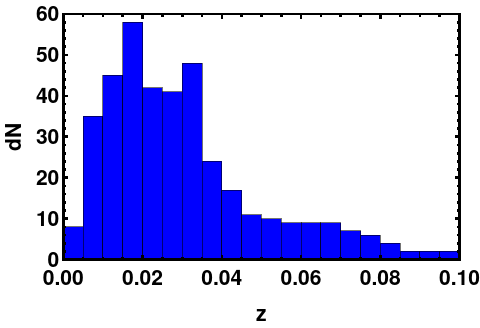}
			\caption{{\it Upper left :} Mollweide projection, in galactic coordinates,  of the distribution of the Pantheon SNIa with redshift $0.01<z<0.05$. {\it Upper right :} histograms of the number of counts as a function of redshift. The second and the third rows are the same as the first, but for the CF3g and CF3sn samples respectively.}
				\label{Fig_1}
		\centering
	\end{center}
\end{figure}

\begin{figure}
\begin{center}
\includegraphics[scale=0.5]{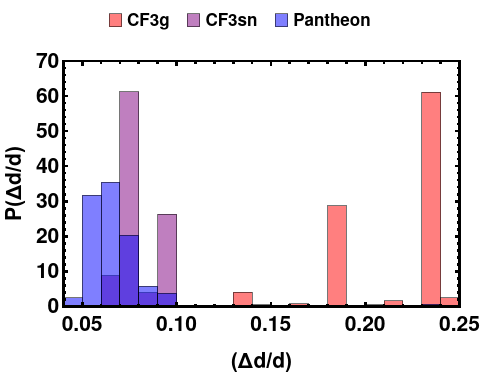}
\caption{The probability distribution of the relative error in the distances  ($\Delta d/d\approx\Delta \mu \ln 10/5$) measured in samples CF3g, CF3sn and Pantheon in the interval $0.01<z<0.05$.}
\label{Fig_erro}
\centering
\end{center}
\end{figure}

\section{Data}  \label{sec:data}

Here we briefly describe the  samples  of redshift-independent distances used to estimate $\eta$. 

\subsection{The Pantheon sample}
The Pantheon SNIa compilation \cite{Scolnic2018} is comprised of 1048 objects lying in the interval $0.01 < z < 2.26$. The catalog was assembled using data from the Supernova Legacy Survey (SNLS) \cite{Guy}, the Sloan Digital Sky Survey (SDSS) \cite{Smith, Sako}, Pan-STARS1 (PS1) \cite{Scolnic2018}, \cite{Riess1999, Jha, Hicken2009a, Hicken2009b,Hicken2012} the Carnegie Supernova Project (CSP) \cite{Contreras} and various surveys made possible by the Hubble Space Telescope (HST), namely CANDLES/ CLASH \cite{Rodney, Graur, Riess2018}, GOODS \cite{Riess2007} and SCP \cite{Suzuki}.

Although many previous investigations of kinematical anisotropies rely on the analysis of  the full Pantheon sample, in this study we  follow a pretty conservative approach   and consider only  a SNIa subsample with  maximal spatial uniformity. To this end, we require the data to sample in a fairly  isotropic manner the sky and also to display sufficient homogeneity in the radial coordinate. The goal is to minimize possible biases and systematic effects induced by incomplete spatial  sampling  without sensibly reducing the statistical constraining power of the data. 

In Fig \ref{Fig_1}  (the first row) the angular  distribution of the Pantheon data is shown together with the 
differential number counts profile $dN(z)$ as a function of redshift. There are no objects in the very local patch of the Universe,   below $z=0.01$, and the  sample becomes quickly anisotropic as soon as  the redshift is larger than $z=0.05$. It also becomes in-homogeneous, i.e. the scaling of $dN$ is not anymore proportional to $z^2$, for  $z \gtrsim 0.04$. 
As a consequence, a  trade-off threshold value $z=0.05$ is chosen  for selecting the SNIa sample to be  used in this study. 
This leaves us with a subsample containing 158 SNIa.

\subsection{The Cosmicflows-3 data}

For the purposes of our analysis, 
we complement the Pantheon supernovae  sample with the Cosmicflows-3 catalog \cite{Tully_2016}.
 This is an all-sky galaxy catalog comprised of 17669 
nearby  galaxies $z\leq 0.116$
 for which redshift-independent distances are inferred using the 
 the
correlation between galaxy rotation and luminosity (Tully-Fisher  law),
or the  Fundamental Plane methods.

This galaxy catalog  offers three key advantages: 
a completely independent  way of estimating galaxy  distances and a richer collection of distance moduli $\mu$ (nearly a hundred times more than those contained in the supernovae catalog). It is this large statistical figure that helps beating down the large error with which the individual galaxy distances are estimated.
 These uncertainties are shown in Fig. \ref{Fig_erro}, where they are also compared to the typical inaccuracies that characterize the SNIa distances. The latter  are at least a factor of 2 smaller than those based on the Tully-Fisher  or the Fundamental Plane estimates (so each is four times or more valuable in a statistical weighting scheme). 
An additional benefit is that the association of galaxies to groups allows local non-linear contributions to the observed redshift to be averaged away.
In this way, the redshift is less sensitive to the local gravitational field at sub-megaparsec scales and more directly reflects the large-scale properties of space-time.

The angular and radial distribution of the Cosmicflows-3 galaxies is shown in Fig. \ref{Fig_1} (the second and the third rows). They are fairly evenly distributed in both redshift  and position in the sky except for redshifts greater than 0.05. So, we don't include the galaxies beyond this redshift since the sample becomes too sparse and covers the sky anisotropically. We also exclude from the sample, galaxies with redshift less than 0.01, in order to facilitate the comparison with the results obtained from the Pantheon data. We, therefore, focus our analysis on the Cosmicflows-3 subsample which is constrained in the range $0.01<z<0.05$ and includes 13661  galaxies. We add in this respect that the purpose of our analysis is to obtain a coarse-grained description of the expansion rate. In the periphery of the Local Group, the geometry and dynamics of the metric are instead dominated by a few large nonlinear structures, and, as a consequence, the galaxies poorly track the large-scale gravitational field we are interested in.

Within this redshift range, the CF3 catalog contains 286 galaxies hosting a SNIa (Fig. \ref{Fig_1} third row) for which the distance modulus is known using the standard candle method.  
Although systematically homogenized, this compilation of SNIa-based distances  remains fundamentally heterogeneous, with distance moduli derived from different light curve fitters.  Although this dataset contains the Pantheon  as a subsample, we  use it as a control sample to check the robustness of the results we obtain using the Pantheon dataset alone.  In what follows, we will refer to the CF3 subsample with SNIa-based distances as CF3sn and indicate the complementary set with  the acronym CF3g.

\begin{figure*}
\includegraphics[scale=0.37]{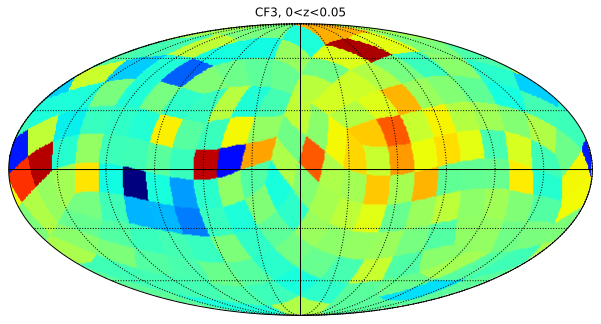}
\includegraphics[scale=0.37]{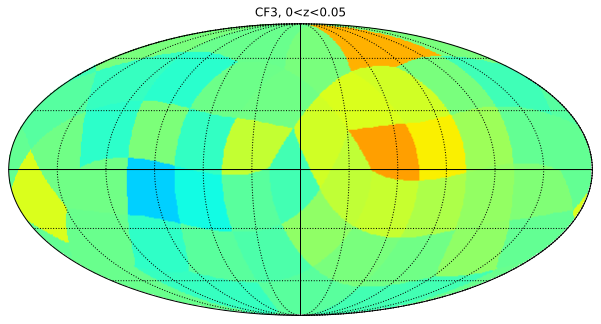}
\includegraphics[scale=0.37]{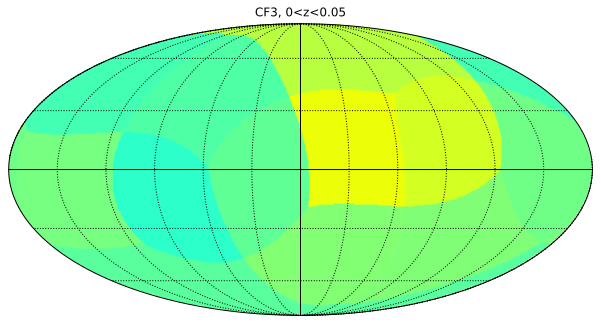}
\\
\includegraphics[scale=0.8]{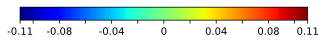}
\\
\includegraphics[scale = 0.37]{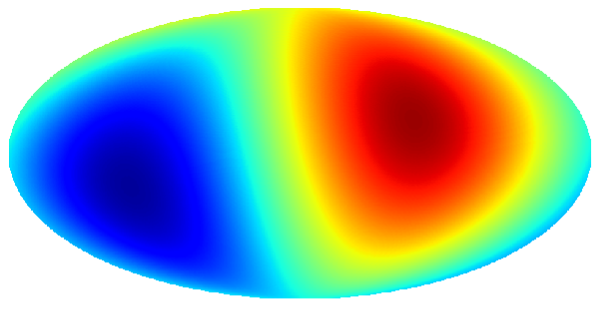}
\includegraphics[scale = 0.37]{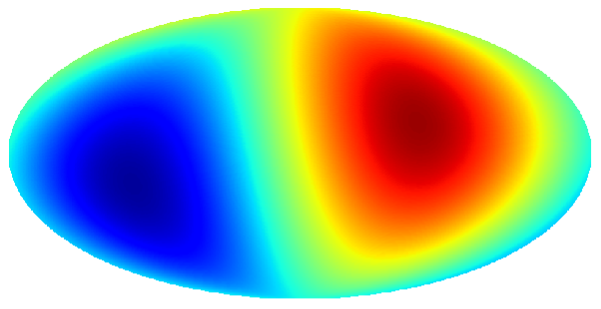}
\includegraphics[scale = 0.37]{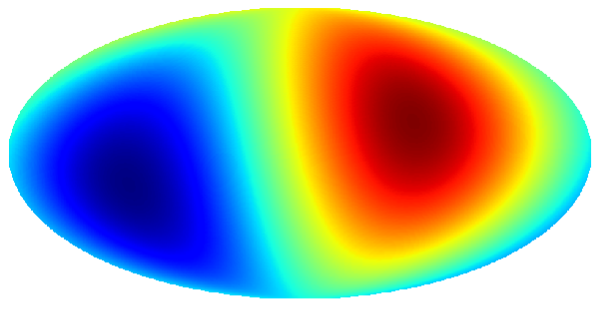}
\includegraphics[scale = 0.37]{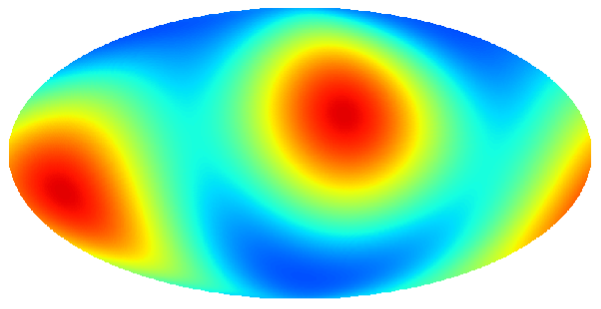}
\includegraphics[scale = 0.37]{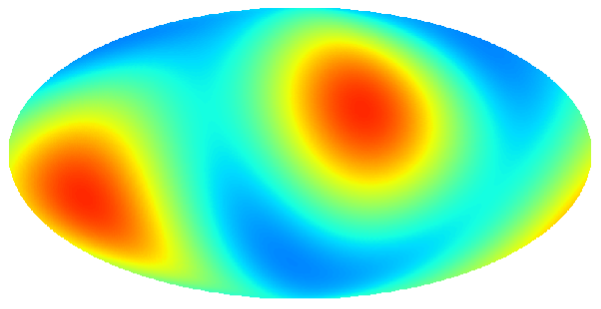}
\includegraphics[scale = 0.37]{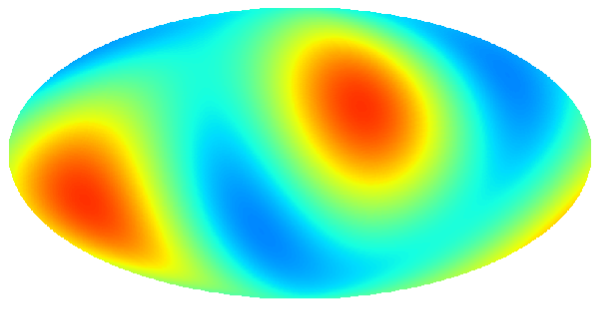}
\includegraphics[scale = 0.37]{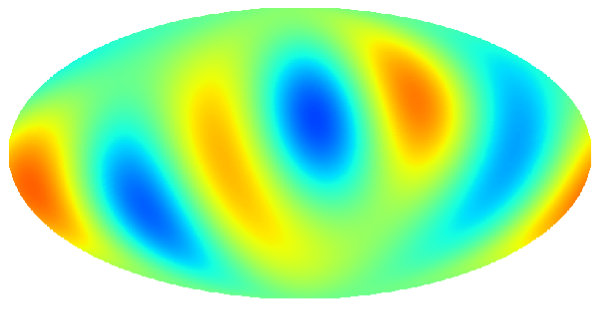}
\includegraphics[scale = 0.37]{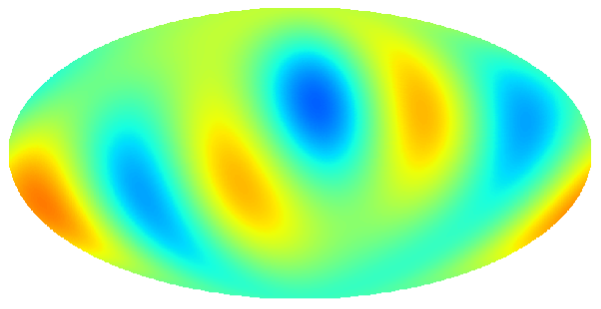}
\;\;\;\;\;\;\;\;\;\;\;\;\;\;\;\;\;\;\;\;\;\;\;\;\;\;\;\;\;\;\;\;\;\;\;\;\;\;\;\;\;\;\;\;\;\;\;\;\;\;\;\;\;\;\;\;\;\;\;\;\;\;\;\;\;\;
\\
\includegraphics[scale = 0.8]{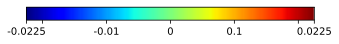}
\centering
\caption{{\it Upper panel:}
the angular $\eta$ field traced by the  CF3. From left to right are shown different resolution maps corresponding to the tessellation of the sky with 192, 48 and 12 HEALPix cells. 
The dipole $\eta_1$ (second panel from the top), quadrupole $\eta_2$ (third panel from the top), and octupole $\eta_3$ (bottom panel) components are also shown.}
\label{Fig_hpixall48}
\end{figure*}

\begin{table*}[]
\scalebox{0.85}{
\begin{tabular}{|c|c|c|c|c|c|c|c|c|c|c|c|c|c|c|c|c|}
\hline
Sample                                                              & $N_{pix}$  & $l_d$      & $b_d$     & \begin{tabular}[c]{@{}c@{}}$\hat{C}_1$ \\ ($10^{-4}$)\end{tabular} & \begin{tabular}[c]{@{}c@{}}$\frac{\eta_{1_{max}}-|\eta_{1_{min}}|}{2}$\\ ($10^{-2}$)\end{tabular} & \begin{tabular}[c]{@{}c@{}}p-value\\ ($\%$)\end{tabular} & $l_q$      & $b_q$      & \begin{tabular}[c]{@{}c@{}}$\hat{C}_2$\\ ($10^{-4}$)\end{tabular} & \begin{tabular}[c]{@{}c@{}}$\frac{\eta_{2_{max}}-|\eta_{2_{min}}|}{2}$\\ ($10^{-2}$)\end{tabular} & \begin{tabular}[c]{@{}c@{}}p-value\\ ($\%$)\end{tabular} & $l_t$     & $b_t$    & \begin{tabular}[c]{@{}c@{}}$\hat{C}_3$\\ ($10^{-4}$)\end{tabular} & \begin{tabular}[c]{@{}c@{}}$\frac{\eta_{3_{max}}-|\eta_{3_{min}}|}{2}$\\ ($10^{-2}$)\end{tabular} & \begin{tabular}[c]{@{}c@{}}p-value\\ ($\%$)\end{tabular} \\ \hline \hline
\begin{tabular}[c]{@{}c@{}}CF3\\ {[}0.01, 0.05{]}\end{tabular}      & 192  &  $291\pm15$  & $12\pm7$  & $3.0\pm1.6$                                                  & 1.6                                                                                               & 0.03                                                        & $323\pm16$ & $9\pm4$    & $2.4\pm0.9$                                                 & 1.9                                                                                               & 0.12                                                       & $289\pm24$ & $16\pm15$ & $0.1\pm0.4$                                                 & 1.2                                                                                               & 30.44                         
\\ \hline \hline
\begin{tabular}[c]{@{}c@{}}CF3\\ {[}0.01, 0.05{]}\end{tabular}      & 48    &  $283\pm6$  & $12\pm5$  & $5.3\pm0.8$                                                  & 1.9                                                                                               & $< 0.01$                                                        & $310\pm11$ & $4\pm8$    & $0.9\pm0.3$                                                 & 1.1                                                                                               & $< 0.01$                                                        & $284\pm7$ & $12\pm5$ & $0.5\pm0.2$                                                 & 1.3                                                                                               & $0.01$                                                        \\ \hline
\begin{tabular}[c]{@{}c@{}}CF3g\\ {[}0.01, 0.05{]}\end{tabular}     & 48      & $286\pm7$  & $4\pm6$   & $7.0\pm1.0$                                                  & 2.0                                                                                               & $< 0.01$                                                        & $338\pm8$  & $22\pm5$   & $1.1\pm0.4$                                                 & $1.3$                                                                                             & $< 0.01$                                                        & $255\pm9$ & $11\pm5$ & $0.7\pm0.2$                                                 & 1.5                                                                                               & 0.01                                                        \\ \hline \hline
\begin{tabular}[c]{@{}c@{}}CF3\\ {[}0.01, 0.05{]}\end{tabular}      & 12  &  $285\pm5$  & $11\pm4$  & $5.1\pm0.8$                                                  & 1.9                                                                                               & $< 0.01$                                                         & $308\pm7$  & $1\pm7$    & $1\pm0.3$                                                   & 1.1                                                                                               & $< 0.01$                                                         & -         & -        & -                                                           & -                                                                                                 & -                                                        \\ \hline
\begin{tabular}[c]{@{}c@{}}CF3g\\ {[}0.01, 0.05{]}\end{tabular}     & 12   &  $296\pm6$  & $18\pm5$  & $4.0\pm0.6$                                                  & 1.7                                                                                               & $< 0.01$                                                        & $323\pm34$ & $2\pm17$   & $1.3\pm0.4$                                                 & 1.4                                                                                               & $< 0.01$                                                        & -         & -        & -                                                           & -                                                                                                 & -                                                        \\ \hline
\begin{tabular}[c]{@{}c@{}}CF3sn\\ {[}0.01, 0.05{]}\end{tabular}    & 12      & $322\pm23$ & $-8\pm18$ & $3.7\pm1.5$                                                  & 1.5                                                                                               & 0.27                                                      & $343\pm15$ & $-8\pm10$  & $1.7\pm1.4$                                                 & 1.7                                                                                               & 2.90                                                      & -         & -        & -                                                           & -                                                                                                 & -                                                        \\ \hline
\begin{tabular}[c]{@{}c@{}}Pantheon\\ {[}0.01, 0.05{]}\end{tabular} & 12      &  $334\pm42$ & $6\pm20$ & $3.5\pm2.7$                                                  & $1.6$                                                                                             & 4.37                                                        & $337$      & $-5$      & $0.6\pm1.9$                                                 & 1.6                                                                                               & 33.33                                                       & -         & -        & -                                                           & -                                                                                                 & -                                                        \\ \hline \hline
\begin{tabular}[c]{@{}c@{}}CF3\\ {[}0.01, 0.03{]}\end{tabular}     & 12      & $279\pm5$  & $12\pm5$  & $7.8\pm1.0$                                                  & $2.3$                                                                                             & $< 0.01$                                                        & $310\pm8$  & $11\pm6$   & $2.9\pm0.6$                                                 & 1.9                                                                                               & $< 0.01$                                                      & -         & -        & -                                                           & -                                                                                                 & -                                                        \\ \hline
\begin{tabular}[c]{@{}c@{}}CF3\\ {[}0.03, 0.05{]}\end{tabular}     & 12     &  $301\pm15$ & $10\pm14$ & $1.1\pm0.7$                                                  & $1.0$                                                                                             & 0.03                                                     & $277\pm28$ & $-12\pm11$ & $0.9\pm0.3$                                                 & 1.0                                                                                               & 2.04                                                      & -         & -        & -                                                           & -                                                                                                 & -                                                        \\ \hline
\end{tabular}}
\caption{Parameters of the spherical harmonic decomposition. Central values are not the bare value returned by the spherical harmonic estimator but are obtained  after subtracting systematic effect determined by means of 1000  Monte Carlo simulations (the bias factors given in TABLE \ref{Tab_3} in  Appendix \ref{app_MC}).
Statistical errors are also determined using Monte Carlo simulations (see TABLE \ref{Tab_4} in Appendix \ref{app_MC}). We quote also the direction for the dipole, and direction of the peaks of the $\ell=2$ and $3$ multipoles which are close to the direction of the dipole. The p-value is computed by using 10000 Monte Carlo simulations. For Pantheon, the error in the direction of the quadrupole is not quoted  because it is larger the 180 degrees.}
  \label{Tab_2}
\end{table*}

\section{Results}\label{sec_results}
 
In this section, we present and comment on the results obtained by applying the formalism  to various datasets. The relevant parameters of  the SH analysis  of the expansion rate field are quoted in TABLE \ref{Tab_2}. Note that  for each entry in this table we do not report the bare value returned by the spherical harmonic estimator, but that obtained  after subtracting the systematic  bias factors quoted in TABLE \ref{Tab_3} of Appendix \ref{app_MC}. The quoted statistical uncertainties  are those estimated numerically by means of 1000 Monte Carlo simulations, although,  as we have highlighted in the previous section, analytical estimations are effectively comparable (see TABLE \ref{Tab_4} in Appendix \ref{app_MC}).

FIG. \ref{Fig_hpixall48} shows the $\eta$ field for the CF3 sample  tessellated according to different resolutions (192, 48 and 12 pixels). 
Smoothing mainly affects the errors with which the relevant SH parameters are estimated, with the error generally decreasing as the number of pixels decreases (as judged on the basis of systematic and statistical errors determined by Monte Carlo simulations,  see tables \ref{Tab_3} and  \ref{Tab_4} in Appendix \ref{app_MC}).  
The central values of the SH parameter, instead, are statistically stable: they  fluctuate from one reconstruction to the next, 
but the discrepancies are within what is  expected from a random sampling of a common underlying Gaussian distribution. The results are thus globally independent of the pixelization strategy  adopted.

However, a valid pixelization scheme for one sample may not be suitable for reconstructing the signal in another catalog. 
A 48-pixel smoothing is for example less than optimal for sizing even the dipole of the sparse SNIa sample. Only when the analysis is performed  by tessellating the sky at lower resolution (12 pixels) we can unpack the information contained in the lowest multipoles. The price to pay is that now the higher multipoles  cannot be estimated: already the  multipole $\ell=3$ becomes now too noisy and therefore unresolved.

The strongest contribution to the signal is provided by the dipole term, whose maximal intensity is about 1\% of  the signal locked in the normalizing term $\log H_0$,  i.e. $\sim 4.5\%$ of $H_0$.
The power locked in the dipole (as determined using the 
the CF3g sample) is  $\hat{C}_1= (4.0 \pm  0.6)\cdot 10^{-4}$, an estimation characterised by a high signal-to-noise ratio (snr  $\sim 6.6$). 
Consistently, this value is in excellent agreement (well within  1$\sigma$) with that estimated from both  the CF3sn and the Pantheon sample ( $(3.7\pm 1.5)\cdot 10^{-4} $ and $(3.5\pm 2.7)\cdot 10^{-4}$ respectively). 

Due to the increased statistical power, the angular position of the dipolar anisotropy axis is better resolved when  the whole CF3 sample  is analyzed ($l_d = 285 \pm 5, b_d = 11 \pm 4$). The 95\% confidence level (C.L.) region falls within the 68\% C.L. uncertainty region defined by any one of the SNIa samples. We remark that this direction is not far away from that of the Shapley Concentration which also roughly coincides with that  of the Great Attractor \cite{Tully:2014gfa}. We will further develop the physical implications of this preferential axis when we discuss bulk flows in section \ref{sec_bulkf}.

We then ask whether this long-range dipole correlation found in each data sample is statistically significant.  To answer the question we compare our hypothesis (occurrence of a truly physical anisotropic dipole) against the probability of the occurrence of a dipole, as a statistical fluctuation, in a  model where the fluctuations in $H_0$ are uncorrelated. To this end, we perform Monte Carlo simulations  to reconstruct the  $p$-value statistics
to invalidate our null-hypothesis against observed data.
Specifically,  we consider a model of  the expansion rate fluctuation $\eta$ that contains only the monopole  and no  higher order terms and  generate 10000 Monte Carlo mock catalogs simulating each data set. This is done  by replacing the model distances with a fictitious one randomly drawn from a Gaussian distribution $G(\eta,\sigma_{\eta})$, where $\sigma_\eta$ is the observational error quoted for each object in the various data samples. 
For each simulation, we calculate the power in the resulting dipole   ($\hat{C}_1^{sim}$) and  compare  it with the observed one $\hat{C}_1^{obs}$.  We then estimate the frequency $p$ with which   $\hat{C}_1^{sim} > \hat{C}_1^{obs}$. A standard rule of thumb consists in rejecting the null hypothesis, {\it i.e.} refute the statistical significance of the signal we observe,  if $p>5\%$. On the contrary, we find that $p$ is virtually zero, {\it i.e.}  $p<10^{-4}$ for CF3g and also critically low  $p =0.3\%$ and $4.4\%$ for CF3sn and the Pantheon sample respectively (see TABLE \ref{Tab_2}), thus confirming the non-accidental nature of the dipolar anisotropy. 

We further note that the dipole  in the expansion rate fluctuations  is consistently tracked by the various samples of galaxies and supernovae.  Its intensity and direction agree fairly well and 
fluctuations from one catalog to the next  are within what is  expected from a random sampling of a common underlying value. 
Since the distances and the associated measurement errors are reconstructed using different and independent methods and calibrations for galaxies and supernovae, it is difficult to interpret these signals as simple statistical fluctuations and to relegate this coincidence to the role of a fluke. These results, on the contrary, seem to suggest that both the galaxy sample and the SNIa sample trace the same anisotropic background expansion rate, despite differences in the uncertainties with which the   distance moduli are determined and in the sampling frequency of large-scale structures.

\begin{figure*}

\includegraphics[scale = 0.5]{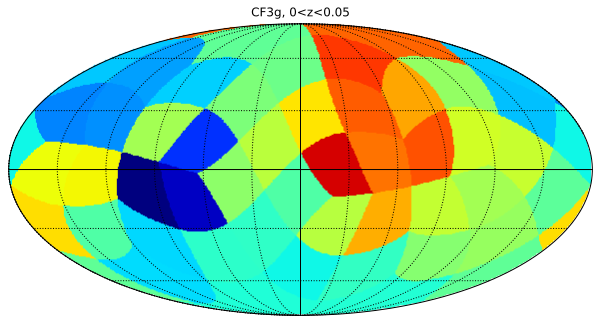}
\includegraphics[scale = 0.5]{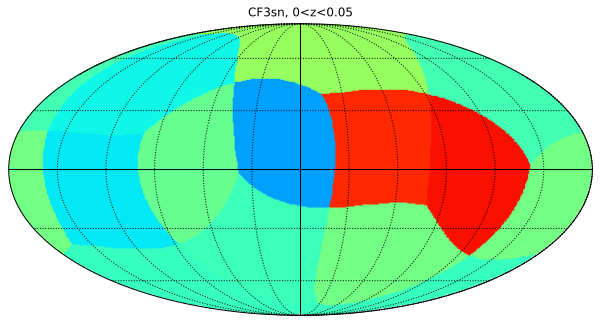}
\\
\includegraphics[scale = 0.8]{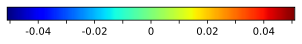}
\\
\includegraphics[scale = 0.5]{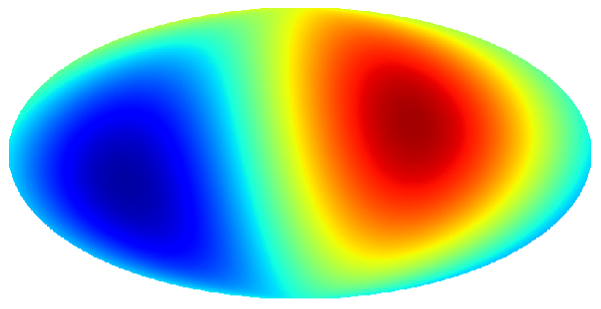}
\includegraphics[scale = 0.5]{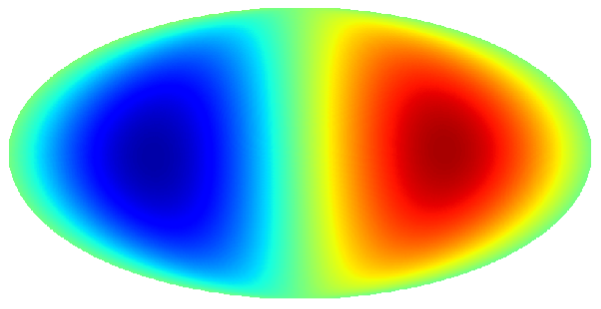}
\\
\includegraphics[scale = 0.5]{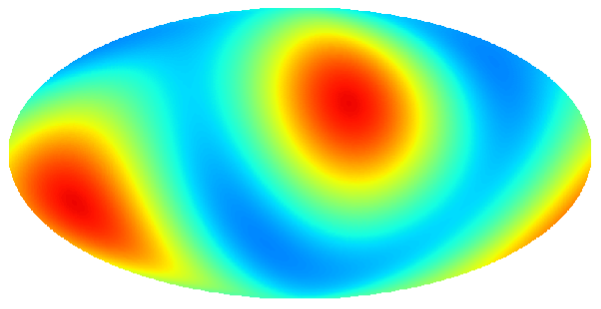}
\includegraphics[scale = 0.5]{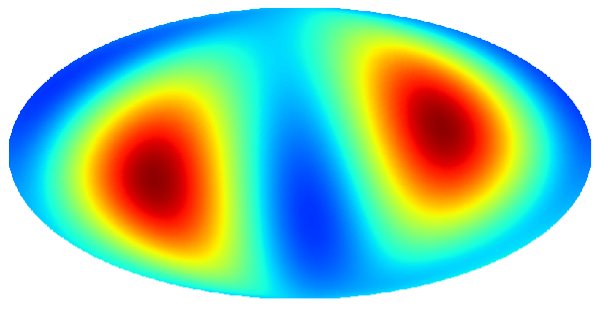}
\\
\includegraphics[scale = 0.5]{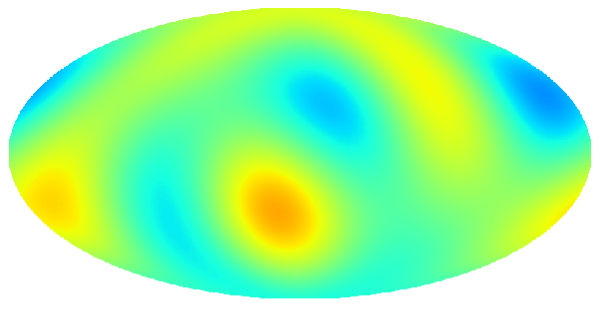}
\;\;\;\;\;\;\;\;\;\;\;\;\;\;\;\;\;\;\;\;\;\;\;\;\;\;\;\;\;\;\;\;\;\;\;\;\;\;\;\;\;\;\;\;\;\;\;\;\;\;\;\;\;\;\;\;\;\;\;\;\;\;\;\;\;\;\;\;\;\;\;\;\;\;\;\;\;\;\;\;\;\;\;\;\;\;\;\;
\\
\includegraphics[scale = 0.8]{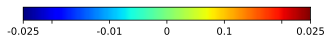}
\centering
\caption{ {\it Left panel:}
the $\eta$ field  tessellated into 48 pixels and traced by the CF3g sample. 
The dipole $\eta_1$ (second panel from the top), quadrupole $\eta_2$ (third panel from the top), and octupole $\eta_3$ (bottom panel) components are also shown.
{\it Right panel:} same as above, but now the expansion rate fluctuation field is tessellated into 12 pixels and traced by the  CF3sn sample.}
\label{Fig_hpixall12}
\end{figure*}

We find that the contribution of the quadrupole component to the anisotropies observed in the $\eta$ expansion field is significant both in terms of its amplitude, which is comparable to that of the dipole ( $\it{max}(\eta_2)\sim \it{max} (\eta_1)$), and in terms of its statistical significance (measured by the signal-to-noise ratio ($snr \sim 3.25$ for CF3g  and $snr \sim 1.2$ for CF3sn). 
Again, a p-value analysis of
its significance performed simulating 1000 mock catalogs, still
confirms that the chances of this signal being a statistical artifact are
minimal (the probability of reconstructing a quadrupole larger than 
the observed one due to a random fluctuation is effectively zero for CF3g and smaller than $3.3\%$ for CF3sn). The probability is reduced even more dramatically if the test is run to answer the question; what is the likelihood of finding a dipole and quadrupole of similar intensity to that observed in the case where the underlying expansion field is uniform?
This result, both independently and in combination with what was found for the dipole term, confirms that the probability of the observed anisotropy being a statistical artifact is very small.

More intriguingly we find that the maximum of the quadrupole signal is aligned with the dipole direction (see FIG. \ref{Fig_hpixall48}).  This peculiar alignment is consistently and independently confirmed by both  galaxy (CF3g) and SNIa (CF3sn) samples (see FIG. \ref{Fig_hpixall12} and TABLE \ref{Tab_2}).  

We highlight the fact that  inferences made with the different samples are consistent, but there is one difference that deserves attention and further investigation: there is no evidence of a quadrupole component in the Pantheon data as judged from the  amplitude of the $\hat{C}_2$ power, both in term of  its  $snr$  and of its $p$-value. 
This result confirms a similar null detection obtained by \cite{Dhawan}  although  in a different redshift range  and using an alternative method which is not based on the Fourier decomposition of the signal but on the maximum likelihood adjustment of quadrupolar coefficients. The possible reasons for this lack of signal will be analysed in section \ref{axialsym}.

\begin{figure*}
\centering
\includegraphics[scale = 0.5]{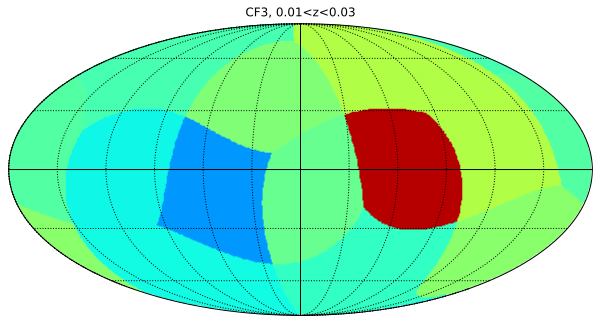}
\includegraphics[scale = 0.5]{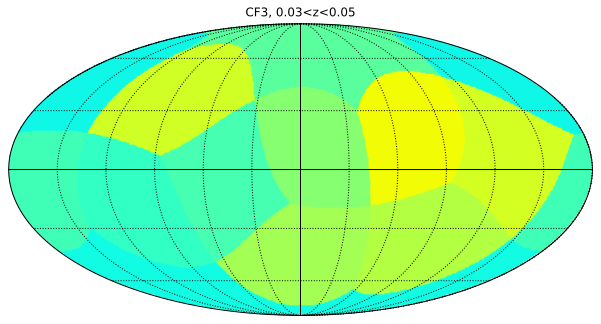}
\\
\includegraphics[scale = 0.8]{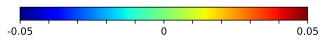}
\\
\includegraphics[scale = 0.5]{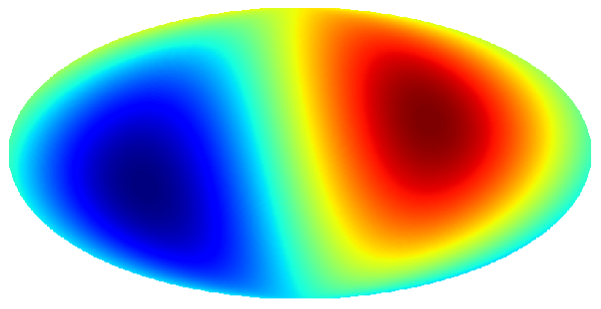}
\includegraphics[scale = 0.5]{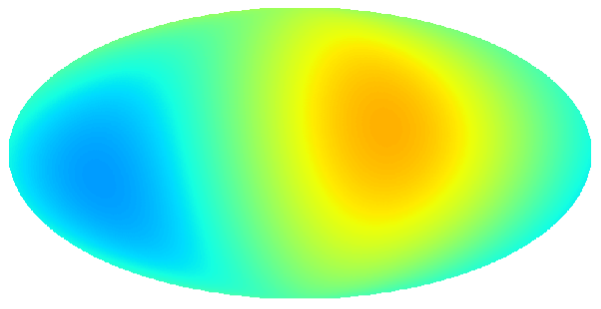}
\includegraphics[scale = 0.5]{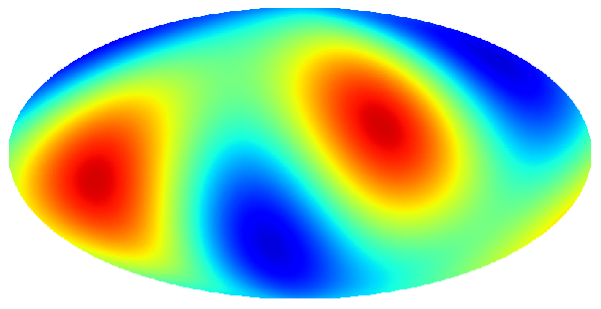}
\includegraphics[scale = 0.5]{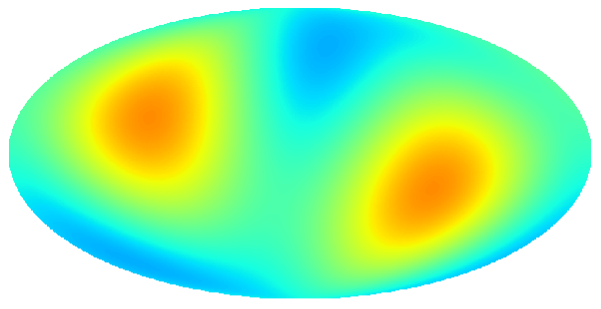}
\includegraphics[scale = 0.8]{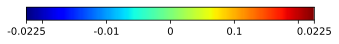}
\centering
\caption{{\it Left panels:} The $\eta$ field (upper),  its  dipolar (center) and quadrupolar (lower) components  traced by the CF3 sample in the redshift interval $0.01\leq z \leq 0.03$. {\it Right panels}
The same as before but for galaxies  in the deeper  redshift range $0.03 \leq z \leq 0.05$.}
\label{Fig_zmulti}
\end{figure*}

Interestingly, we find  evidence ($snr \sim 2.5$) that the contribution of the octupole is also not negligible, at least when the whole CF3 sample is analyzed with a 48 HEALPix pixel smoothing. The power stocked in this component is roughly half that in the quadrupole although the intensity peaks at a value comparable to the maximum intensity of the quadrupole. 
Even more unexpected is the fact that the direction of the maximum of the octupole component ($l_t=284\pm 7 , b_t=12\pm 5$) appears to be aligned with that of the dipole ($l_d =283\pm 6 , b_d=12\pm 5$) and of the quadrupole ($l_q= 310\pm 11,b_q=4\pm8$) (see TABLE \ref{Tab_2}).

The fact that we cannot confirm the octupolar signal independently in the 192 and 48 pixels maps suggests the need to pursue  another  route to quantify its amplitude and direction, and also  to ensure the general robustness of our results. 
 We have thus  determined the value of the SH parameters (direction of the dipole and power spectrum coefficients)  not by Fourier calculating them via eq. (\ref{closedex}) but by considering them as free fitting parameters to be determined by means of  a maximum likelihood analysis.
 In the ideal case of zero noise, the best approximation theorem   ensures that the Fourier coefficients are also those that minimize the difference (in the L2 norm) between the signal and its decomposition on a finite orthonormal basis.
 This statistical approach is independent from any tessellation scheme
 adopted to convert the estimator of $\eta$ (cf. eq. \ref{eta3D2}) into a stochastic field. Indeed, we directly minimize the difference between the discrete random variable $\hat{\eta}({\bf r})$ and the SH model  (\ref{etamodo}). 
 In the presence of errors, however, it does not protect against power spilling in and out from multipolar coefficients of different orders.  
 Method and results are detailed  in Appendix \ref{app_MC} (see tables \ref{Tab_SHfit2} and \ref{Tab_SHfit3} where we quote the least-square best fitting amplitudes together with the $68\%$ c.l.)
 
 Overall, the central values and errors deduced by means of this statistical procedure  are in excellent agreement with the Fourier determinations reported in TABLE (\ref{Tab_2}) and provide independent confirmation of the soundness of our findings. Interestingly, the detection of the octupolar component is confirmed, at least for the whole CF3 sample. The best fitting amplitude $\hat{C}_3^{fit}=0.6\pm0.2$ is consistent with what is determined by the Fourier analysis ($\hat{C}_{3}=0.5\pm 0.2$), and also  the direction of the axis of maximum intensity is in the general direction of the dipole ($l \sim 287, b \sim 9)$. 
 Also the minimum value of the normalized $\chi^2$  statistic is of relevance.
 This implies that the 2D model (\ref{etamodo}) provides an adequate fit to data that depend, in principle, on the $3D$ spatial position. This fact further supports our claim on the validity, in the limit $z<<1$, of the ansatz (\ref{ansatz}). 

The advantage of a SH analysis of anisotropies  over other approaches is that  the fact that Fourier  coefficients represent projections on an orthonormal basis implies that adding additional higher order terms to the SH expansion does not alter the characteristics (amplitude and direction) of the low-order components. 
It is however critical to assess whether their curious structure is the result of a fortuitous averaging coincidence, {\it i.e.} the accidental combination of different patterns at different depths,  or it is a persistent feature independent from the radial depth of the survey. 

To investigate this issue  we repeat the analysis  separately in two redshift intervals $0.01 \leq z \leq 0.03$ and $0.03 \leq z \leq 0.5$. These are the smallest subvolumes that still provide detections with acceptable {\it snr} and low risk of misinterpretation (low p-value), if only for the CF3 sample. Results are shown in Fig. \ref{Fig_zmulti}. 

The direction of the dipole and the quadrupole reconstructed in the two volumes consistently point in the same direction (to within about $1\sigma$) and is also in excellent agreement with the results found for the whole samples. 
This  confirms that the alignment phenomenon is not a random overlap
but rather physical in nature. There is, instead a significant change in the power spectrum amplitude of the multipoles. The $\hat{C}_1$ component decreases by more than about $ 5.5\sigma$ as the redshift of the sample doubles. The same decrement with depth ($\sim 3 \sigma$) is observed for the amplitude of the  $\hat{C}_2$ coefficient.

As a further check, we investigated whether quadrupole or octupole components are spuriously generated by the anisotropic and inhomogeneous distribution of galaxies and supernovae. To this end, we simulate an expansion rate fluctuation field that has a dipole component of the same intensity as that detected by the data, but with zero quadrupole and octupole. We then perturb the model's redshift-independent  distances by adding Gaussian noise that mimics observational errors. 
Finally, we apply the SH analysis pipeline and examine the detection rate of a quadrupole or octupole of the same intensity as those measured from the real data. 
The power in both channels ($\hat{C}_2$ and $\hat{C}_3$) is on average an order of magnitude lower than that inferred from the CF3 dataset. For example,  10000 Monte Carlo replicas of the CF3 sample, analyzed with a smoothing of 48 pixels, give, on average,  $\hat{C}_2=0.1 \pm 0.1$ and  $\hat{C}_3=0.09 \pm 0.06 $. Moreover,  the probability of simulated values even more extreme than those actually observed is effectively zero $(p<10^{-4})$. Furthermore, if a dipole and a quadrupole of the same amplitude as the measured ones are inserted into the $\tilde{H}_0$ map itself, instead of $\eta$, we find, that the systematic and statistical errors as determined by the analysis of the Monte Carlo  simulations induce on average a signal $\hat{C}_3=0.1\pm0.06$. Also the p-value for an even more extreme octupole than the one measured is  negligible ($p=0.01\%$).    

\subsection{Axial symmetry of the multipoles}\label{axialsym}

\begin{figure}
\begin{center}
\includegraphics[scale=0.5]{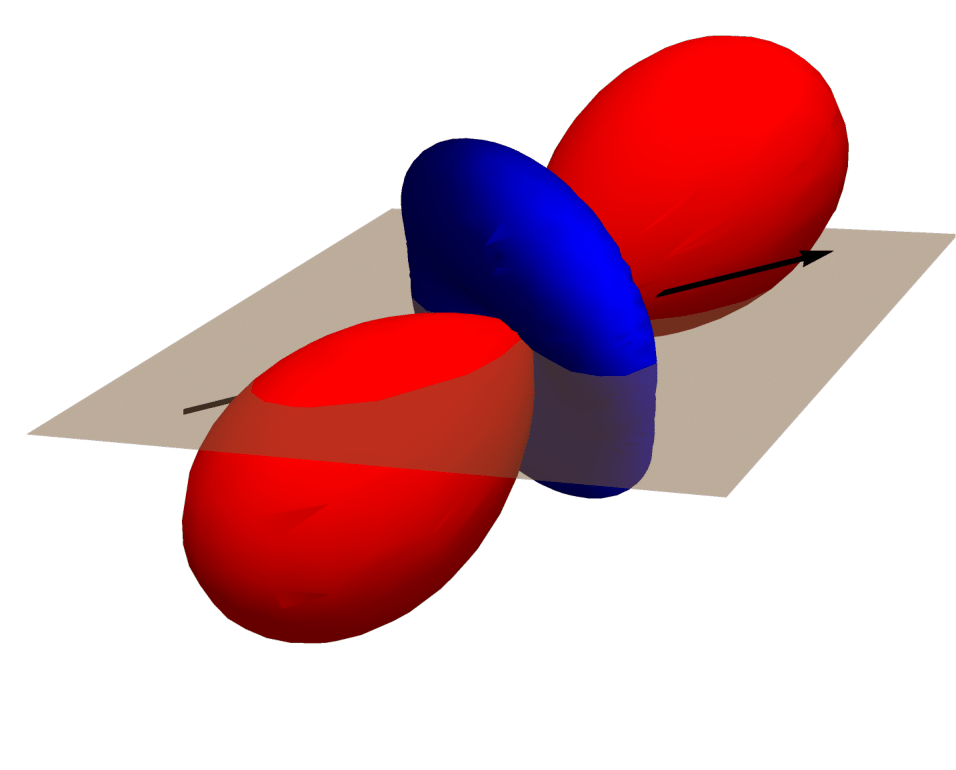}
\\
\includegraphics[scale=0.65]{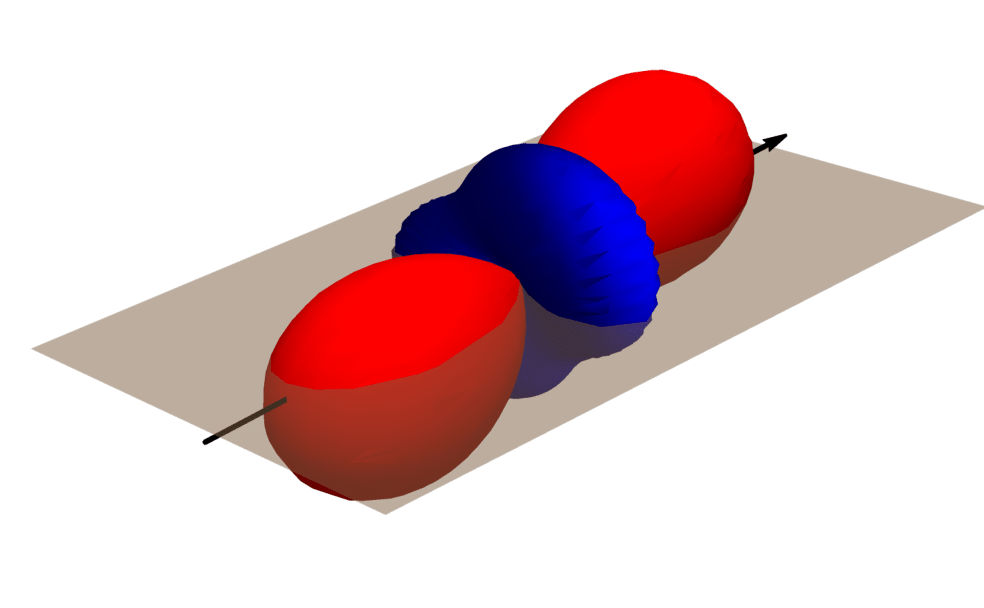}
\caption{The 3D structure of the multipole component $\eta_2$ of the CF3g sample  (upper panel) and for the CF3sn sample  (lower panel). The  radial coordinate represents the absolute  value for $\eta$; the red and blue colors indicate  positive and negative signs respectively.  For reference, the gray surface represents the orientation of the galactic plane while  the black arrow gives the direction of the dipole as reconstructed for each sample.}
\label{Fig_struc3d}
\end{center}
\end{figure}

In FIG. \ref{Fig_struc3d} we show the 
3D structure of the quadrupole  component of the expansion rate fluctuations $\eta$ reconstructed using either the galaxy or the supernovae sample. This figure offers a different perspective on the dipole-quadrupole alignment.  It shows that both  quadrupoles independently reconstructed using galaxy and supernova data   present an axially symmetric configuration which strongly polarizes in the direction defined by the dipole.

This additional symmetry, although physically unexpected since it seems to imply extreme fine-tuning in the local distribution of matter fluctuations, makes it possible to simplify, at least mathematically, the analysis of the anisotropies of the eta field. 
The fact that  the direction ($l_d, b_d$) along which the multipoles aligne (Apex direction)  defines not only a preferred axis, but indeed an  axis of symmetry implies that   the expansion rate fluctuation field $\eta$ effectively depends only on a single variable, the polar  angle $\alpha$   between 
between the line-of-sight of an object  and the direction  ($l_d, b_d$).
A simple expansion of the field in Legendre polynomials $P_\ell(\cos\alpha)$, as opposed to the full spherical harmonics  machinery, is thus enough for seizing the essential functional behavior of the $\eta$ field.

\begin{figure*}
\begin{center}
	\includegraphics[scale=0.5]{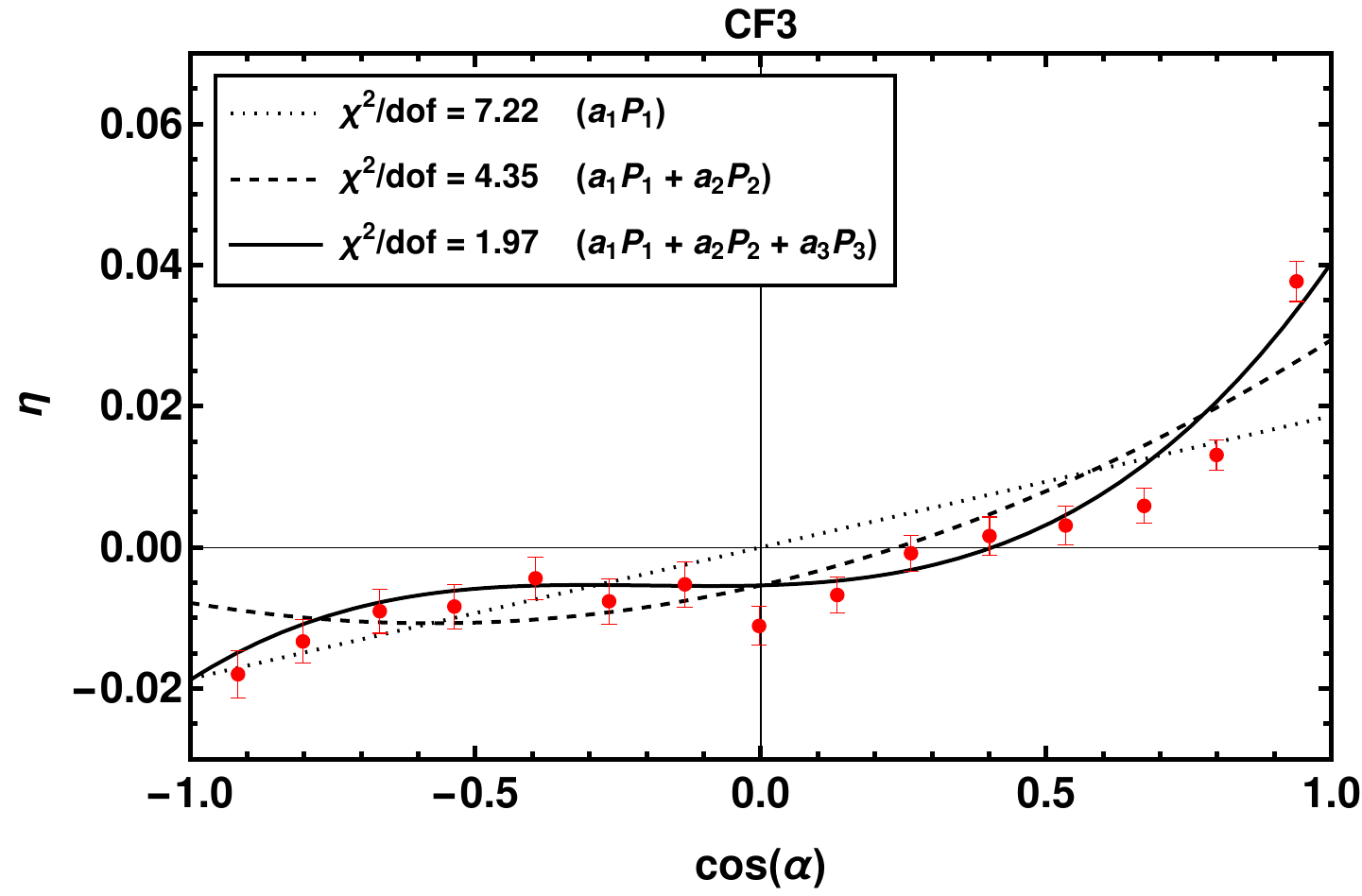}
	\includegraphics[scale=0.5]{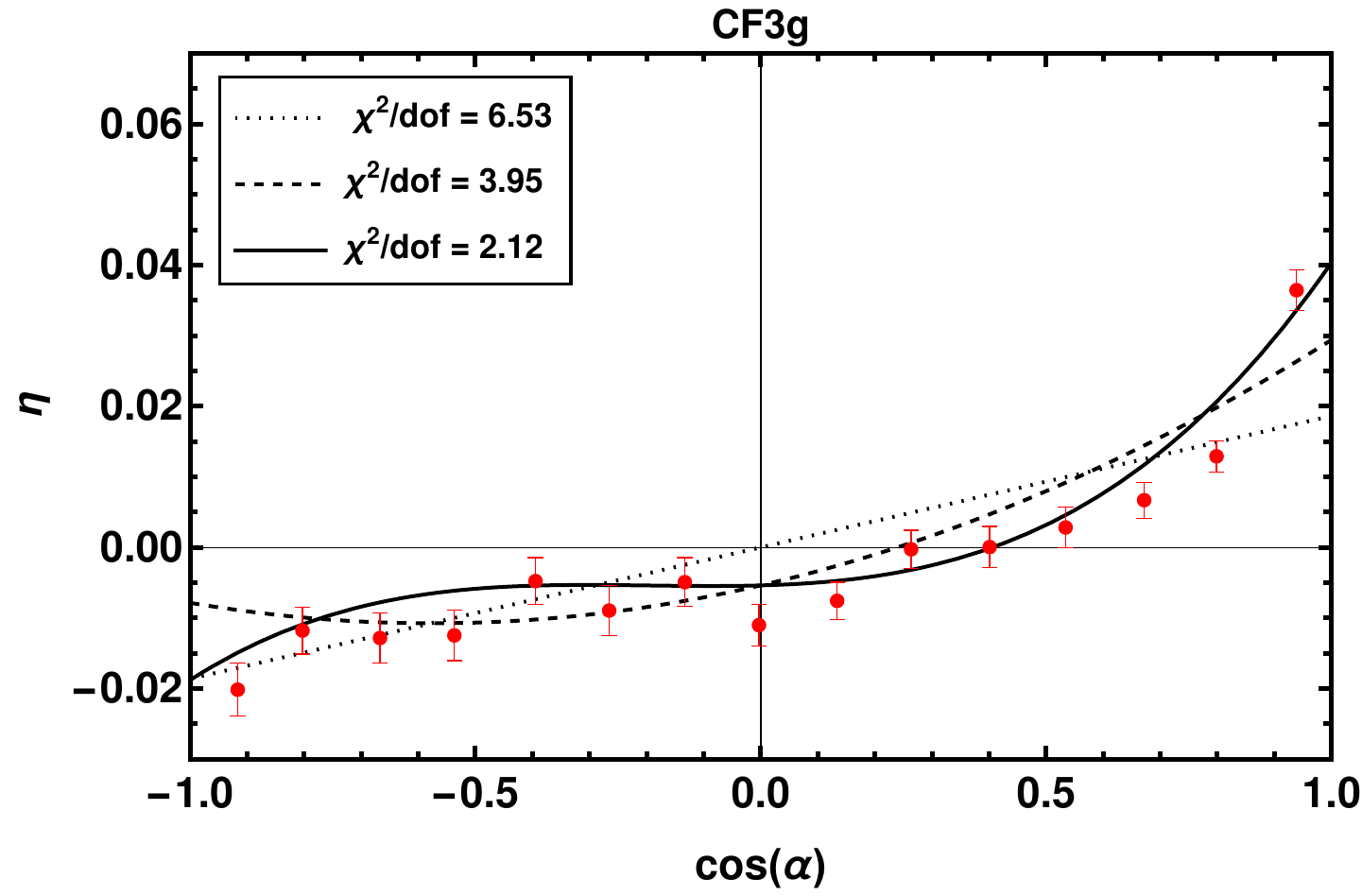}
	\includegraphics[scale=0.5]{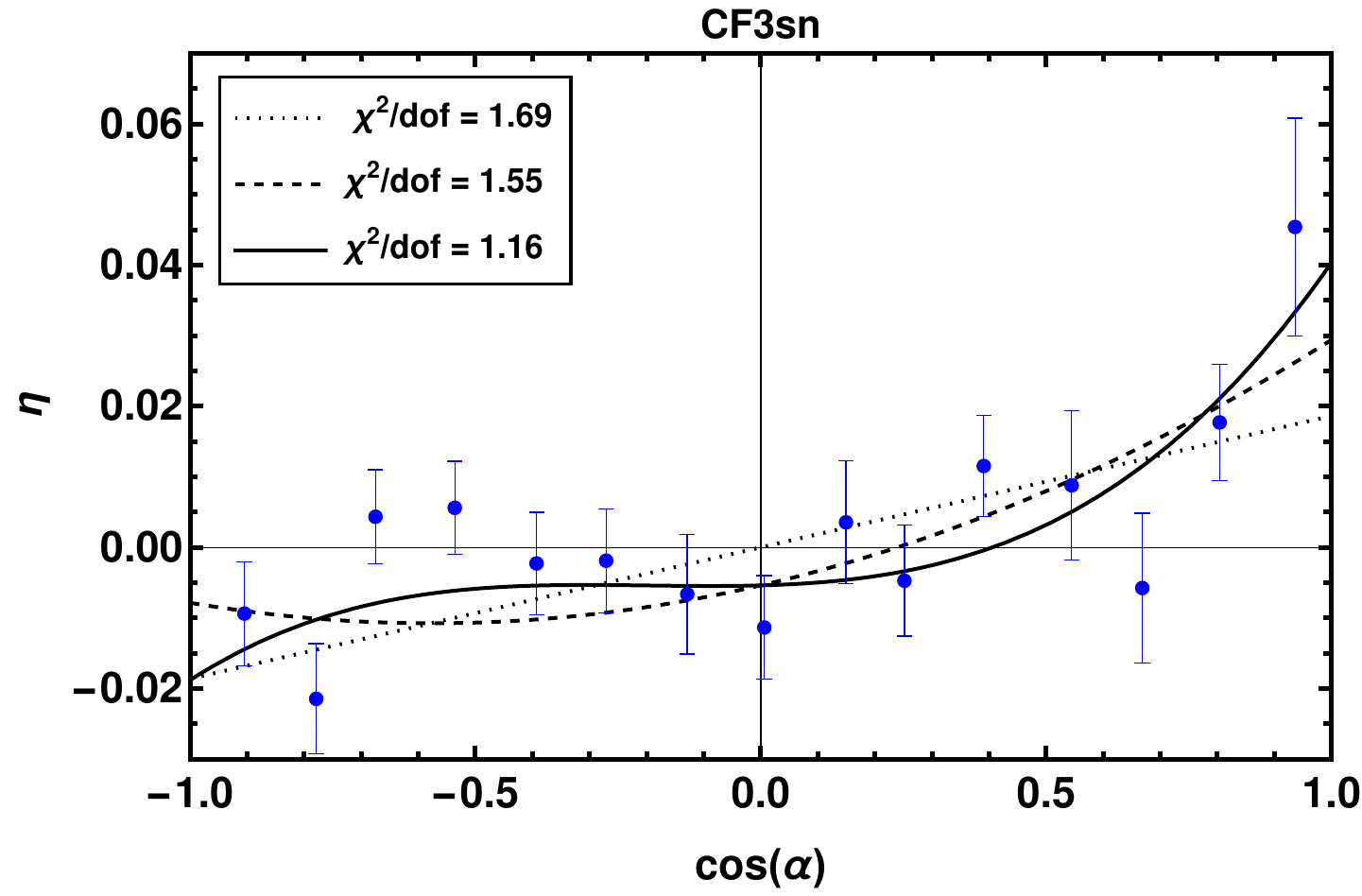}
	\includegraphics[scale=0.5]{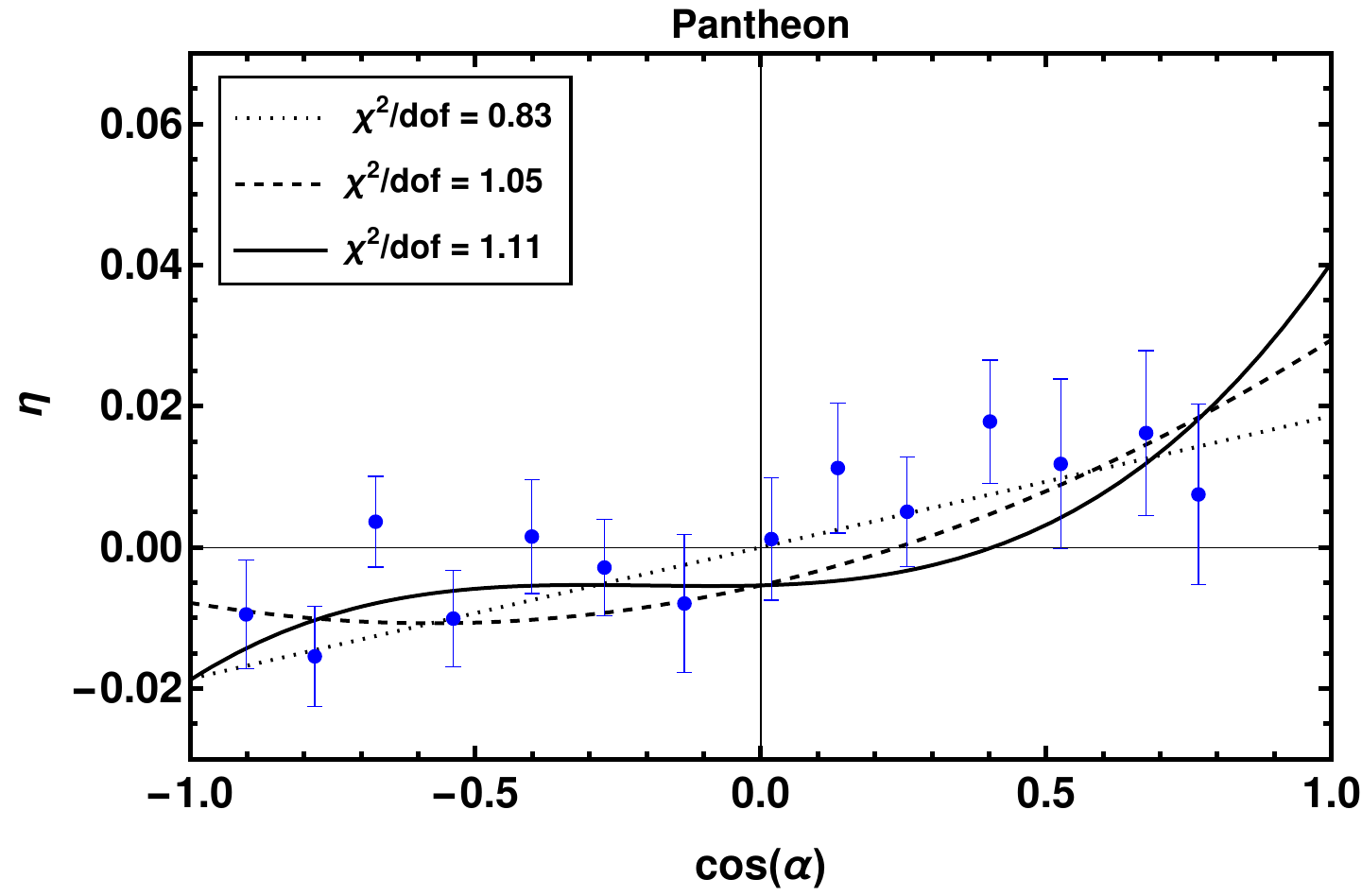}
	\caption{{\it Upper left:} average value of $\eta$, for the CF3 sample,  in open spherical sectors of identical width $\Delta \cos \alpha=2/15$ and angular separation $\alpha$ from the direction $l=285, b=11$. The  dotted line corresponds to the  dipolar model ( $a_1=1.9 \cdot 10^{-2}$), the dashed line includes also the contribution of a quadrupole term ( $a_2=1.1  \cdot 10^{-2}$). The  effects of adding the octupole term ( $a_3=1.1  \cdot 10^{-2}$)  are shown by the solid line. These models are compared to CF3g data  ({\it upper right}), the CF3sn data ({\it lower right}) and the Pantheon data ({\it lower right}). In this latter case, Note the absence of data in the rightmost bin, i.e., along the direction in which the low-order multipoles of the CF3 sample align. }
        \label{CF3_ax_sym_15bins}
		\centering
	\end{center}
\end{figure*}

\begin{figure*}
\begin{center}
	\includegraphics[scale=0.5]{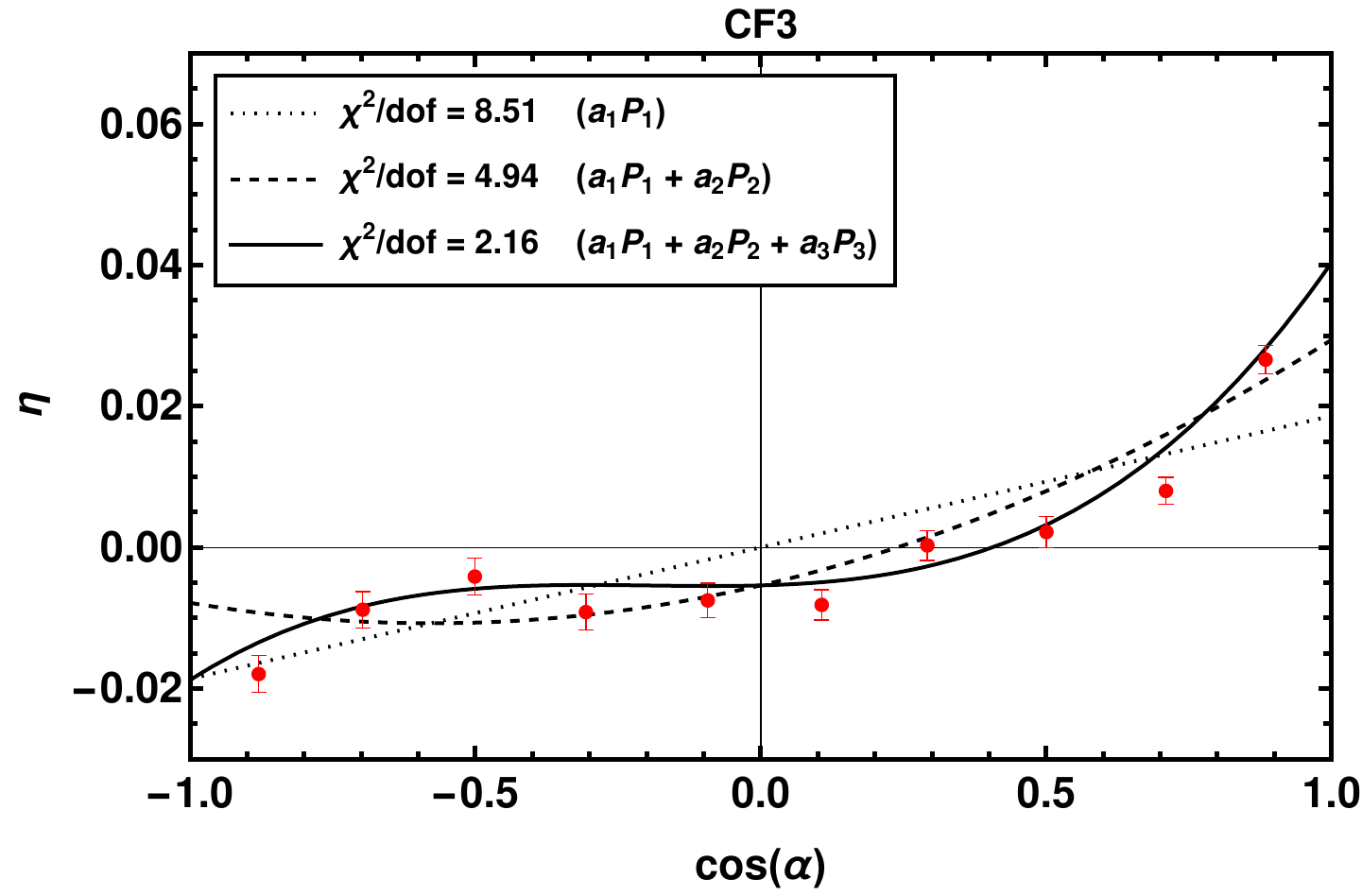}
	\includegraphics[scale=0.5]{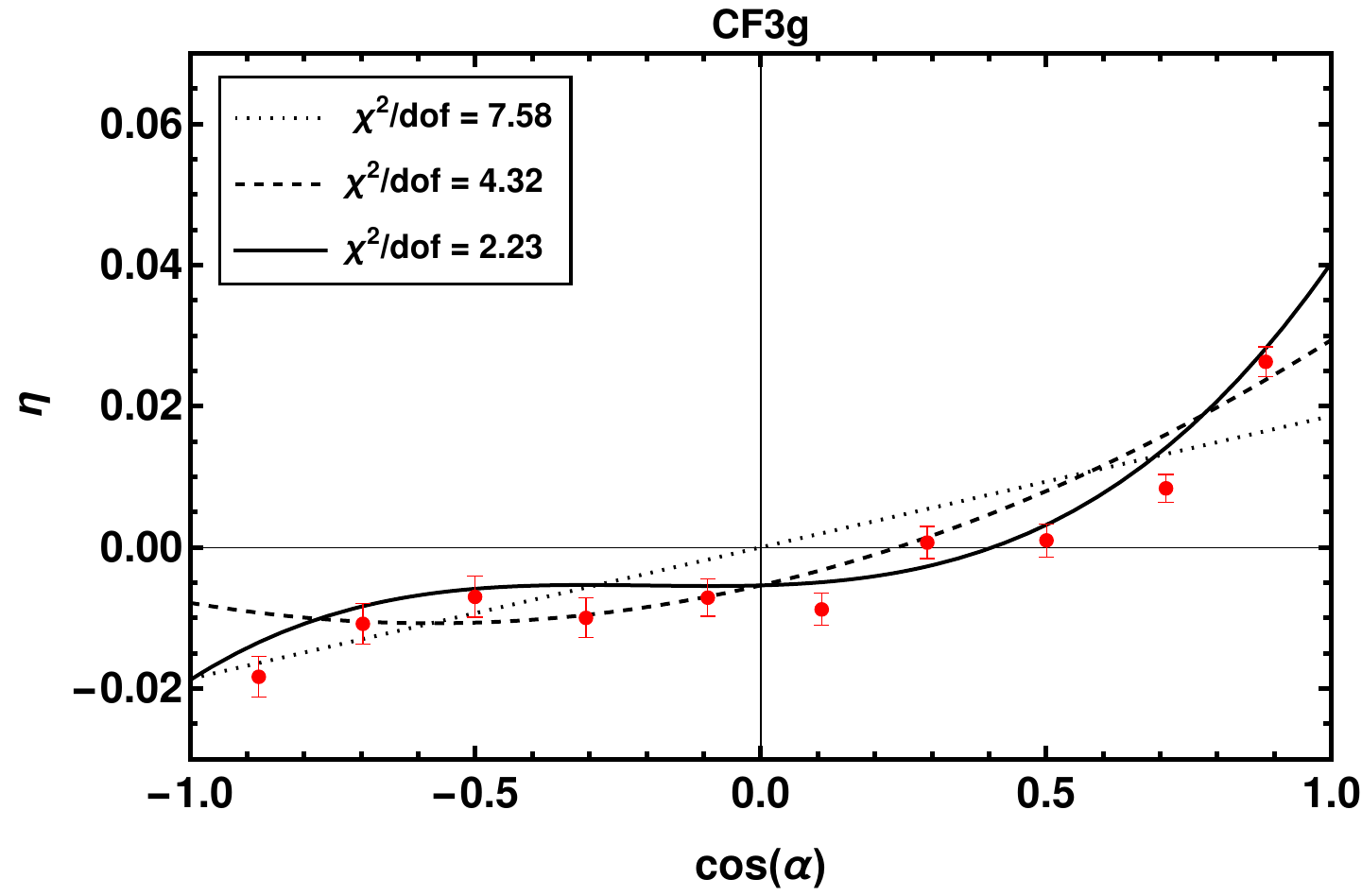}
	\includegraphics[scale=0.5]{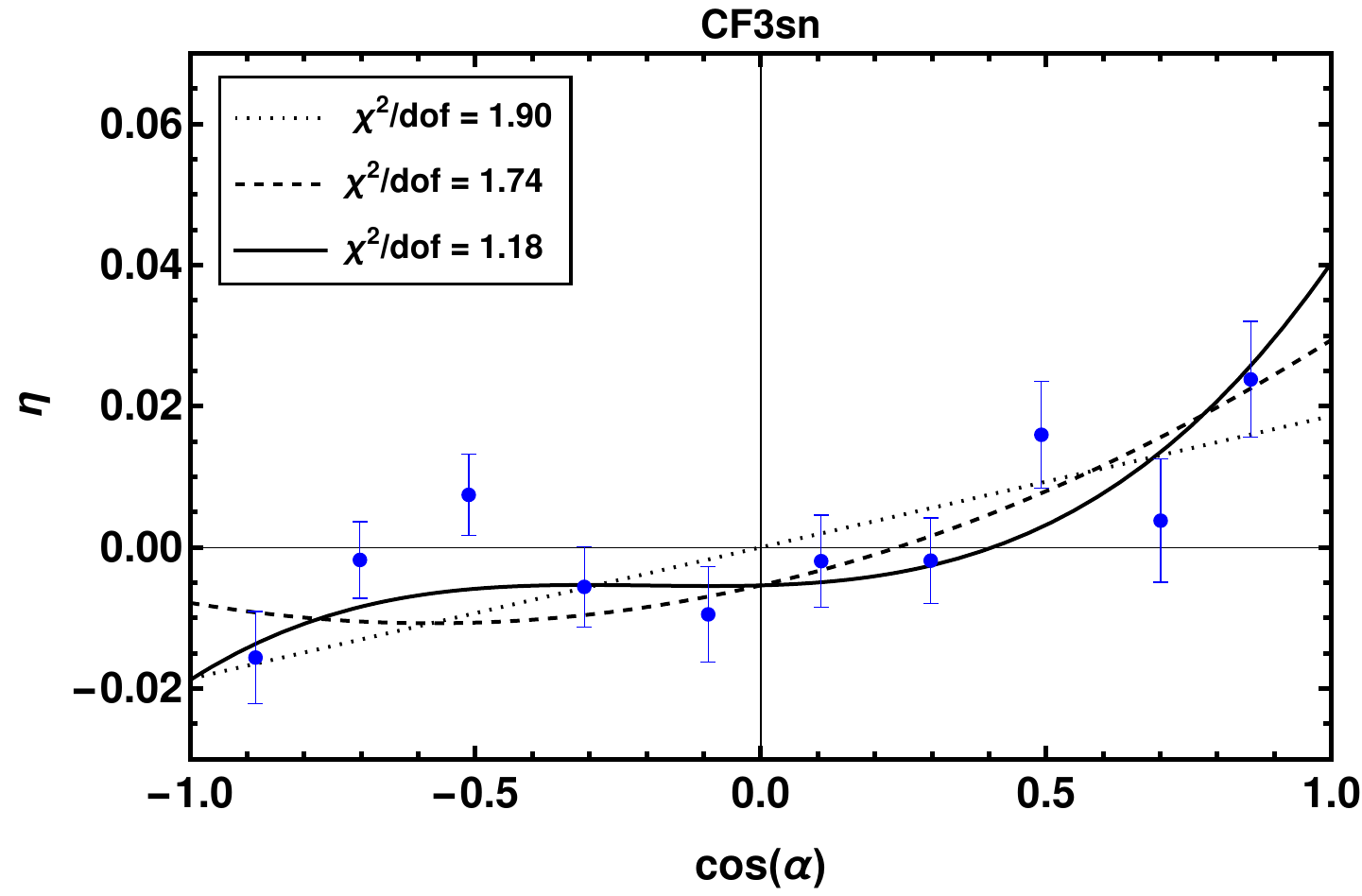}
	\includegraphics[scale=0.5]{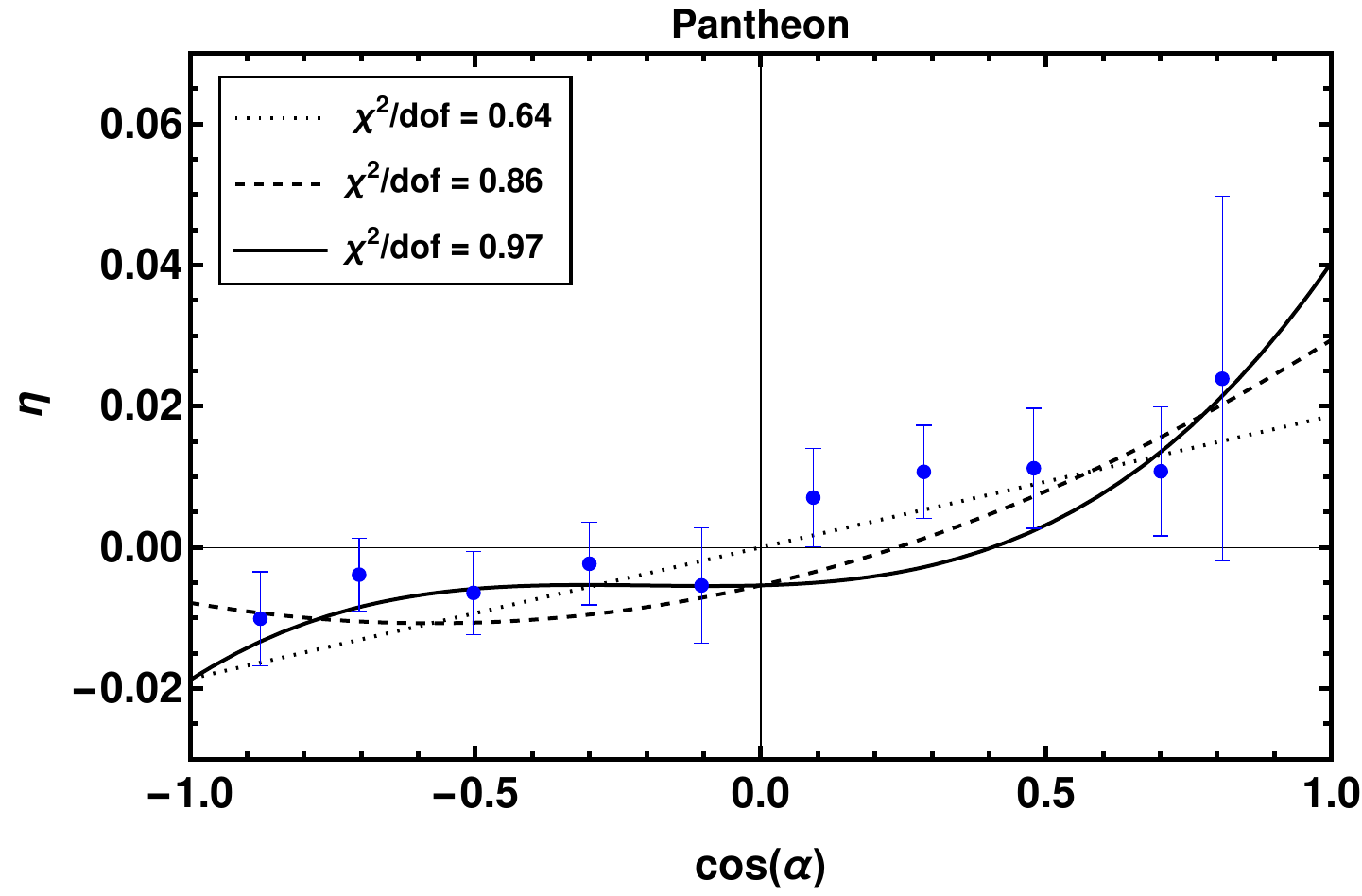}
	\caption{The same as figure \ref{CF3_ax_sym_15bins}, with the same curves, but the number of bins is 10 here.}
        \label{CF3_ax_sym_10bins}
		\centering
	\end{center}
\end{figure*}

The average value $\eta(\alpha)$ reconstructed 
in open spherical sectors  of identical width having the center on the axis of symmetry and angular separation $\alpha$ from the apex direction is shown in FIG \ref{CF3_ax_sym_15bins} and \ref{CF3_ax_sym_10bins}  for the galaxy and SNIa samples. 
		 The Fourier coefficients of the expansion 
\begin{equation}\label{expleg}
		 \eta(\alpha)=\sum_{\ell=1}^{3} a_\ell P_\ell(\cos\alpha),
\end{equation}		 
are computed as 
\begin{equation}
	a_\ell=\frac{2\ell +1}{N_{bins}}\sum_{i=1}^{N_{bins}}\eta(\alpha_i)P_\ell(\cos\alpha_i),
 \label{alPl1}
\end{equation}
		 and the results are quoted in TABLE \ref{table2}.
		 Again, since the Legendre basis is orthogonal,
		 the inclusion of the higher moments does not modify the value already calculated for the lower order terms.
		 As a check of the stability of the estimation, we have also calculated the Legendre coefficients  as
		 $a_\ell=\sqrt{\frac{2\ell +1}{4\pi}}a_{\ell 0}$ where 
		 the coefficients $a_{\ell 0}$ is the SH component (cf. eq. \ref{fouco}) reconstructed  after  rotating the $z$ axis into the position defined by the symmetry axis.
		 We find that the results from the two methods are   effectively indistinguishable.

\begin{table*}[]
\begin{tabular}{|c|c|c|c|c|c|c|c|c|}
\hline
Sample                                                           & $N_{bins}$ & \begin{tabular}[c]{@{}c@{}}$a_1$\\ ($10^{-2}$)\end{tabular} & \begin{tabular}[c]{@{}c@{}}$a_2$\\ ($10^{-2}$)\end{tabular} & \begin{tabular}[c]{@{}c@{}}$a_3$\\ ($10^{-2}$)\end{tabular} & $\chi^{2}_{1}/dof$ & $\chi^{2}_{2}/dof$ & $\chi^{2}_{3}/dof$ & \begin{tabular}[c]{@{}c@{}}$v_b$\\ (km/s)\end{tabular} \\ \hline \hline
\begin{tabular}[c]{@{}c@{}}CF3\\ $0.01<z<0.05$\end{tabular}      & 15         & $1.9\pm0.1$                                                 & $1.1\pm0.2$                                                 & $1.1\pm0.2$                                                 & 7.22               & 4.35               & 1.97               & $318\pm22$                                             \\ \hline
\begin{tabular}[c]{@{}c@{}}CF3g\\ $0.01<z<0.05$\end{tabular}     & 15         & $2.0\pm0.1$                                                 & $1.0\pm0.2$                                                 & $1.0\pm0.2$                                                 & 6.55               & 4.07               & 2.23               & $334\pm24$                                             \\ \hline
\begin{tabular}[c]{@{}c@{}}CF3sn\\ $0.01<z<0.05$\end{tabular}    & 15         & $1.7\pm0.4$                                                 & $1.2\pm0.6$                                                 & $1.4\pm0.6$                                                 & 1.57               & 1.48               & 1.10               & $244\pm64$                                             \\ \hline\hline
\begin{tabular}[c]{@{}c@{}}CF3\\ $0.01<z<0.05$\end{tabular}      & 10         & $1.7\pm0.1$                                                 & $0.9\pm0.2$                                                 & $0.6\pm0.2$                                                 & 8.50               & 4.83               & 2.71               & $292\pm21$                                             \\ \hline
\begin{tabular}[c]{@{}c@{}}CF3g\\ $0.01<z<0.05$\end{tabular}     & 10         & $1.8\pm0.1$                                                 & $0.9\pm0.2$                                                 & $0.5\pm0.2$                                                 & 7.59               & 4.34               & 2.83               & $307\pm23$                                             \\ \hline
\begin{tabular}[c]{@{}c@{}}CF3sn\\ $0.01<z<0.05$\end{tabular}    & 10         & $1.3\pm0.4$                                                 & $0.5\pm0.5$                                                 & $0.4\pm0.5$                                                 & 1.53               & 1.40               & 1.18               & $195\pm57$                                             \\ \hline
\begin{tabular}[c]{@{}c@{}}Pantheon\\ $0.01<z<0.05$\end{tabular} & 10         & $1.6\pm0.7$                                                 & $0.1\pm0.7$                                                & $-0.8\pm0.5$                                                & 0.46               & 0.46               & 0.54               & $243\pm110$                                            \\ \hline
\end{tabular}
\caption{The coefficient of Legendre expansion for 15 and 10 bins in $\cos\alpha$ where $\alpha$ is the angle between the open spherical sectors and the apex  direction  $(l,b)=(285,11)$. The error is calculated by the error of average $\eta$ for each bin. $a_{\ell}$ is calculated by eq. (\ref{alPl1}).}
\label{table2}
\end{table*}

The first thing we remark is that the coefficients computed for different samples are all compatible among themselves, within 1-$\sigma$ uncertainty. For this reason, the data from various samples shown in  figures 
\ref{CF3_ax_sym_15bins} and \ref{CF3_ax_sym_10bins}  are compared to the best-reconstructed template, the one with smaller errors, {\it i.e.} that resulting from the analysis of the whole CF3 sample. This helps in judging the consistency of different datasets in tracing the same underlying expansion rate fluctuation field.    
Concerning the contributions to the signal, we see that the dipole term alone, as expected, on the basis of the results of the previous section provides a poor description  of the angular modulation of $\eta$. This is true for both the CF3g and CF3sn samples. An analysis of the goodness of fit with a reduced $\chi^2$ statistic  gives $6.55$ and $1.57$ respectively, i.e. a probability $p \sim 0$ and $p \sim 7\%$ respectively of being wrong in rejecting the dipolar model. Although the latter value is formally higher than the traditional rejection threshold of $5\%$, and thus the risk of rejecting a good model is not negligible, it is equally true that once the quadrupole and octupole terms calculated for the CF3 sample are added to the model, the description of the CF3sn data improves, as evidenced by the systematic decrease in the $\chi^2$ statistic.

As already discussed in section \ref{sec_results} the Pantheon sample provides an exception, in that it does not show sign of a quadrupolar or higher order multipole components. A simple dipole  optimally captures the angular modulation of the fluctuations of the expansion rate $\eta$ ($\chi^2/dof= 0.66)$. Moreover, by including the quadrupole and octupole terms, the goodness of fit systematically degrades. 
The lack of signal is explained by the concomitance of two effects: the larger error bars with which the $\eta$ observable is reconstructed from the sparse Pantheon sample and also, importantly given the alignment between the dipole and quadrupole found in the CF3g and CF3sn samples, the lack of objects along such a critical direction in the Pantheon sample.

\subsection{Bulk motion model}\label{sec_bulkf}

The perturbation theory of the standard cosmological model provides a framework for interpreting our results. 
If peculiar velocity are random and uncorrelated in a given angular direction,  the average expansion field  $\eta$ vanishes in that direction. Consider instead a peculiar velocity field $v_b$  which is constant in  both direction and amplitude over a typical scale $R$. If we  choose $\alpha$ as measuring the angle between its direction and the line of sight,  the expansion field is predicted to vary  
as (cf. eq. \ref{lpte})
\begin{equation}
\eta(\alpha) =\frac{v_b}{\ln 10} \langle (1+z)/z \rangle \cos \alpha.
\label{prede}
\end{equation}
Suppose that  $\langle (1+z)/z\rangle$, the average  over the volume
subtended by open spherical sectors of angular separation $\alpha$ from the direction of the bulk motion, does not depend on $\alpha$, which is a fairly good approximation for large samples. Then, by comparing (\ref{prede}) with (\ref{expleg}), 
we deduce that the bi-dimensional expansion  field $\eta$ is compatible with being the sky projected 
realisation of a three-dimensional  bulk flow model.  The amplitude of the bulk velocity  follows  from the amplitude of the dipolar parameter $a_1$ (the coefficient of the expansion on the Legendre basis $P_1$)  
\begin{equation}
v_b= \frac{a_1 \ln 10}{\langle  (1+z)/z \rangle }.
\label{buve}
\end{equation}
Assuming $500$ km/s  as a  typical value for the peculiar velocity of galaxies, we expect  the latter relation to apply fairly well  for  objects with  $ z\ge 0.01$, those considered in our analysis.
Note that in this picture, the amplitude of the bulk is controlled by the amplitude of  the dipolar parameter $a_1$ and also by the depth of the survey volume.

\begin{figure}
\begin{center}
	\includegraphics[scale=0.43]{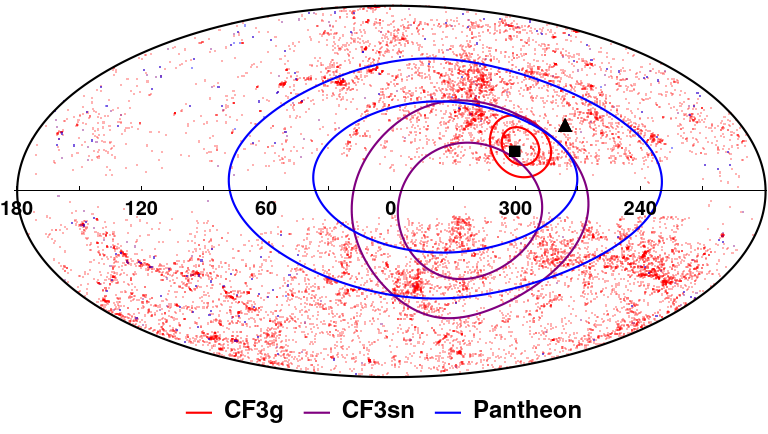}
			\caption{
			Likelihood contours ($1 \sigma$ and $2 \sigma$) for the direction of the dipole reconstructed using 12 pixels for the CF3g (red lines), CF3sn (purple lines) and Pantheon (blue lines) samples. For reference,  the direction of the motion of the barycenter of the Local Group with respect to the CMB $(l,b)\approx (270,27)$ 
			\cite{Fixsen}  is marked (black triangle) together with the direction of its bulk component $(l,b)\approx (299,15)$  (Black square).}
                                 \label{Fig_lbdir}
		\centering
	\end{center}
\end{figure}

The direction of the bulk motion for the three samples is shown in Figure \ref{Fig_lbdir}. This direction results from separating in the SH analysis the direction of the dipole from that of other higher order multipoles. 
However,  due to the alignment of the lower multipoles, 
the direction of the bulk coincides fairly well with the direction of maximum anisotropy in the $\eta$ maps.

It is interesting to note that the direction of the bulk flow of the CF3g sample
agrees remarkably  well ( $\sim 4 ^{\circ}$ apart)  with that of the bulk component of  the Local Group velocity ($\sim 455$ km/s in the direction $l \sim  299, b \sim  15$). This latter is obtained subtracting from the velocity of the LG with respect to the CMB ($\sim 631$ km/s in the direction $l \sim 270, b\sim 27$  \cite{Fixsen}) the local perturbations  due to the infall of the LG onto the Virgo cluster ($\sim 185$ km/s in the direction $l\sim 284, b\sim 74$ (e.g. \cite{Tully2008, mari1998})) 
as well as to the repulsion from the local void  ($259$ km/s in the direction $l \sim 210, b \sim -2$) \cite{Tully2008}).
This alignment suggests that most of the LG motion is indeed generated by sources at a distance not less than the depth of the samples analyzed here, i.e. $z\sim 0.05$.

We also confirm that the bulk motion traced by the CF3g sample points in the general direction in which the most massive concentrations in this region of the sky are located, namely Hydra-Centaurus ($\sim 302, l\sim21$) and the Shapley supercluster $(l\sim 311, b\sim32)$ (which are about $6$ and $19$ degrees away, respectively).

The amplitude of the bulk flow  estimated from the CF3sn  ($v_b=195\pm 57$km/s) is comparable with that measured using the Pantheon sample ($243 \pm 110$km/s) and both values are statistically compatible with the result obtained using the CF3g sample ($307 \pm 23$km/s), although the large error on the latter sample makes agreement almost a foregone conclusion.    These different samples also agree on the direction of the velocity field which is best determined using the total CF3 sample $(l=285\pm 5, 11\pm 4$).

 Comparing our results with those in the literature, we find $2 \sigma$  overlap with those of \cite{Boruah_2020}, who, using a different compilation of redshift-independent  distances covering approximately the same volume of the CF3 sample, found $v_{b}=252 \pm 11$, $l=293 \pm 5, b=14 \pm 5$. They also compare favorably with those of 
  \cite{Hong_2014} ($v_{b}=292 \pm 28$ km/s, $l=296 \pm 16, b=19 \pm 6$),  \cite{turnbull:2011ty} ($v_{b}=249 \pm 76$ km/s, $l=319\pm 18, b=7 \pm 14$)   and \cite{Nusser_2011}   ($257 \pm  44$ km/s $ l=279, b=10$).
  Regarding the results of \cite{Scrimgeour_2015}, 
  although there is general agreement on the bulk amplitude, the directions are misaligned by about $40$ degrees.

\begin{figure*}[]
\begin{center}

\includegraphics[scale=0.46]{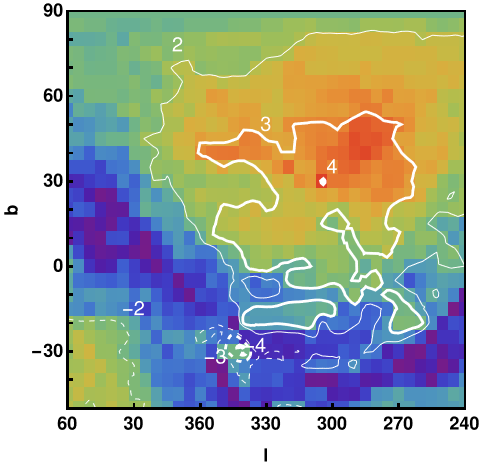}
\includegraphics[scale=0.46]{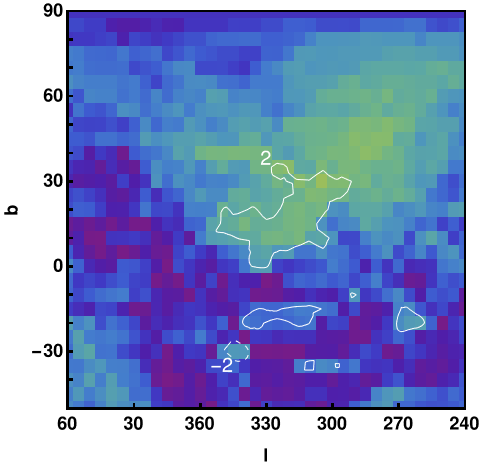}
\includegraphics[scale=0.46]{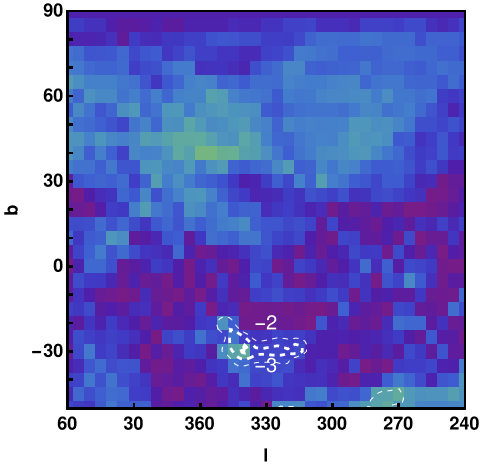}
\\
\includegraphics[scale=0.5]{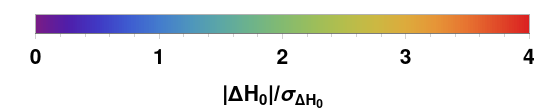}
\\
\vspace{7mm}
\includegraphics[scale=0.37]{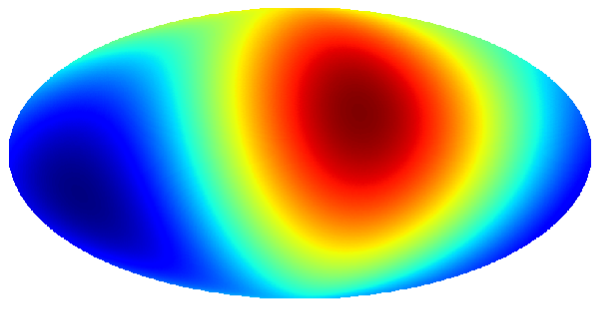}
\includegraphics[scale=0.37]{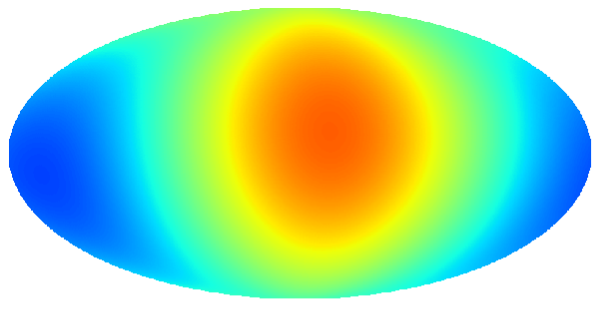}
\includegraphics[scale=0.37]{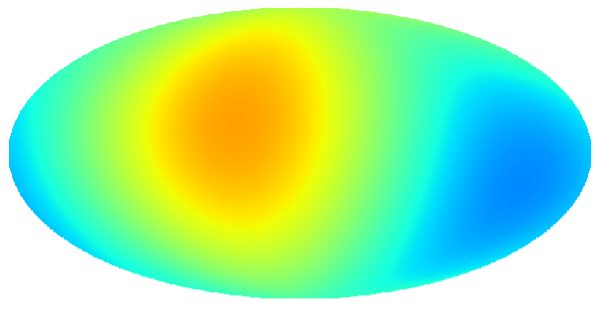}
\\
\includegraphics[scale=0.8]{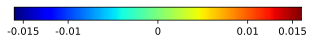}
\caption{{\it Upper left panel:} Solid and dashed lines   represent positive and negative isocontours of $\Delta H_0=H^{apex}-H_0^{antiapex}$ calculated using Pantheon SNe with  angular separation $\leq 60^{\circ}$ from an axis of coordinates $(l,b)$. Different thicknesses correspond to different amplitudes, as indicated by the labels. Isocontours are superimposed to the smoothed signal-to-noise map. {\it Bottom left panel:}  the dipole component (12 pixel map) of the expansion rate fluctuation calculated using the observed redshifts.  
{\it Central panels:} as above, but after subtracting from the redshift of each object the peculiar velocities listed in the Pantheon catalog. {\it Right panels:} as above,  but  after correcting the redshift using the $\eta(\alpha)$ model (\ref{expleg}) with Fourier coefficients computed from the CF3g data (and quoted in the fifth row of TABLE \ref{table2}). 
 The maximum antipodal anisotropy detected  are  $(l,b,\Delta H_0)=(305,30,4.1\pm1.1),(305,30,2.4\pm1.1)$ and $(360,40,1.9\pm0.9)$ respectively. The best reconstructed  dipolar amplitudes ($\Delta H_0=\eta_{1 (max)}H_0 \ln10$) instead  are  $(334,6,2.6\pm1.0)$,  $(341,11,1.6\pm1.1)$ and $(29,7,1.3\pm1.2)$ respectively. }
 
\label{Fig_corp}
\centering
\end{center}
\end{figure*}

Finally, note that in addition to the bulk component, the velocity field also exhibits a quadrupole modulation. The magnitude of this additional contribution is simply given by $v_q=a_2/a_1v_b$. As discussed above, this component is not negligible, at least for samples CF3g and CF3sn.

\subsection{Effects of anisotropies on the Hubble diagram}

As a consequence of the  alignment of the maximal intensities of its dipole, quadrupole and octupole components, the expansion field displays an `apex’ towards which the rate of expansion is  significantly higher than average and an anti-apex where the expansions is coherently lower than the monopole component.    This preferred axis characteristically shows up  in the Hubble diagram. 

FIG. \ref{Fig_corp} shows the difference between the best estimates of $H_0$ deduced from the Hubble diagram analysis in two antipodal directions.  In practice this is achieved by fitting the relation $\mu=5\log(z/H_0)+25$ separately
to the distance modulus of the Pantheon objects falling in two cones of total width $120^{\circ}$ (about $25 \%$ of the sky)  centered on the observer and whose axes point in antipodal directions (apex and antiapex). 
Objects in sky regions where the expansion rate is larger (apex direction) do show systematically lower  values of the distance modulus. 
The maximal deviation observed is  $ \Delta H_0=4.1 \pm 1.1$ km/s/Mpc  in the direction $(l=305, b=30)$. 
For the sake of comparison, the difference   is 
$\Delta H_0 = H_{0}^{apex}-H_{0}^{antiapex}=5.1\pm0.4$ km/s/Mpc for the CF3g sample (in the direction $(l=295, b=5)$), $\Delta H_0 = 5.9\pm 1.2$ km/s/Mpc for the CF3sn dataset (in the direction $(l=280, b=-5)$) all values compatible with that obtained for Pantheon to better than $1 \sigma$.  

In the standard cosmological model, anisotropies in the expansion rate are accounted for by  peculiar motions. When estimating $H_0$ using the Hubble diagram, it is essential to correct the observed redshifts to remove their contributions.  This is done by using velocity maps derived from the linear perturbation  analysis of large-scale mass fluctuations. 
This correction can be applied to the Pantheon catalog since, for each entry, the radial velocities (derived from the analysis of \cite{Carrick_2015} ) are given.
FIG \ref{Fig_corp} shows the maximal antipodal anisotropy $\Delta H_0$  
after such a correction is implemented. 

As it can be seen, factoring out the non-cosmological  contribution induced by peculiar motions reduces the  amplitude of the dipolar anisotropy  (now the maximal change between antipodal directions,  $\Delta H_0= 2.4 \pm 1.1$, is reduced in both amplitude and signal to noise) . Although this removal is in the direction of making the results more consistent with what the standard model predicts (see, for a similar conclusion, \cite{Appleby:2014kea}), the remaining anisotropic signal still has a non-trivial structure that makes it incompatible with being a residual fluctuation of random nature. 
In fact, the largest residual deviations are also the most statistically significant, as can be seen by contrasting the map of the residual signal  with the signal-to-noise ratio maps (see the upper central panel of FIG \ref{Fig_corp}). 
The former clearly shows that  for a given distance, redshifts are systematically higher than expected over a large and contiguous region of sky.

A second remarkable fact is that the direction of the dipolar anisotropy  in the Pantheon sample  is not affected by removing the peculiar velocity distortions. 
A spherical harmonic decomposition of the expansion rate fluctuation field $\eta$  (using a  12 pixel tessellation and cosmological redshifts) confirms that the angular structure of the fluctuation is preserved and only its amplitude is rescaled. Indeed, the lower central panel of 
FIG. \ref{Fig_corp} shows that  the 
dipole map reconstructed after correcting the redshifts for peculiar velocities still displays 
an anisotropy axis pointing in the direction $(l=341,b=11)$. This latter is only $22$ degrees away from the direction of the expansion rate reconstructed using the observed (uncorrected) redshifts  ($l=334, b=6 $), and  still close to the direction of the bulk  component of LG motion, which is in turn close to that of the CMB dipole.
However, the  power locked in this component is $\hat{C}_1=1.4\cdot 10^{-4}$
nearly a factor 3 lower than what measured by neglecting peculiar velocity effects ($\hat{C}_1=3.9\cdot 10^{-4})$. 

We  checked whether this residual systematicity was the result of incorrect redshift determinations in the Pantheon catalog. In this respect,  we redid the analysis using the redshift catalog  made available by \cite{Steinhardt} which should also provide updates on the angular position of the host galaxies. The systematic is  still detected, but this time  with an even greater significance. The maximal variation of the Hubble constant between antipodal directions before applying any peculiar velocity correction  is  
$\Delta H_0 =
(4.5 \pm  1.1)$ km/s/Mpc in the direction $(l=305, b=30)$, 
After correction, both the systematic difference  
 ($\Delta H_0 = (2.9 \pm  1.1)$ km/s/Mpc in the direction $(l=305, b=30)$) and  the residual dipolar power($\hat{C}_1=1.7\times10^{-4}$) are larger than in the original Pantheon catalogue.

 This residual signal might indicate flaws in our understanding of gravity on a local scale and might even motivate the exploration of new physics beyond the standard model. However, we will show in the following that these results rather highlight the need for further efforts to calibrate SN data if they are to be used to constrain the value of $H_0$ even more accurately. Mapping the observed redshift $z$ of supernovae into the cosmological redshift $z_c$ by subtracting from each object the peculiar  velocity reconstructed at that spatial position by linear perturbation theory is indeed a far from trivial task. In addition to the uncertainties and systematics inherent to the reconstruction scheme adopted and to the samples of galaxies used to trace the matter density fluctuations, such a reconstruction is model-dependent.  It inevitably involves the choice of a gravitational theory and therefore the  value of cosmological parameters such as, for example,  the matter density parameter $\Omega_m$.

The observable $\eta$, on the other hand, provides a neat  way to take  into account systematic perturbations in the expansion rate in a completely model-independent manner. We have shown that, as expected, galaxies and supernovae  trace, within observational errors, the same underlying anisotropies.   
When estimating $H_0$ via the analysis of the Hubble diagram of  supernovae, it is thus sufficient to remove the $\eta$ fluctuations
estimated from the independent galaxy catalog of redshift-independent distances. 
Indeed  the estimated  amplitude of the distance modulus  is
\begin{equation}
\mu = 5\log \left ( \frac{z}{H_0} \right ) +25-5 \eta
\label{mueta}
\end{equation}
where $z$ is the observed redshift of the SNIa data and where $\eta$ is approximated using the Legendre expansion (\ref{expleg}) with Fourier coefficients computed from the CF3g data (and quoted in the fifth row of TABLE \ref{table2}).

The difference between the best fitting $H_0$ recovered in various antipodal directions by means of eq. \ref{mueta}  is shown  in the upper right panel of FIG. \ref{Fig_corp}. 
The distribution of $\Delta H_0$ values  has an amplitude that is now not only reduced but also more centered on zero. 
Also the spherical harmonic analysis of the residual expansion rate fluctuation field $\eta({\rm pantheon})-\eta({\rm CF3g})$ shows that the axial anisotropy  is now effectively removed: the residual dipolar modulation points in the direction ($l\sim 40, b \sim 15)$) and the  power locked in this component is $\hat{C}_1=0.8\times10^{-4}$, about twice as  small  than the residual signal obtained after correcting redshifts with the peculiar velocity field model.

The advantages of this debiasing scheme are not insignificant. First, we achieve better anisotropy subtraction with a three-parameter model (cf. eq. \ref{expleg}) than using 
corrected distance modules using numerical grids of peculiar velocities. 
In addition, correcting for $\eta$ fluctuations has a significant impact on the estimate of $H_0$. 
The Hubble parameter that best fits the Pantheon sample (in the range $0.01<z<0.05$) is larger by $2 \sigma$ than that determined on the basis of observed redshifts alone ($\delta H_0 = 0.7$ km/s/Mpc). 
 Also the  normalized $\chi^2$ (for 157 degrees of freedom)
improves in a statistically significant way  (from 1.03 to 0.96), signaling that data are better  described by a  linear redshift-distance relation.

\subsection{The scale of anisotropies}

How far does this anisotropic pattern extend?
For galaxy samples, the determination of redshift-independent distances  becomes problematic at high redshift. Only SN data give access to the deep regions of the Universe.  However, the SN sample is very incomplete in its coverage of the sky, a fact that prevents the extension of the study of angular anisotropies beyond the $z=0.05$ limit. 

Many studies seem to agree that the expansion rate is slightly higher in the direction of the generic CMB dipole (see \cite{Aluri:2022hzs} for a review) even at redshifts much higher than those we investigated. 
However, the reported magnitude of this anisotropy depends on the sample analyzed and, for the same sample, seems to depend on the technique used to measure it (e.g., \cite{DengWei2018}). 
The statistical significance of deviations from uniformity is also at stake. For example, with regard to studies of supernovae samples, some papers either indicate their significance, and thus the potential biasing effect on the inference of cosmological parameters (\cite{Krishnan:2021jmh}), or refute them as pure statistical fluctuation (\cite{Zhao:2019azy, Andrade:2018eta}).

\begin{figure}[]
\begin{center}
\includegraphics[scale=0.95]{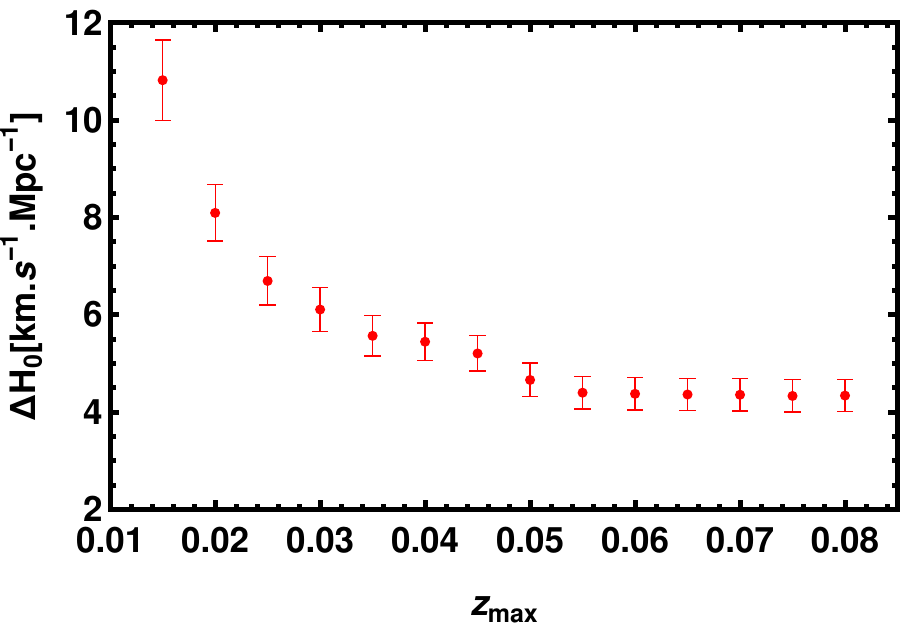}\\
\includegraphics[scale=0.9]{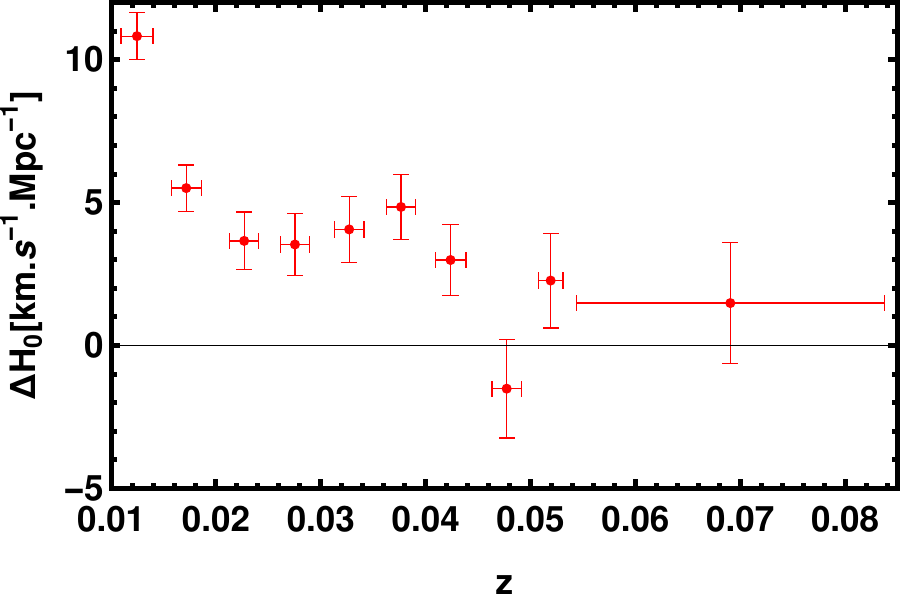}
\caption{Difference between the best fitting antipodal values of $H_0$  for  CF3 objects. The Hubble diagram is constructed using  objects within $60^{\circ}$  from the apex $(l=285, b=11)$ and antiapex directions. Points represent estimates in the cumulative intervals $0.01<z<z_{max}$ (upper panel) and in differential intervals (lower panel).}
\label{FIG_hev_CF3}
\centering
\end{center}
\end{figure}

Fig \ref{FIG_hev_CF3} shows the difference between the best fitting $H_0$ in the apex and antiapex direction also including all objects in the CF3 sample (also those  beyond the $z=0.05$ cut imposed for our analysis, i.e. $N=739$ objects in the apex direction and $N=159$ in the antiapex direction). It is fair to say that 
 the problem of characterising the extent of the radial scale of the local anisotropy  remains unanswered by current data. 
 
It is clear that, from the point of view of the Standard Model, the anisotropy of the Hubble diagram should be suppressed as a function of distance if the amplitude of the peculiar motions has an upper limit. Therefore, it would be surprising if the anisotropy of the expansion rate fluctuation extended consistently beyond $z \sim 0.1$.
However, as detailed above, even if the contribution of the peculiar  velocities is modeled and subtracted, the structure of the residual anisotropy in the expansion rate fluctuations remains unchanged. This is a fact that should not be overlooked, as it could be an indication of possible shortcomings of the standard gravitational paradigm.

\section{Conclusion}\label{sec:conclusions}

The failure to converge on a consensus value of the Hubble's constant triggered investigations into the reliability of geometric descriptions of the local spacetime that deviate from the standard cosmological metric. The question that arises is whether metrics with lower symmetries, while still simple, provide a reliable description of the data in the local 
patch of the Universe where global uniformity is violated. 

We address this problem, from a new angle, trying  to go beyond the standard perturbative 
approaches by which non-uniformities in the local expansion rate are accounted for in the standard model.   To this end, we have designed an observable  that captures, in a statistically unbiased way, the average angular fluctuations in the local expansion rate. The observable $\eta$ is fully model independent not requiring any knowledge about the metric,
the gravity model or  the amount of matter in 
the Universe. On the contrary, it is sensitive, and therefore instrumental to understand local violations of the cosmological principle, i.e. how far the spatial distribution of matter is from
being homogeneous and how far the kinematics of the cosmic expansion is from being isotropic.
In the limit of small fluctuations, a perturbative expansion around the FRW expectation  shows that $\eta \propto \delta H_0/H_0$ i.e. it measures the relative fluctuations in the Hubble parameter.  For this reason, the monopole  of the expansion rate field vanishes and the $\eta$ observable is  practically insensitive to the actual value of $H_0$.

We have estimated the $\eta$ field using catalogs of redshift-independent distances such as the  Cosmicflows-3 galaxy sample and the Pantheon sample of supernovae.  We have then compressed  the resulting signal  into independent spherical harmonics components to better analyse the structure of the anisotropies in the linear redshift-distance relation.
 Particular care was paid to assess the reliability of the reconstructions, either analytically, computing the variance of relevant SH parameters, either numerically by means of Monte Carlo simulations. This last approach has the advantage of also correcting the systematic bias induced by the scarcity and incompleteness of the data used in our analysis as well as by the pixelization strategies applied to them.

All data samples analyzed  consistently suggest the existence of a preferred axis in the local universe ($0.01<z<0.05$),  in  the direction $(l,b)=(285 \pm 5, 11 \pm 4)$, along which the local redshift-distance relation displays a dipolar pattern.
Within the standard model of cosmology, notably its linear perturbative extension, this dipolar modulation of the angular  expansion rate can be understood as the imprint of a large-scale bulk flow model. 
We find that galaxies  inside the survey volume  $r<150 h^{-1 }\rm Mpc$ coherently moves  with an average speed  $v_b=(299 \pm 22)$ km/s. This direction  agrees fairly well 
  with the estimated direction of the bulk component of  the Local Group velocity 
  ($l,b) \approx  (299,15)$ \cite{Tully2008} and confirms  that the LG  participates
in a bulk flow that extends out to distances of at least  $z=0.05.$

Interestingly, the SH analysis suggests that a simple dipole provides a poor representation of the  angular fluctuations in the local expansion rate.  A more sophisticated description  is necessary if we are to properly model the  anisotropies  in the redshift-distance relation.
We find, that  about $50\%$ of the anisotropic signal is contributed by a quadrupole component. This is independently 
confirmed by the CF3g and CF3sn samples, which also consistently show that the 
 axis of maximal expansion of the quadrupole is aligned with the direction of the dipole.  This intriguing feature persists when the sample is split in two and the analysis is repeated in two separate spherical shells of different depths. This evidence allows us to exclude the possibility that the observed alignment is a casual coincidence, that is,  a volume-dependent fluke.  
The analyses of both samples also agree on the shape of the quadrupole. It is a rather axisymmetrical configuration, with no indication of prominent secondary axes.

The findings about the alignement of quadrupole  in the general direction of the dipole  independently confirms  results of \cite{Lilje_1986} or,  more recently,  of \cite{Hoffman:2017ako} who, using  smaller samples, showed that the eigenvector expansion of the shear tensor (a proxy for the quadrupole moment) is aligned in the direction of the Hydra-centaurus/Shapley superclusters. We extend the validity of these results  over a  volume four times larger through a separate analysis the galaxy and SN samples. Furthermore we  find that the octupole is also involved in the alignment.
Despite the large errors induced by the sparsity of the sample and the relatively large  coarse graining of the multipole reconstruction scheme, there is indeed  tentative evidence for the detection of an octupole signal, at least for the CF3 sample of galaxies. Its intensity is of similar amplitude to that of the quadrupole. Its configuration is also quite peculiar, with its  axis of maximal intensity colinear with the axes of the dipole and the quadrupole. 

We note, as a curiosity, that, the direction where the quadrupole of the expansion field has a maximum does not coincide with that where the quadrupole of the CMB temperature fluctuations reaches its maximum. Also their configuration is  different: planar for the CMB and axisymmetric for the local expansion field. 

There are several independent indications confirming the robustness of our findings about  the geometry of the  multipolar structure of the expansion rate fluctuation field: a) the  anisotropies detected in different samples agree both in amplitude and direction, b) the probability that they are random fluctuations is ruled out by montercarlo analyses c) the random alignment of the independent dipole, quadrupole and, tentatively, octupole moments is statistically improbable.  Given that the amplitude of the dipole, quadrupole and octupole components of the fluctuation field scale differently with distance from the gravitational sources ($R, R^3 and R^5$, respectively) one would expect that their peculiar configuration is due either to an extraordinary fine-tuning in the alignment of the sources themselves or to the dominant role of a single structure dominating the gravitational dynamics in the local universe. We note that indeed \cite{Kocevski:2005kr} finds evidence that the LG lies approximately on the line joining the barycenter of three large cosmic structures, the Shapley concentration and the Great Attractor on one side, and a large galactic underdensity in the diametrically opposite direction.  We do not dwell on this and other possible explanations in this article, but will address this in a further study.  

According to the  standard model,  the departures from uniformity are the manifestation of deviations from co-mobility induced by local gravitational fields. 
After factoring out from the Pantheon sample the contribution of these peculiar velocities (reconstructed by applying prescriptions of the theory of linear perturbations to the observed fluctuations in the spatial distribution of galaxies), we observe that the $H_0$ perturbations, although reduced in amplitude, do not distribute as random (Gaussian) residuals.  Curiously, the structure of the expansion rate field  still present an axial anisotropy in the  same general direction as the CMB dipole. We find that the power in the residual dipole component, as quantified by the coefficient $\hat{C}_1$  is reduced only by  about $50\%$, and that, as a consequence, the Hubble parameter still has a sinusoidal modulation of amplitude $\Delta H_0=1.6 \pm 1.1 $km/s/Mpc between the  apex and antiapex directions (as opposed to $\Delta H_0= 2.6\pm1.0$ before any peculiar velocity corrections). 

This residual systematicity does not indicate the failure of the standard model, but rather the need to improve methods of reconstructing peculiar velocities. In this regard we show how to exploit  the expansion rate fluctuation field $\eta$  to subtract redshift anisotropies in a  fully model-independent way. 
In practice, we correct the distance moduli of the Pantheon sample with a neat three-parameter Legendre expansion formula (see eq. (\ref{expleg})) calibrated (and not fitted) using information extracted from the analysis of the  CF3g galaxy sample. 
Despite its simplicity, the approach proves effective: not only is the dipole pattern suppressed by a factor of about five in power, but even its direction is now offset from the general direction of the CMB dipole. The low signal-to-noise ratio of the residuals is also consistent with the virtual absence of any systematicity.
As a result, the best-fit Hubble parameter is 0.7 km/s/Mpc higher than if the distance modulus were calculated with the observed redshifts, and also the goodness of fit, as measured by the $\chi^2_{\nu}$, improves in a statistically significant way.

In addition to the similarities among results inferred from independent samples, there are also differences that require further study. Of these, the most intriguing is the fact that no quadrupole nor octupole component shows up in the Pantheon catalog. This is essentially due to the fact that this catalog contains virtually no objects in the direction where the $l=1,2,3$   multipole aligns.
This also explains why the axisymmetric $\eta$ model calibrated using the CF3g and CF3sn data sets, and which incorporates quadrupolar and octupolar contributions, also effectively explains the purely dipolar anisotropies characteristic of the pantheon data sample. 
It would therefore be very useful to look for SN candidates in this critical region of the sky if one wants to advance towards a more solid understanding of the anisotropies in the local expansion rate. 

To conclude, a key objective has been to show, as a proof-of-concept, the potential power of a new observable, the expansion rate fluctuation $\eta$, both to study  the structure of the  
anisotropies in the redshift-distance relation and to minimize  eventual systematics in
the  locally inferred value of $H_0$. 
 It is thus necessary, in follow-up papers, to build on the current formalism by doing a more intensive data analysis. 
This will include the use of updated and expanded data sets, including Cosmicflows-4 \cite{2020ApJ...902..145K}, Pantheon+ \cite{Riess:2021jrx}. It will also involve parallel analysis of large suites of N-body simulations of local patches of the universe to assess the typicality of the results we have found by taking into account the cosmic variance. 
It will be  also interesting to explore the theoretical consequences of our analysis and interpret the salient features of the local expansion field in terms of alternative metrics (e.g. \cite{Clarkson2010, Macpherson2022}) with the goal of extending  the predictive power of the FRW model to sub-uniform cosmic scales. 

 \vskip 1.truecm

 \noindent {\bf Acknowledgements}.
We would like to thank  Pierre Fleury, J. Larena, Roy Maartens, Federico Piazza and Licia Verde for useful discussions. We also thank the M2 student Pierre Ciccodicola who participated in the initial phase of this work. This study was partially supported  by the \textit{Institut  Physique de l'Univers} (IPHU Grant No. 013/2020)  and by the Programme National GRAM of CNRS/INSU with INP and IN2P3 co-funded by CNES.

\appendix

\section{Dealing with  HEALPix matrices}\label{app_mat}

We estimate the SH coefficients  $a_{\ell m}$ by passing the HEALPix iteration scheme and exploiting, instead, the closed-form expression given by eq. (\ref{closedex}). To compute it in practice,   it is necessary to compress the SH $a_{\ell m}$ coefficients into a 1D matrix, that is, to suppress a labeling index. There are many ways to do this; we do it so that the  row vector $\boldsymbol{a}$ contains as elements

\begin{equation*}
    \begin{aligned}
    \boldsymbol{a}^{T}= & \{a_{00},a_{10},a_{20},\ldots,a_{\ell_{max}0},a_{11},a_{21},\ldots,a_{\ell_{max}1},\ldots,a_{\ell_{max}\ell_{max}},\\& a_{1-1},a_{2-1},\ldots,a_{\ell_{max}-1},\ldots,a_{\ell_{max}-\ell_{max}}\}
    \end{aligned}
\end{equation*}
If we identify with the index $j$ (starts from $1$) each vector component, the relations between $j$ and $l$ and $m$ are 

\begin{equation*}
    m_j=\begin{cases}
          -\llbracket \sqrt{\left(\ell_{max}+\frac{1}{2}\right)^2+2(\ell_{max}+1-j)} -\frac{1}{2}-\ell_{max} \rrbracket \quad , \, j\leq\nu \\
          \llbracket \sqrt{\left(\ell_{max}+\frac{1}{2}\right)^2-2(j-\nu)} -\frac{1}{2}-\ell_{max} \rrbracket \quad ,\, j>\nu \\
     \end{cases}
\end{equation*}

where $\nu$ is the number of $a_{\ell m}$ with $m\geq 0$ until $\ell_{max}$, so $\nu=\frac{1}{2}(\ell_{max}+1)(\ell_{max}+2)$, and $\llbracket\rrbracket$ is the floor function.

\begin{equation*}
    \ell_j=j-\mathcal{N}_{m_j}+\ell_{max}
\end{equation*}
and
\begin{equation*}
    j=\mathcal{N}_{m}+\ell-\ell_{max}
\end{equation*}

where $\mathcal{N}_m$ is

\begin{equation*}
\mathcal{N}_m=\begin{cases}
          m(\ell_{max}+\frac{1}{2})+\ell_{max}+1-\frac{m^2}{2} \quad , \, m\geq0 \\
          (\ell_{max}-m)(\ell_{max}+\frac{1}{2})+\ell_{max}+1-\frac{1}{2}(\ell_{max}^2+m^2) \quad ,\, m<0 \\
     \end{cases}
\end{equation*}
The elements of the matrix $\boldsymbol{A}$ (cf. eq. (\ref{matrixA})) are
thus
\begin{equation*}
    A_{jp}=\frac{4\pi}{N_{pix}}Y_{\ell_j,m_j}^{*}(\theta_p,\phi_p)
\end{equation*}
and its dimension is given by  $(\ell_{max}+1)^2\times N_{pix}$. It follows that 
\begin{equation*}
    \hat{a}_{\ell_j m_j}^{(0)}=\sum_{p=1}^{N_{pix}}A_{jp}\eta(p)
\end{equation*}

Since  $\boldsymbol{S}= \frac{N_{pix}}{4\pi}\boldsymbol{A}^{* T}$, and $\boldsymbol{M}=(\boldsymbol{A}\boldsymbol{S})^{-1}$, we can thus write eq. \ref{closedex} explicitly as

\begin{equation*}
    \hat{a}_{\ell_i m_i}=\sum_{j=1}^{(\ell_{max}+1)^2}M_{ij}a^{(0)}_{\ell_j m_j}.
\end{equation*}
where $\hat{a}_{\ell m}$ are the elements of the vector $\boldsymbol{a}^{(\infty)}$.

\section{Variance of the estimator $\hat{a}_{\ell m}$}\label{app_varalm}

The statistical errors affecting the estimates of the SH coefficients can  be estimated analytically.  We start by noticing that the zeroth order value of the SH coefficients is
given by the equation (\ref{alm0estimate}).
The variance of its real and imaginary parts are 
\begin{equation*}
    V[\Re[\hat{a}_{\ell m}^{(0)}]]=\left(\frac{4\pi}{N_{pix}}\right)^2\sum^{N_{pix}}_{p=1}\sigma^{2}(p)\left(Y_{\ell m}^{*} (\theta_p,0) \cos{(m\phi_p)}\right)^2
\end{equation*}
\begin{equation*}
    V[\Im[\hat{a}_{\ell m}^{(0)}]]=\left(\frac{4\pi}{N_{pix}}\right)^2\sum^{N_{pix}}_{p=1}\sigma^{2}(p)\left(Y_{\ell m}^{*} (\theta_p,0) \sin{(m\phi_p)}\right)^2
\end{equation*}
where 
\begin{equation*}
    \sigma^2(p)=\frac{1}{\sum^{N_{g(\text{in pixel p})}}_{i=1} \frac{1}{\delta_{i}^{2}}}
\end{equation*}
is the variance affecting the estimate of $\eta(p)$, the expansion rate fluctuation  in pixel $p$. Note that complex conjugation is not anymore effective since $Y_{\ell m} (\theta_p,0)$ is real.
Since also $\eta$ is a real field,  then $a_{l-m}=(-1)^m a_{\ell m}^{*}$, which implies that the SH coefficients are not  independent random variables. To overcome this issue we express 
  $a_{\ell m}^{(\infty)}$ as a linear combination of $a_{\ell m}^{(0)}$, with $m\geq0$. To this end we need to decompose $\boldsymbol{M}$ (see eq. \ref{closedex})  into its real  ( $\boldsymbol{\mathcal{R}}$), and imaginary ($\boldsymbol{\mathcal{I}}$) parts, so that 
\begin{equation*}
\begin{aligned}
    \Re[a_{\ell_i m_i}^{(\infty)}]=&\sum_{j=1}^{\ell_{max}+1}\mathcal{R}_{ij}\Re[a^{(0)}_{\ell_j 0}]\\
    &+\sum_{j=\ell_{max}+2}^{\nu}\bigg[
    \left(\mathcal{R}_{ij}+(-1)^{m_j} \mathcal{R}_{i(j+\nu-\ell_{max}-1)} \right)\Re[a^{(0)}_{\ell_j m_j}]\\&-\left(\mathcal{I}_{ij}+(-1)^{m_j+1} \mathcal{I}_{i(j+\nu-\ell_{max}-1)} \right)\Im[a^{(0)}_{\ell_j m_j}]\bigg]
\end{aligned}
\end{equation*}
and
\begin{equation*}
\begin{aligned}
    \Im[a_{\ell_i m_i}^{(\infty)}]=&\sum_{j=1}^{\ell_{max}+1}\mathcal{I}_{ij}\Re[a^{(0)}_{\ell_j 0}]\\
    &+\sum_{j=\ell_{max}+2}^{\nu}\bigg[
    \left(\mathcal{R}_{ij}+(-1)^{m_j+1} \mathcal{R}_{i(j+\nu-\ell_{max}-1)} \right)\Im[a^{(0)}_{\ell_j m_j}]\\&+\left(\mathcal{I}_{ij}+(-1)^{m_j} \mathcal{I}_{i(j+\nu-\ell_{max}-1)} \right)\Re[a^{(0)}_{\ell_j m_j}]\bigg]
\end{aligned}
\end{equation*} 
We thus  obtain 

\begin{equation*}
\begin{aligned}
V[\Re[\hat{a}_{\ell_im_i}]]=&\sum_{j=1}^{\ell_{max}+1}\mathcal{R}_{ij}^{2}V[\Re[a^{(0)}_{\ell_j 0}]]\\
    &+\sum_{j=\ell_{max}+2}^{\nu}\bigg[
    \left(\mathcal{R}_{ij}+(-1)^{m_j} \mathcal{R}_{i(j+\nu-\ell_{max}-1)} \right)^2 V[\Re[a^{(0)}_{\ell_j m_j}]]\\&+\left(\mathcal{I}_{ij}+(-1)^{m_j+1} \mathcal{I}_{i(j+\nu-\ell_{max}-1)} \right)^2 V[\Im[a^{(0)}_{\ell_j m_j}]]\bigg]
\end{aligned}
\end{equation*}
and 
\begin{equation*}
\begin{aligned}
V[\Im[\hat{a}_{\ell_i m_i}]]=&\sum_{j=1}^{\ell_{max}+1}\mathcal{I}_{ij}^{2}V[\Re[a^{(0)}_{\ell_j 0}]]\\
    &+\sum_{j=\ell_{max}+2}^{\nu}\bigg[
    \left(\mathcal{R}_{ij}+(-1)^{m_j+1} \mathcal{R}_{i(j+\nu-\ell_{max}-1)} \right)^2 V[\Im[a^{(0)}_{\ell_j m_j}]]\\&+\left(\mathcal{I}_{ij}+(-1)^{m_j} \mathcal{I}_{i(j+\nu-\ell_{max}-1)} \right)^2V[\Re[a^{(0)}_{\ell_j m_j}]]\bigg]
\end{aligned}
\end{equation*}
and, in the end, the variance  
\begin{equation}
 \sigma_{\ell m}^{2}\equiv V[\hat{a}_{\ell m}]=V[\Re[\hat{a}_{\ell m}]]+V[\Im[\hat{a}_{\ell m}]]
\label{varalmex}
\end{equation}

\section{Monte Carlo estimation of measurement errors}\label{app_MC}
  We determine the errors, both statistical and systematic, that plague the SH reconstruction by means of Monte Carlo simulations. 
  We consider as input model  the   Fourier coefficients  (up to  $\ell_{max}$) measured from the data, and we use them to simulate a fiducial  $\eta$ field. We them randomly perturb the expansion field, at the angular position of the objects,  by means of a Gaussian noise with mean value  $\eta$ and with standard deviation  $\delta$. We construct in this way a suite of 1000 mock catalogs which are tessellated with HEALPix and Fourier transformed in exactly the same way as the data. This means we apply  the same rotation scheme to fill all the pixels with objects, and the same numerical scheme to calculate the power spectrum (up to $\ell_{max}$). The resulting distribution of the output signals, notably the 
  sky coordinates where the dipole and quadrupole signals are maximal and  the amplitude of the power spectrum coefficients $\hat{C}_\ell$ (up to $\ell_{max}$) is finally reconstructed and analyzed to assess the confidence level with which our analysis pipeline retrieves the fiducial input values.

\begin{figure*}
\centering
\includegraphics[width = 3.5in]{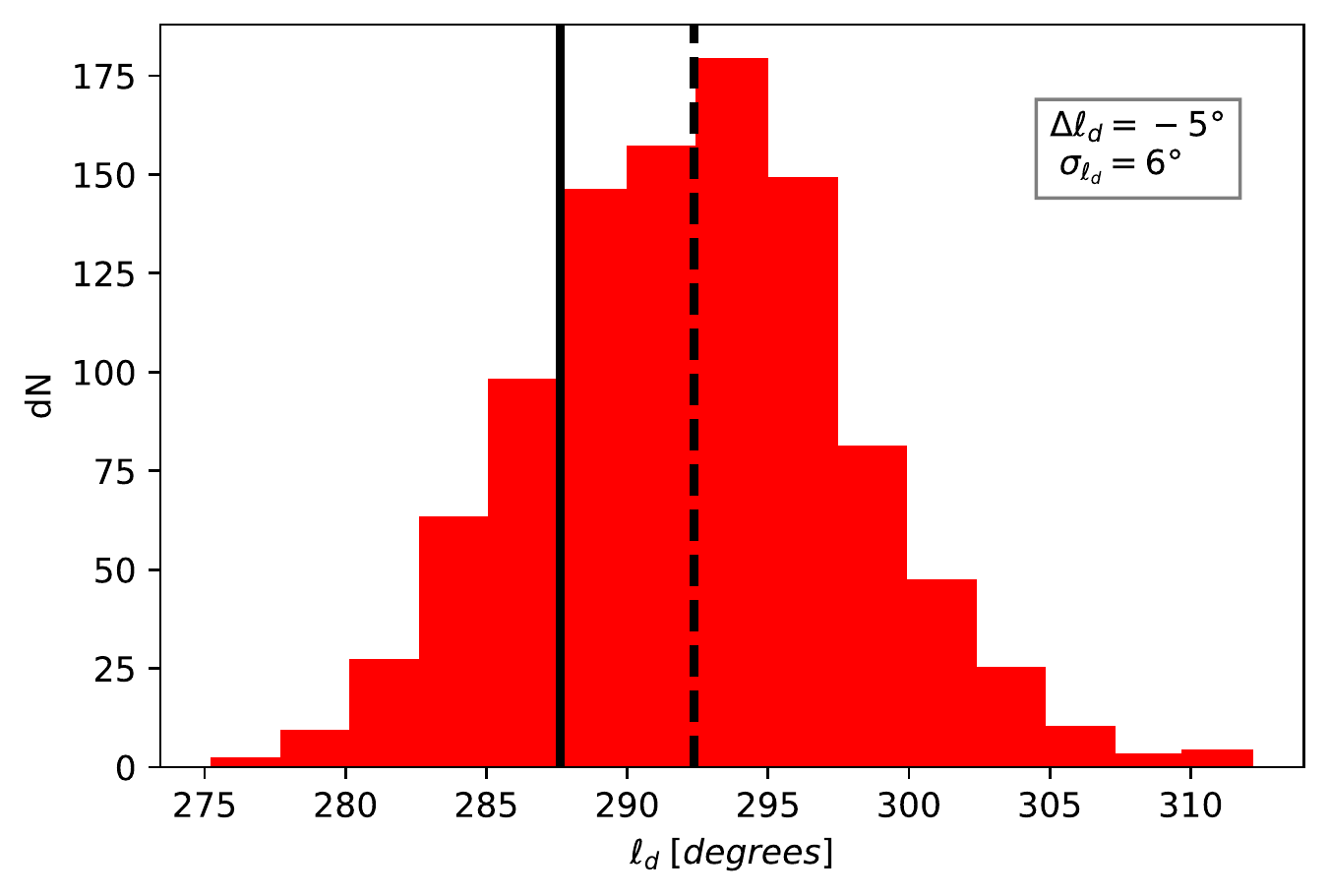}
\includegraphics[width = 3.5in]{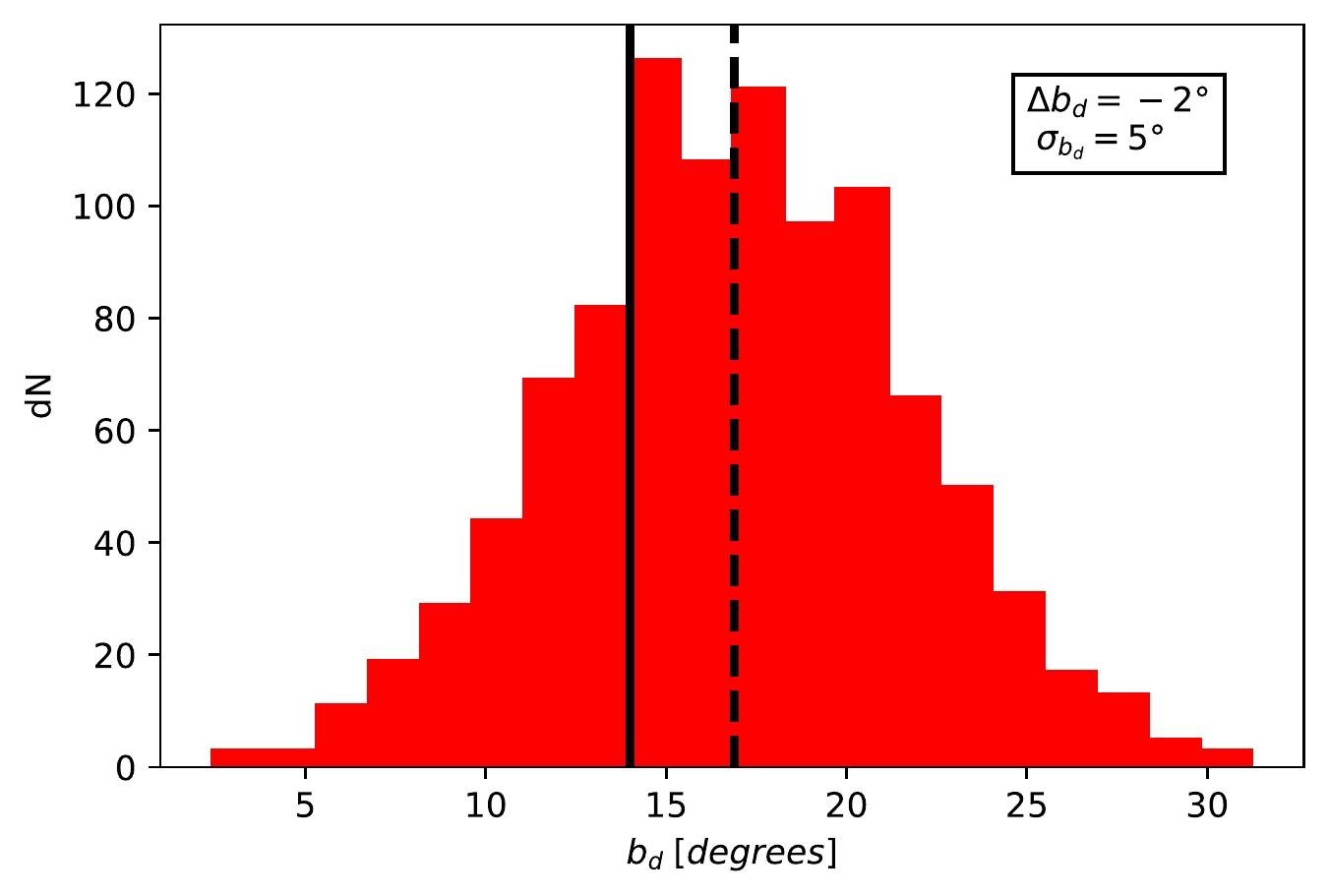}
\includegraphics[width = 3.5in]{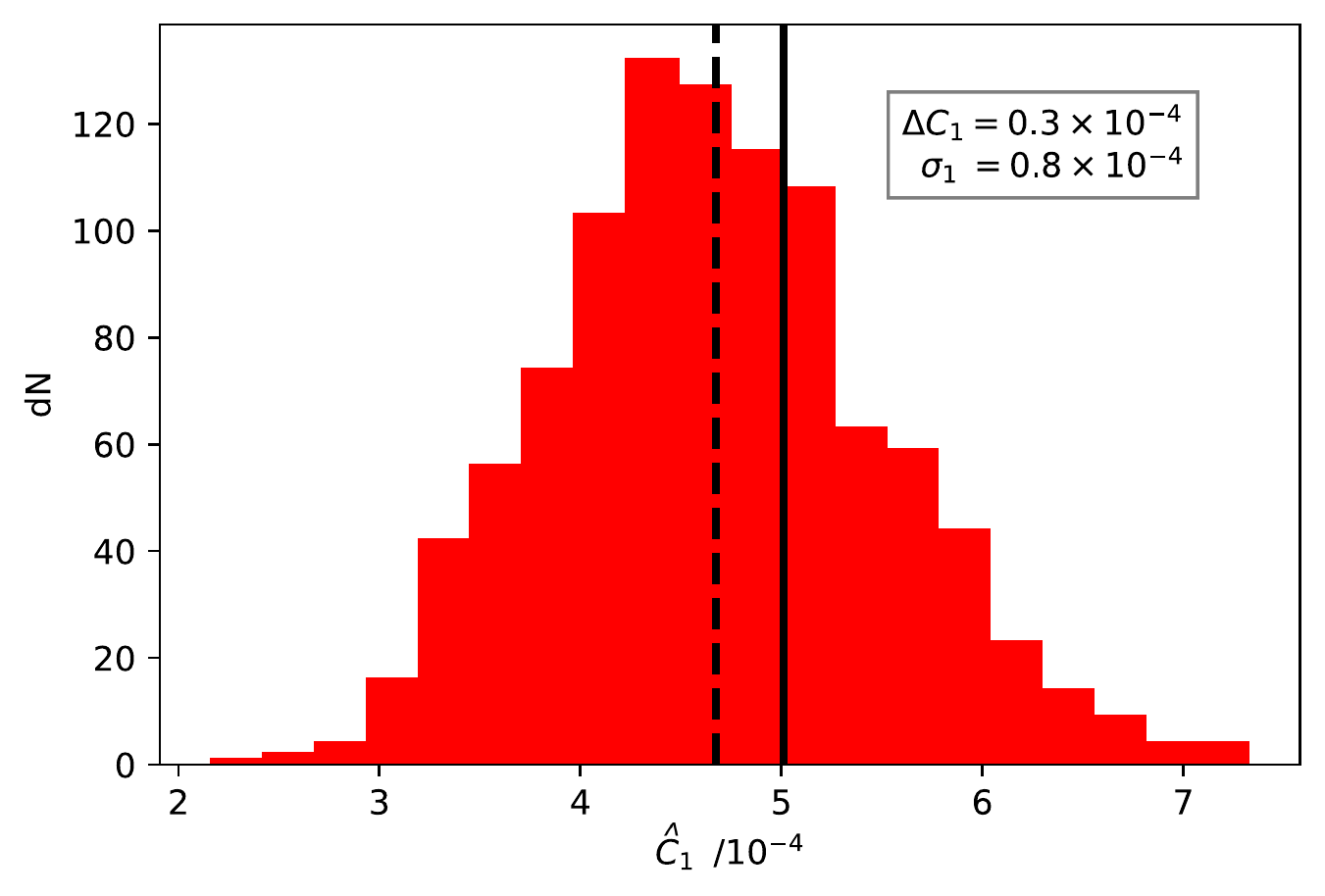}
\includegraphics[width = 3.5in]{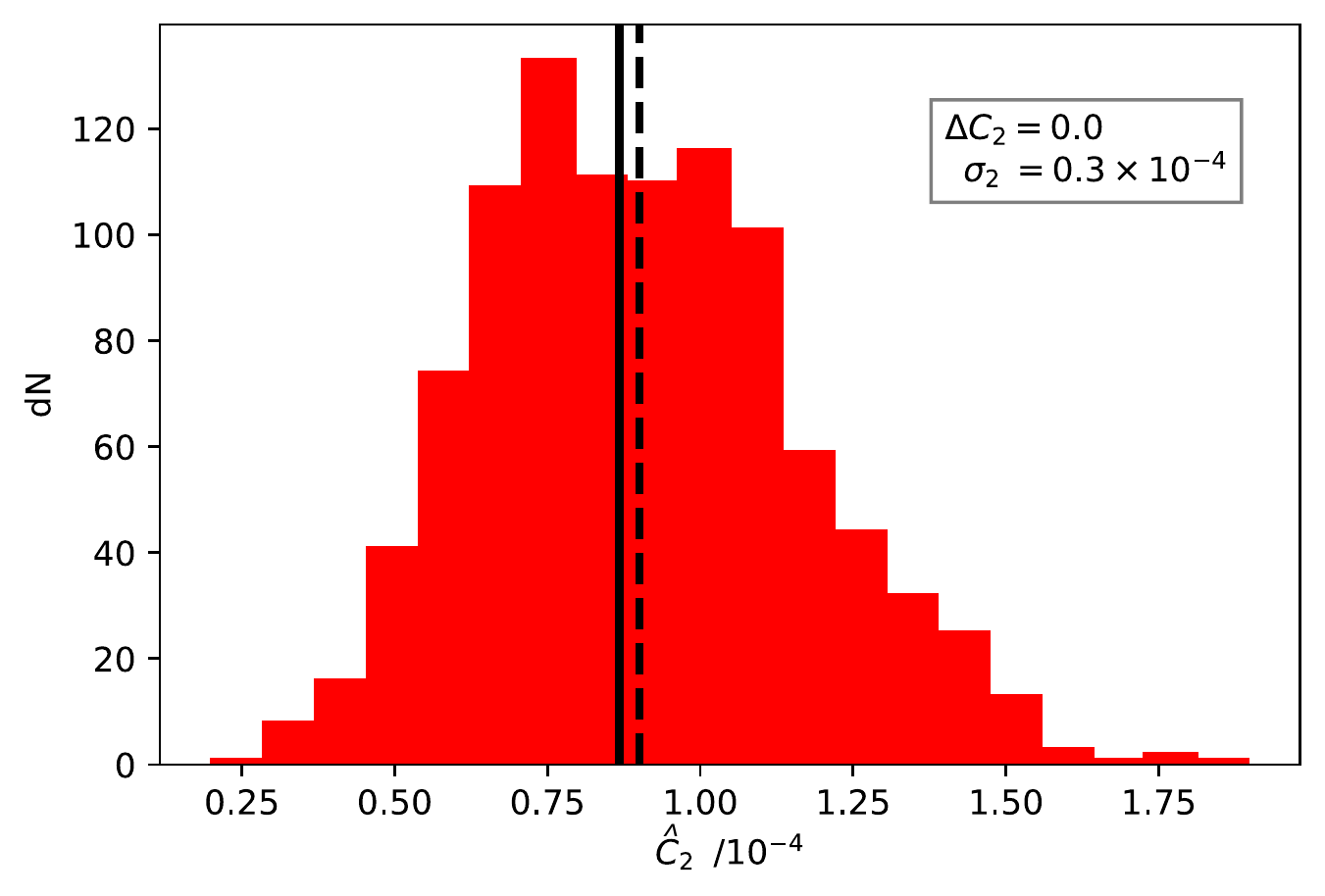}
\includegraphics[width = 3.5in]{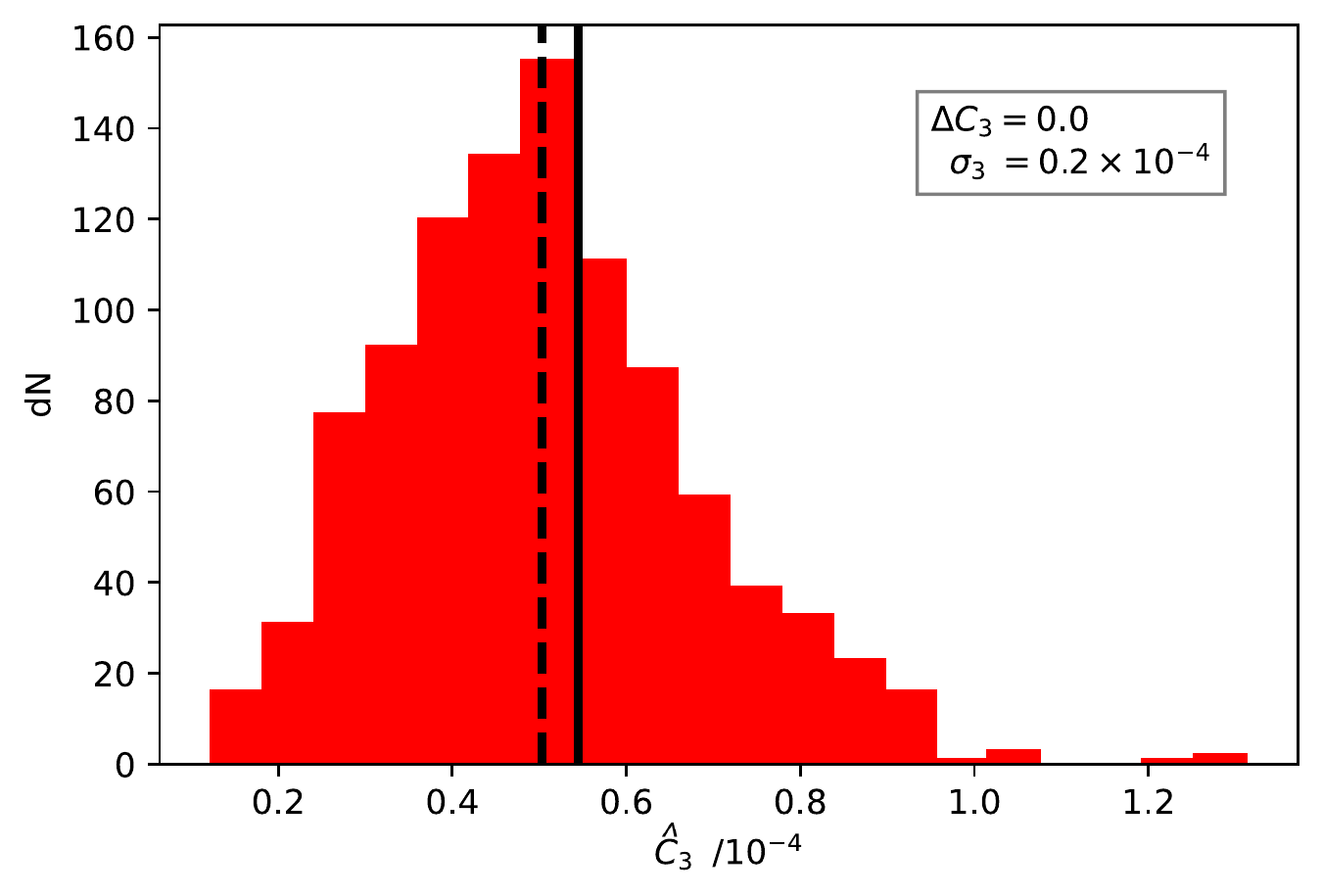}
\includegraphics[width = 3.5in]{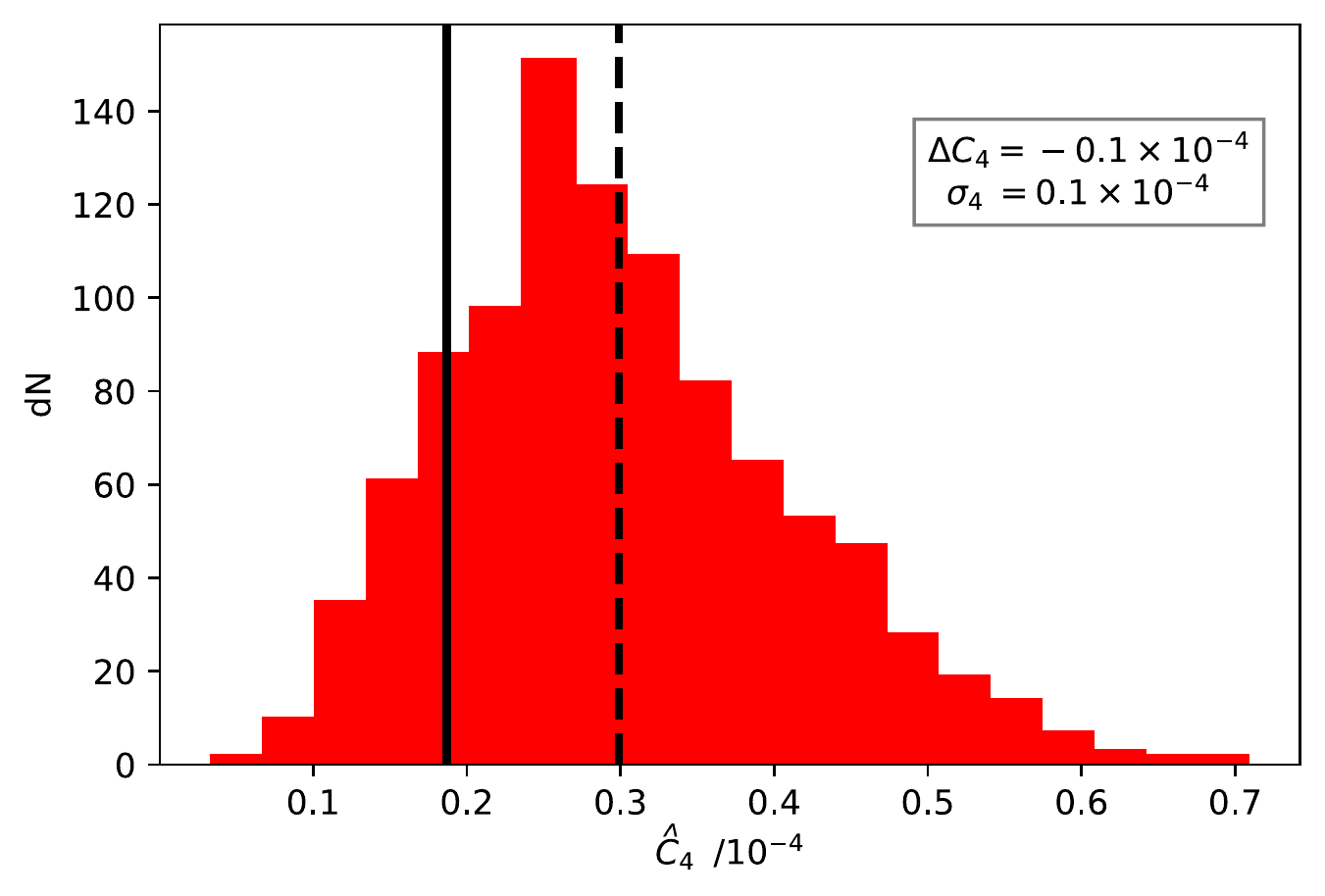}
\centering
\caption{
Distribution of relevant SH parameters recovered from analysing, with the same pipeline used for real data, 1000 Monte Carlo simulations of the CF3 galaxy sample [0.01, 0.05] (48 pixels). The parameters of interest are 
the power spectrum coefficients $\hat{C}_\ell$ and the direction of the dipole (in galactic coordinates). 
The black line displays the simulated input value (fiducial model), while the dashed black line corresponds to the average of the distribution.}
\label{FIG_simCF3}
\end{figure*}

\begin{figure*}
\centering
\includegraphics[width = 3.5in]{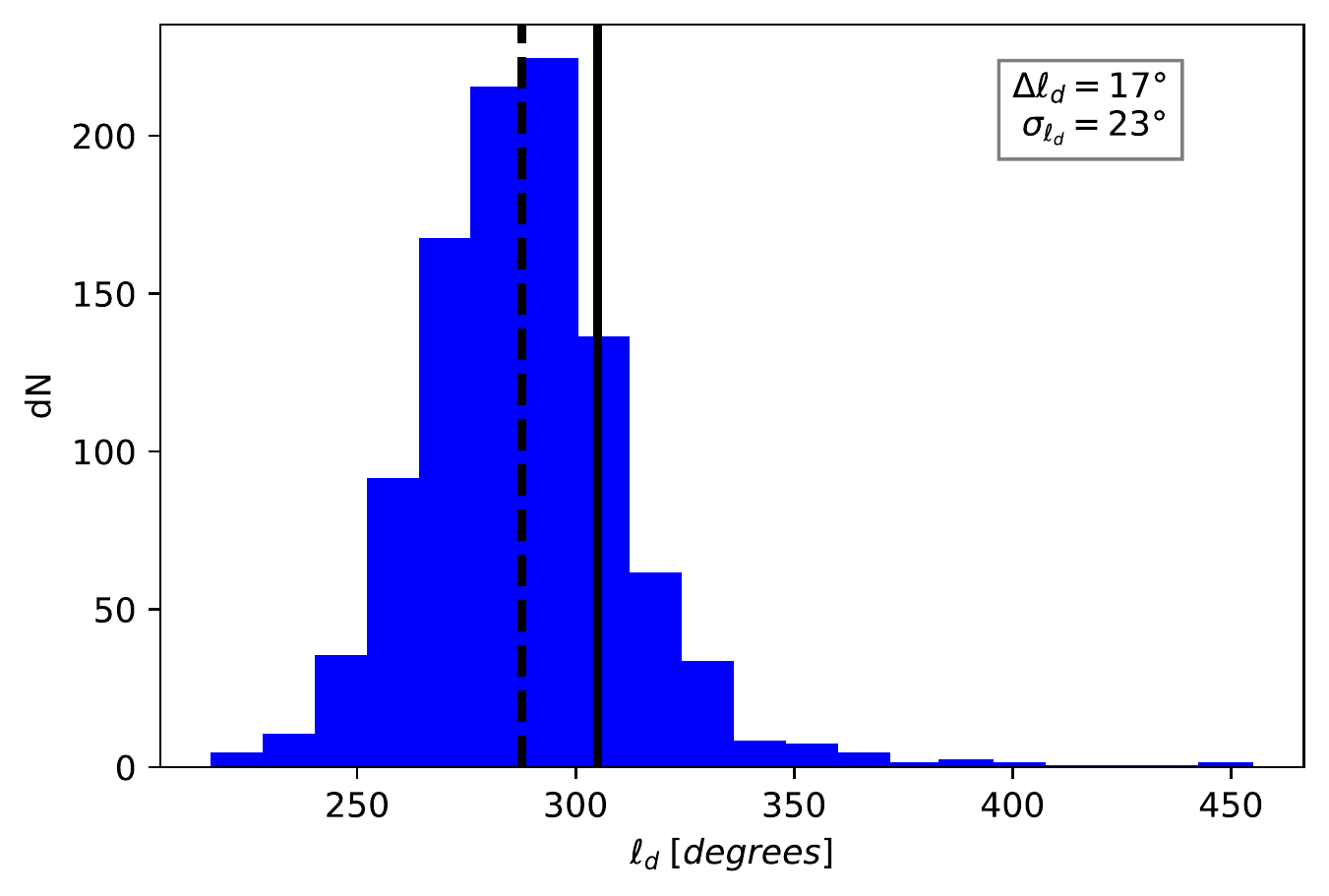} 
\includegraphics[width = 3.5in]{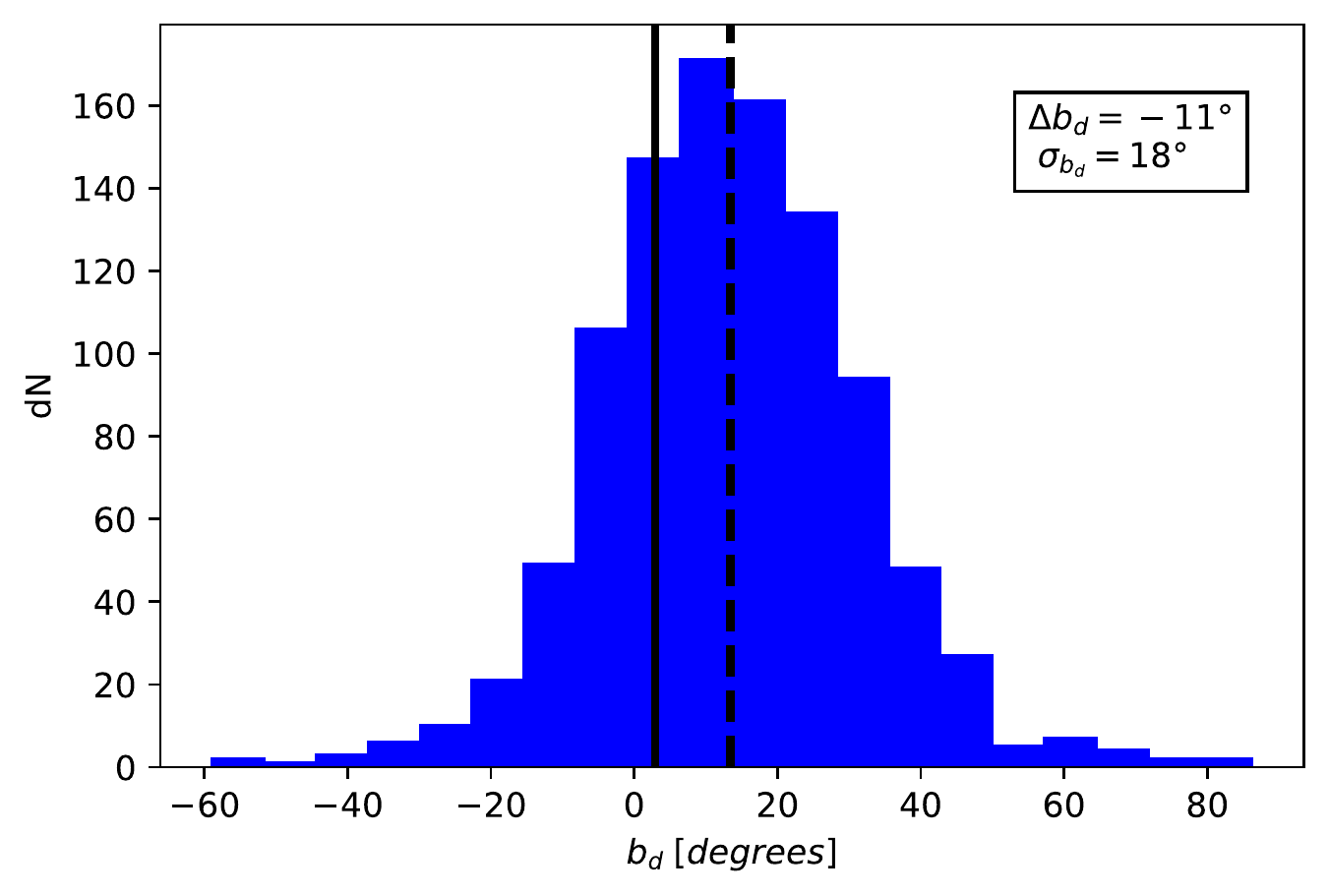} 
\includegraphics[width = 3.5in]{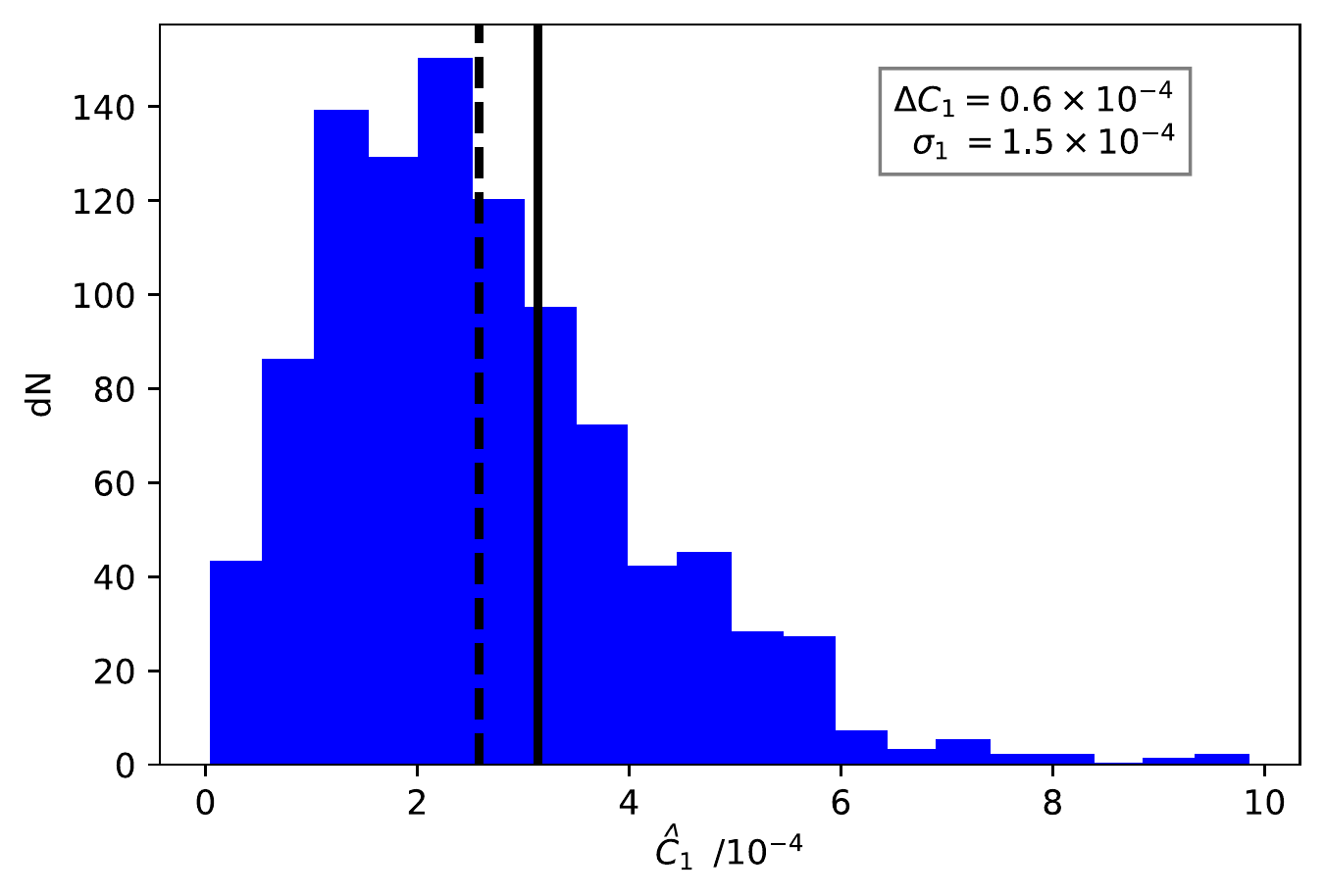} 
\includegraphics[width = 3.5in]{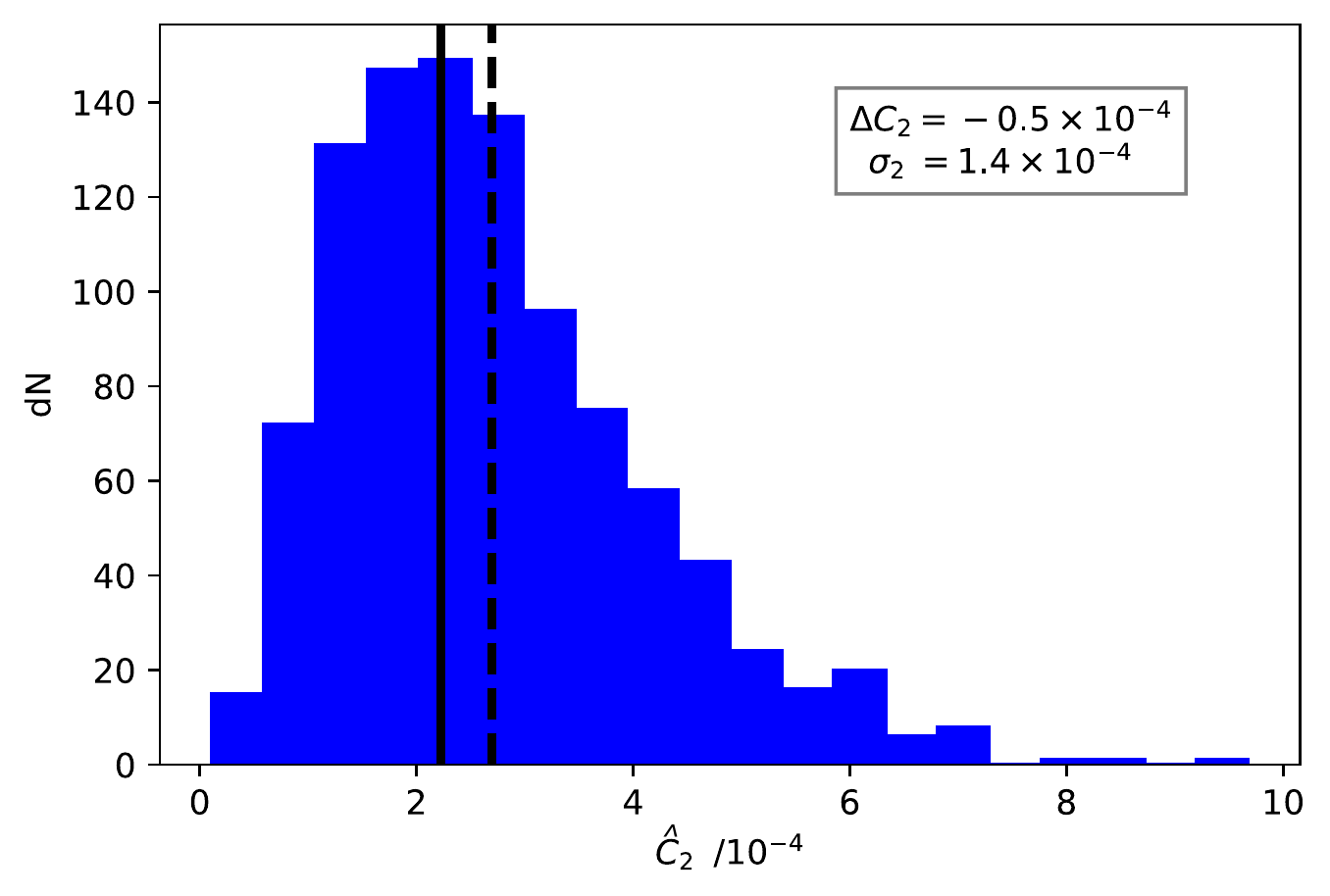} 
\centering
\caption{Same as in FIG. \ref{FIG_simCF3}, but for the CF3sn sample [0.01, 0.05] (12 pixels).}
\label{FIG_simCF3sn}
\end{figure*}

\begin{figure}
\begin{center}
	\includegraphics[scale=0.5]{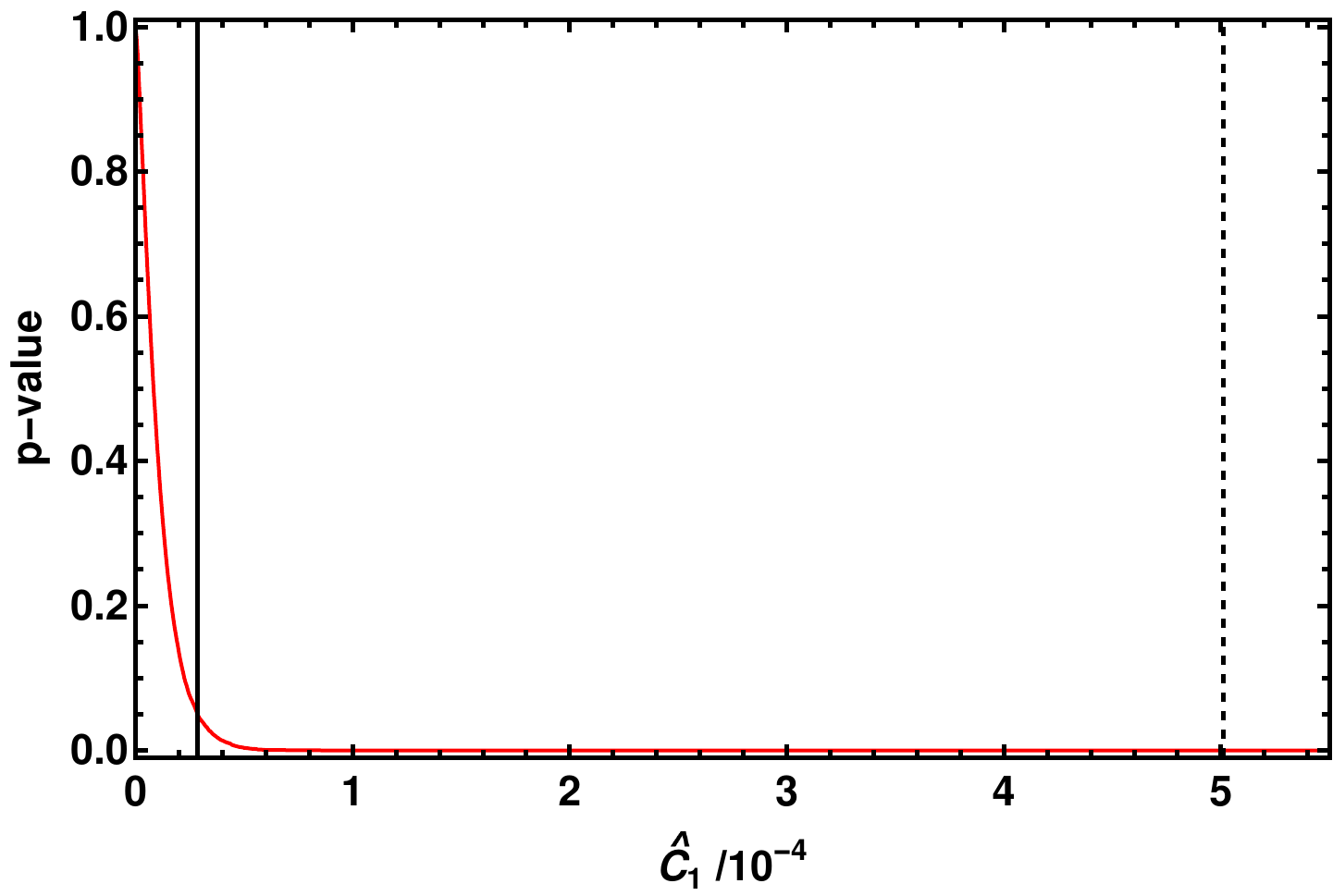}
	\includegraphics[scale=0.5]{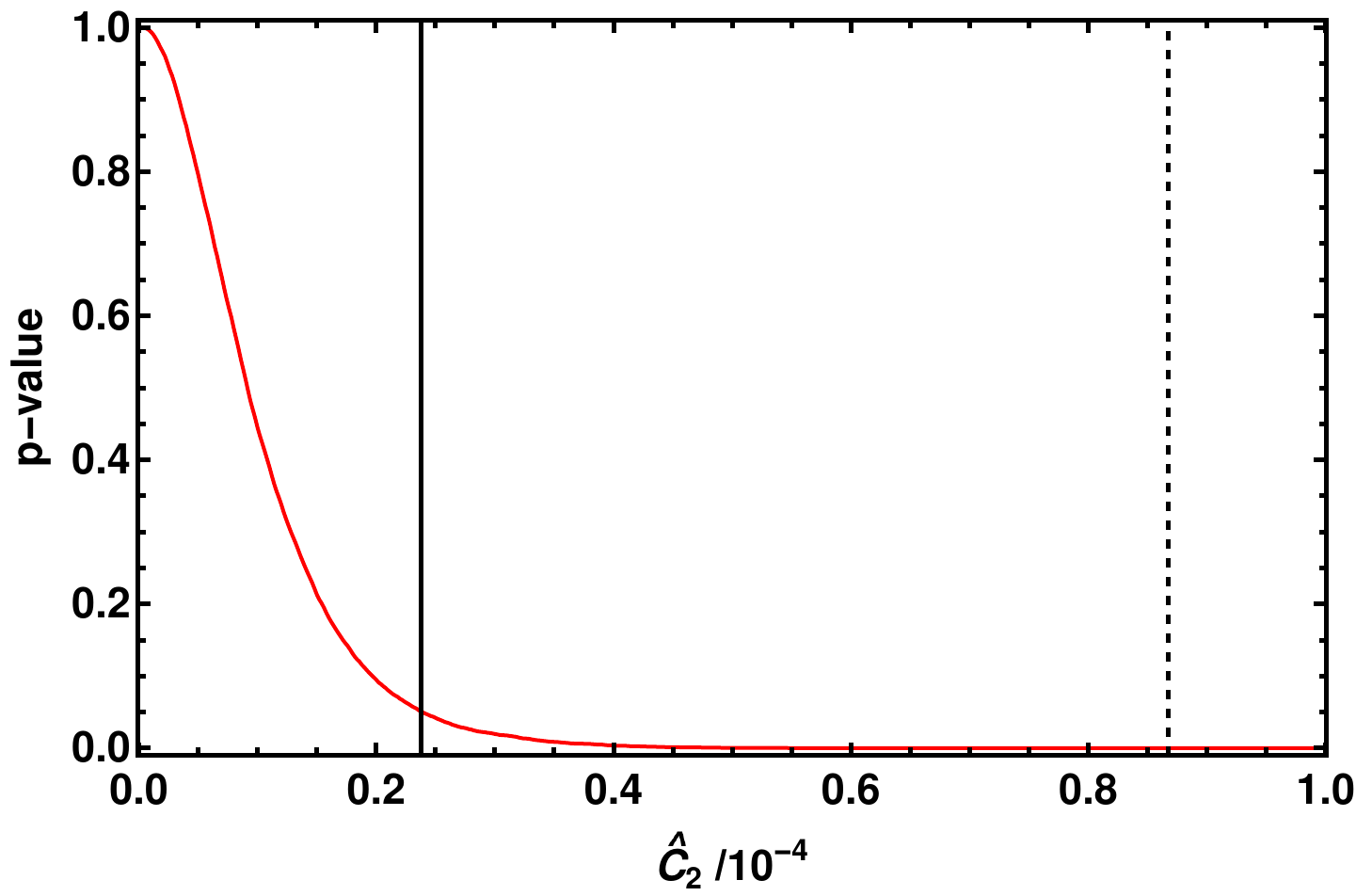}
	\includegraphics[scale=0.5]{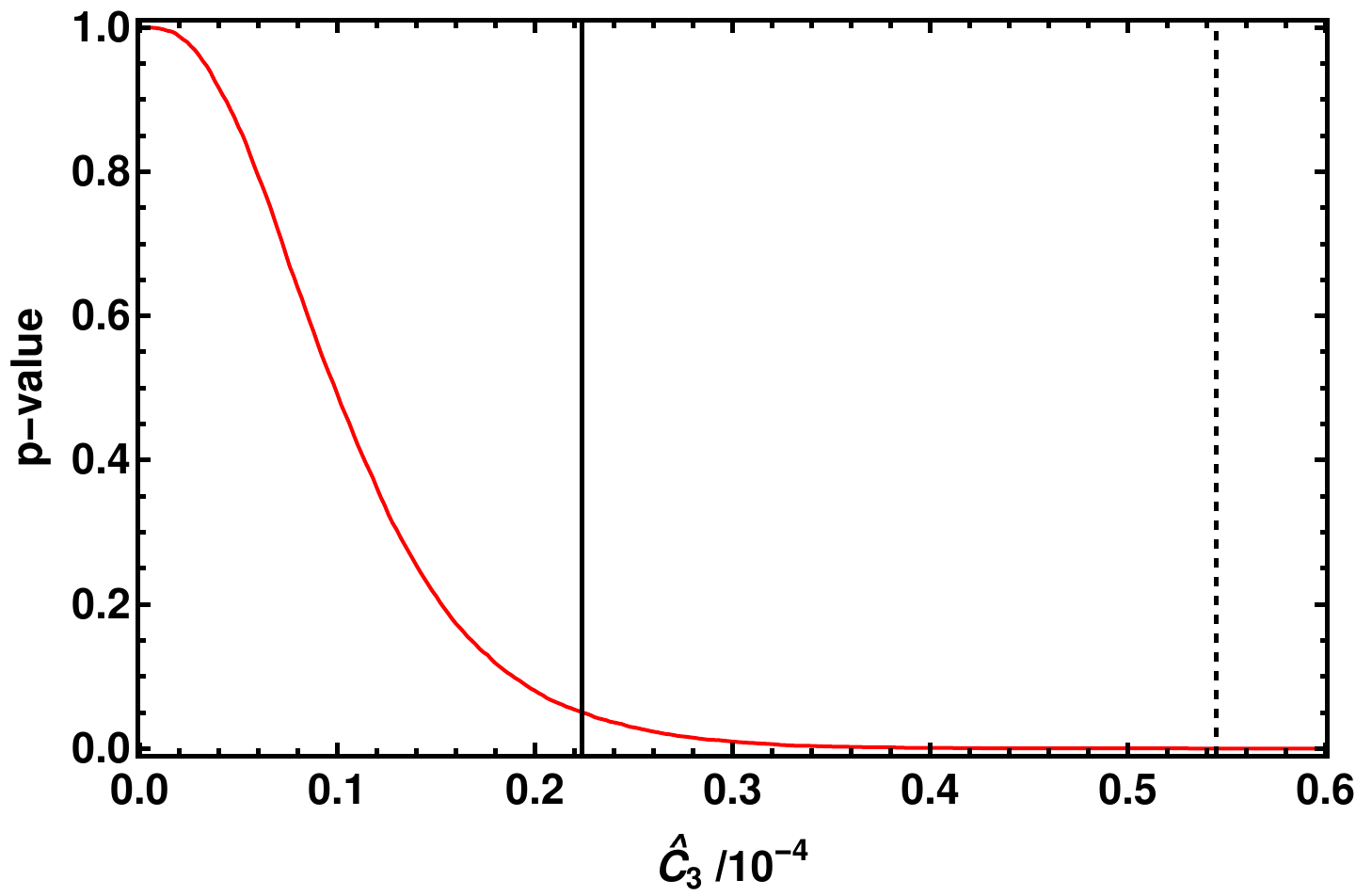}
	\includegraphics[scale=0.5]{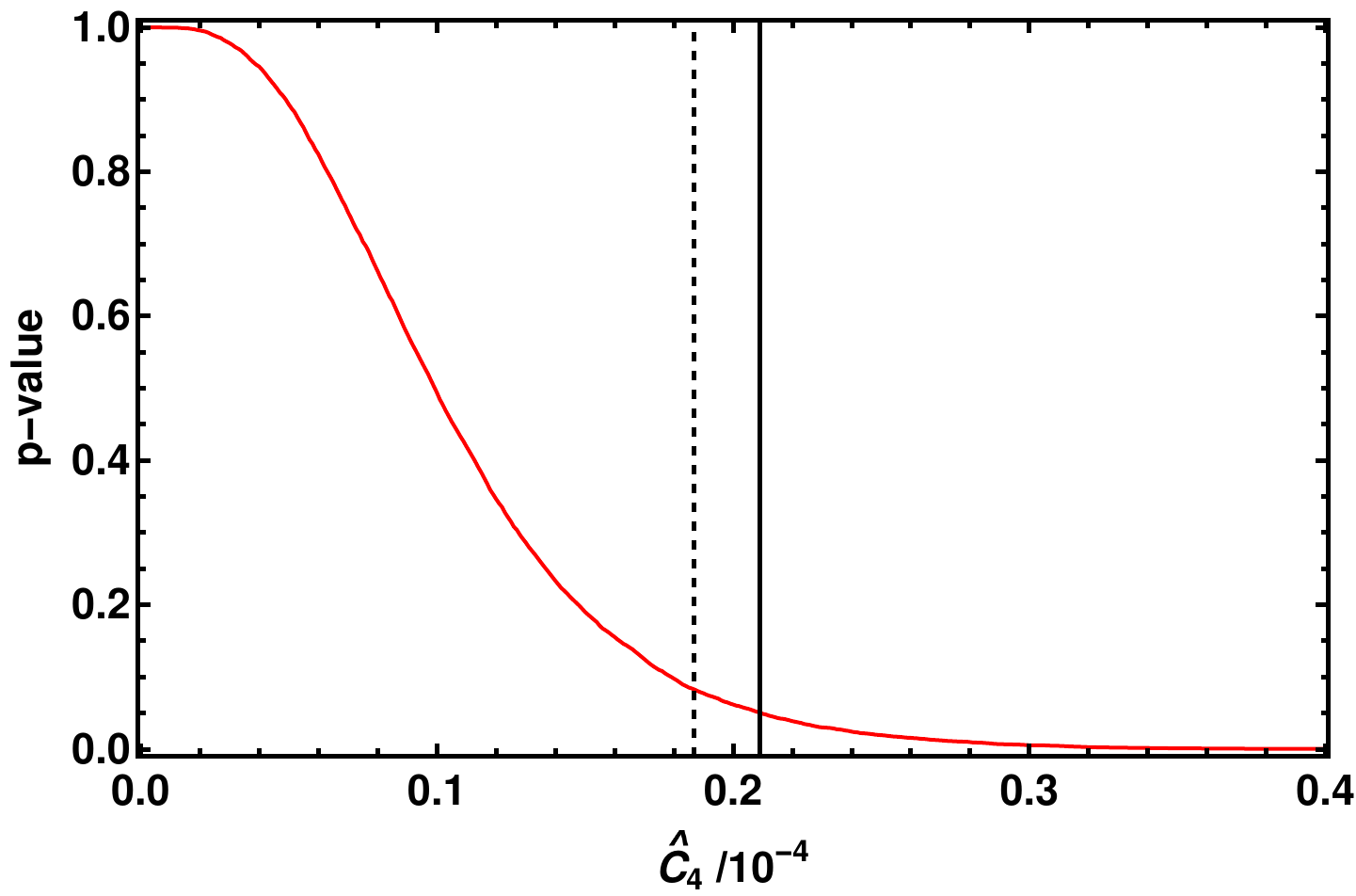}
			\centering
            \caption{Probability (p-values) of obtaining a power spectrum coefficient larger than $\hat{C}_\ell/10^{-4}$, when the underlying model does not have any multipole. The  black dashed line displays the value of $\hat{C}_\ell$ actually measured in the CF3 sample, partitioned into 48 pixels Estimates are based on 10000 Monte Carlo simulations of the CF3 sample. The  p-values for the $\hat{C}_\ell$ actually measured  are $ 0\%, \;0\%, \;0.01\%, \;8.27\%$ 
            respectively for the first four $\hat{C}_\ell$. The black solid line indicates the value of $\hat{C}_\ell$ which has the threshold p-value for acceptance  ($5\%$).}
\label{FIG_fakedetCF3}
		\centering
	\end{center}
\end{figure}

\begin{figure}
\begin{center}
	\includegraphics[scale=0.5]{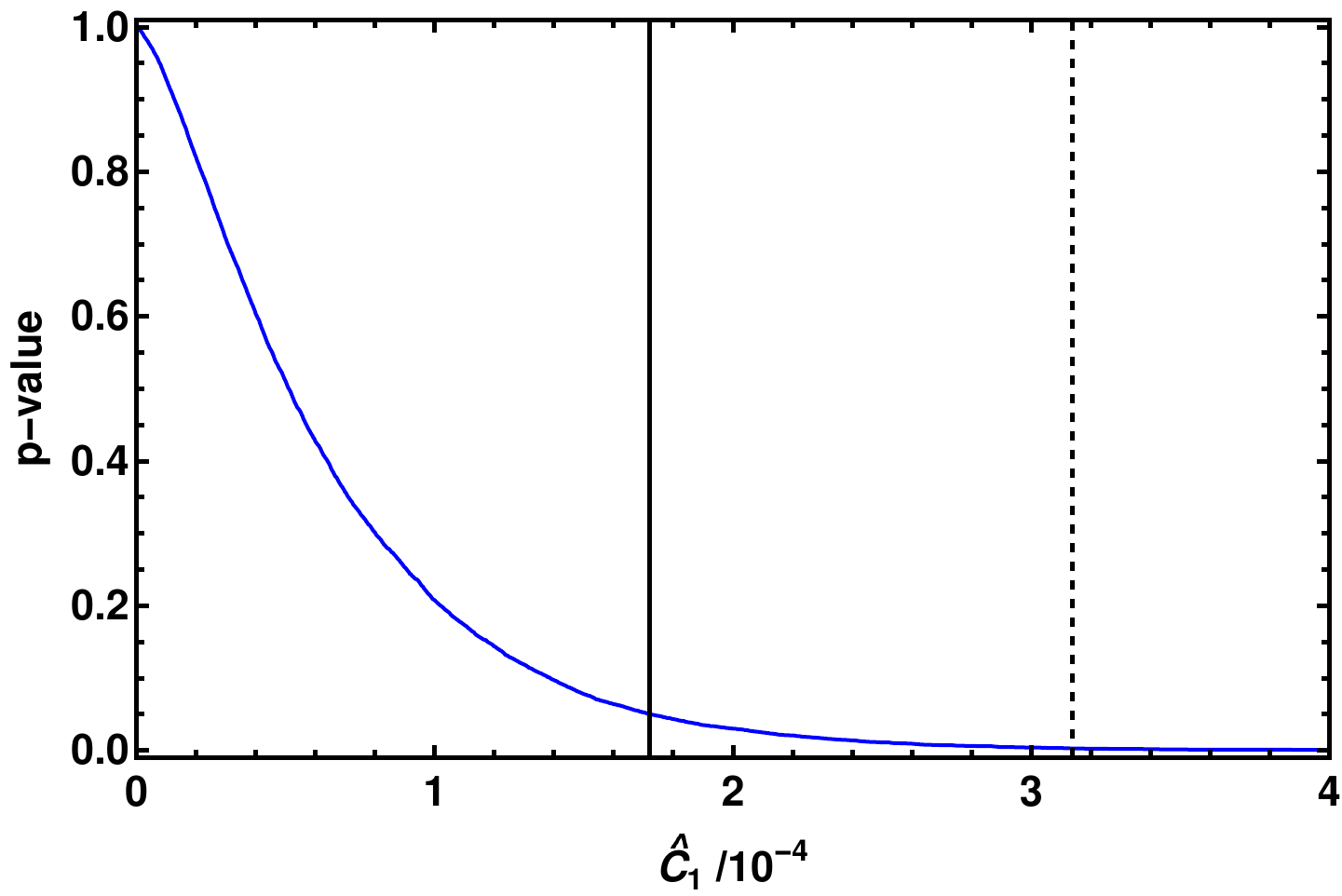}
	\includegraphics[scale=0.5]{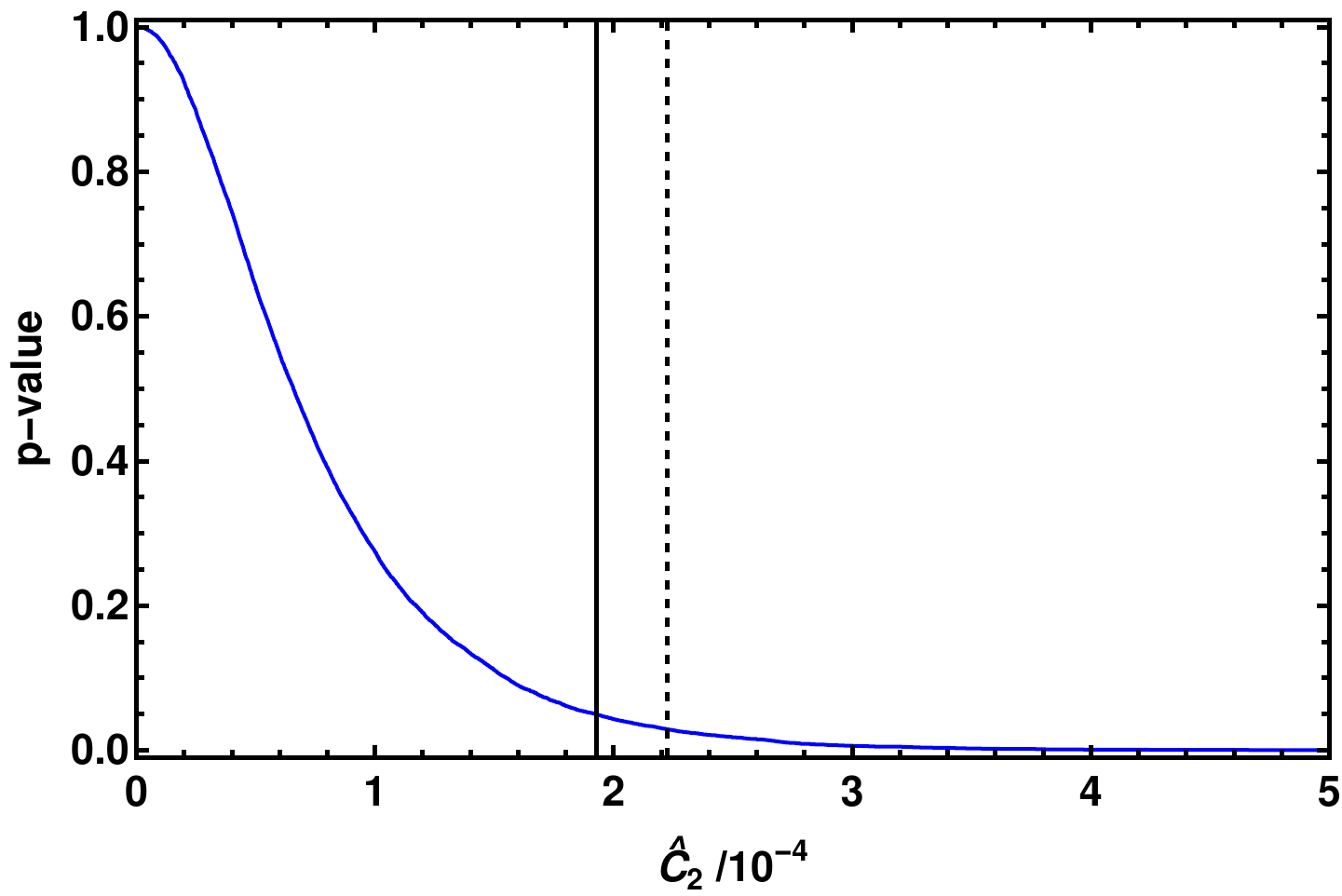}
			\centering
            \caption{Same as in FIG. \ref{FIG_fakedetCF3} but using simulations of the CF3sn sample tessellated  into 12 pixels. 
}
\label{FIG_fakedetCF3sn}
		\centering
	\end{center}
\end{figure}

\begin{table}[]
\scalebox{0.77}{
\begin{tabular}{|c|c|cc|cc|cc|}
\hline
\multirow{2}{*}{Sample}                                             & \multirow{2}{*}{$N_{pix}$} & \multicolumn{2}{c|}{\begin{tabular}[c]{@{}c@{}}$\sigma_{\hat C_1}$\\ ($10^{-4}$)\end{tabular}} & \multicolumn{2}{c|}{\begin{tabular}[c]{@{}c@{}}$\sigma_{\hat C_2}$\\ ($10^{-4}$)\end{tabular}} & \multicolumn{2}{c|}{\begin{tabular}[c]{@{}c@{}}$\sigma_{\hat C_3}$\\ ($10^{-4}$)\end{tabular}} \\ \cline{3-8} 
                                                                    &                            & \multicolumn{1}{c|}{Numerical}                        & Analytical                       & \multicolumn{1}{c|}{Numerical}                        & Analytical                       & \multicolumn{1}{c|}{Numerical}                        & Analytical                       \\ \hline
\begin{tabular}[c]{@{}c@{}}CF3\\ {[}0.01, 0.05{]}\end{tabular}      & 192                        & \multicolumn{1}{c|}{1.6}                              & 1.6                              & \multicolumn{1}{c|}{0.9}                              & 1.0                              & \multicolumn{1}{c|}{0.4}                              & 0.4                              \\ \hline
\begin{tabular}[c]{@{}c@{}}CF3\\ {[}0.01, 0.05{]}\end{tabular}      & 48                         & \multicolumn{1}{c|}{0.8}                              & 0.9                              & \multicolumn{1}{c|}{0.3}                              & 0.3                              & \multicolumn{1}{c|}{0.2}                              & 0.2                              \\ \hline
\begin{tabular}[c]{@{}c@{}}CF3g\\ {[}0.01, 0.05{]}\end{tabular}     & 48                         & \multicolumn{1}{c|}{1.0}                              & 1.2                              & \multicolumn{1}{c|}{0.4}                              & 0.4                              & \multicolumn{1}{c|}{0.2}                              & 0.3                              \\ \hline
\begin{tabular}[c]{@{}c@{}}CF3\\ {[}0.01, 0.05{]}\end{tabular}      & 12                         & \multicolumn{1}{c|}{0.8}                              & 0.8                              & \multicolumn{1}{c|}{0.3}                              & 0.3                              & \multicolumn{1}{c|}{-}                                & -                                \\ \hline
\begin{tabular}[c]{@{}c@{}}CF3g\\ {[}0.01, 0.05{]}\end{tabular}     & 12                         & \multicolumn{1}{c|}{0.6}                              & 0.7                              & \multicolumn{1}{c|}{0.4}                              & 0.4                              & \multicolumn{1}{c|}{-}                                & -                                \\ \hline
\begin{tabular}[c]{@{}c@{}}CF3sn\\ {[}0.01, 0.05{]}\end{tabular}    & 12                         & \multicolumn{1}{c|}{1.5}                              & 1.9                              & \multicolumn{1}{c|}{1.5}                              & 1.4                              & \multicolumn{1}{c|}{-}                                & -                                \\ \hline
\begin{tabular}[c]{@{}c@{}}Pantheon\\ {[}0.01, 0.05{]}\end{tabular} & 12                         & \multicolumn{1}{c|}{2.7}                              & 3.1                             & \multicolumn{1}{c|}{1.4}                              & 1.7                              & \multicolumn{1}{c|}{-}                                & -                                \\ \hline
\begin{tabular}[c]{@{}c@{}}CF3\\ {[}0.01, 0.03{]}\end{tabular}     & 12                         & \multicolumn{1}{c|}{1.0}                              & 1.2                              & \multicolumn{1}{c|}{0.6}                              & 0.7                              & \multicolumn{1}{c|}{-}                                & -                                \\ \hline
\begin{tabular}[c]{@{}c@{}}CF3\\ {[}0.03, 0.05{]}\end{tabular}     & 12                         & \multicolumn{1}{c|}{0.7}                              & 0.7                              & \multicolumn{1}{c|}{0.3}                              & 0.4                              & \multicolumn{1}{c|}{-}                                & -                                \\ \hline
\end{tabular}}
\caption{The standard deviation of the $\hat{C}_\ell$ coefficients 
reconstructed using the various  samples displayed in the first column.
For each sample, and for each $\hat{C}_\ell$ we compare error estimates obtained numerically (by means of 1000 Monte Carlo simulations) and analytically (by means of eq. \ref{varCl1}).}
\label{Tab_4}
\end{table}

  Results for the CF3 and CF3sn samples are shown in FIG. \ref{FIG_simCF3} and
  \ref{FIG_simCF3sn} respectively, and quantitatively summarized in tables \ref{Tab_3} and \ref{Tab_4}.
  Both figures \ref{FIG_simCF3} and
  \ref{FIG_simCF3sn} show that 
  the systematic errors  that affect our analysis, defined as the 
  difference between the input parameter and the 
  average value inferred by means of the Monte Carlo simulations, 
  are always smaller than the statistical error. However,  they are non-negligible and we thus correct the observational measurements presented in TABLE \ref{Tab_2}  for these bias factors.

\begin{table}[]
\begin{tabular}{|c|c|c|c|c|c|c|c|c|c|c|}
\hline
Sample                                                              & $N_{pix}$ & $\Delta l_d$ & $\Delta b_d$ & \begin{tabular}[c]{@{}c@{}}$\Delta C_1$ \\ ($10^{-4}$)\end{tabular} & $\Delta l_q$ & $\Delta b_q$ & \begin{tabular}[c]{@{}c@{}}$\Delta C_2$\\ ($10^{-4}$)\end{tabular} & $\Delta l_t$ & $\Delta b_t$ & \begin{tabular}[c]{@{}c@{}}$\Delta C_3$\\ ($10^{-4}$)\end{tabular} \\ \hline \hline
\begin{tabular}[c]{@{}c@{}}CF3\\ {[}0.01, 0.05{]}\end{tabular}      & 192        & 0           & -2           & -0.5                                                                 & 1           & 0           & -0.1                                                                  & 5           & -2           & -0.4                                                                  \\ \hline \hline
\begin{tabular}[c]{@{}c@{}}CF3\\ {[}0.01, 0.05{]}\end{tabular}      & 48        & -5           & -2           & 0.3                                                                 & -1           & -4           & 0                                                                  & -2           & -3           & 0                                                                  \\ \hline
\begin{tabular}[c]{@{}c@{}}CF3g\\ {[}0.01, 0.05{]}\end{tabular}     & 48        & 0            & -2           & 1.3                                                                 & 5            & 1            & -0.1                                                               & -16          & -2           & 0                                                                  \\ \hline \hline
\begin{tabular}[c]{@{}c@{}}CF3\\ {[}0.01, 0.05{]}\end{tabular}      & 12        & -1           & -4           & 0                                                                   & -5           & 3            & 0.1                                                                & -            & -            & -                                                                  \\ \hline
\begin{tabular}[c]{@{}c@{}}CF3g\\ {[}0.01, 0.05{]}\end{tabular}     & 12        & 0            & -4           & 0                                                                   & 12           & 7            & 0                                                                  & -            & -            & -                                                                  \\ \hline
\begin{tabular}[c]{@{}c@{}}CF3sn\\ {[}0.01, 0.05{]}\end{tabular}    & 12        & 17           & -11          & 0.6                                                                 & 11           & 5            & -0.5                                                               & -            & -            & -                                                                  \\ \hline
\begin{tabular}[c]{@{}c@{}}Pantheon\\ {[}0.01, 0.05{]}\end{tabular} & 12        & 13           & -14          & -1.1                                                                & -            & -            & -1.0                                                               & -            & -            & -                                                                  \\ \hline \hline
\begin{tabular}[c]{@{}c@{}}CF3\\ {[}0.01, 0.03{]}\end{tabular}      & 12        & 0            & -5           & 0.7                                                                 & 1            & -3           & 0.5                                                                & -            & -            & -                                                                  \\ \hline
\begin{tabular}[c]{@{}c@{}}CF3\\ {[}0.03, 0.05{]}\end{tabular}      & 12        & 1            & 1            & -0.3                                                                & 2            & 7            & 0.1                                                                & -            & -            & -                                                                  \\ \hline
\end{tabular}
\caption{The systematic bias factor for different parameters of the SH reconstruction. The systematic error is computed using 1000 Monte Carlo simulations of the catalogs displayed in the first column.}
\label{Tab_3}
\end{table}

In TABLE  \ref{Tab_4} we compare the standard deviation of the $\hat{C}_\ell$ coefficients estimated by Monte Carlo simulations with that obtained by applying eq. \ref{varCl1}.  There is an excellent agreement between the numerical and analytical estimates. The small remaining discrepancy can be mainly attributed to two effects. First, the estimator $\hat{a}_{\ell m}$ is a biased estimator for $a_{\ell m}$ and that is not included in eq. \ref{varCl1}. Also the finite number of Monte Carlo realisations (1000) contributes to the small discrepancies we observe. 

An alternative strategy to test the reliability of the reconstruction  consists in determining the value of the coefficients $a_{\ell m}$ not via a deterministic scheme (cf. eq. (\ref{closedex})) but by means of  a maximum likelihood analysis. We thus look for the set of SH parameters that minimize the quantity

\begin{equation}
    \chi^2=\sum_{i=1}^{N}\left( \frac{\hat{\eta}({\bf r}_i)- \eta(\theta_i, \phi_i)}{\delta_i^2}\right)^2
\end{equation}
where $N$ is, depending on the catalog,  either the number of galaxies or SNIa. 
The theoretical model that describes the expansion rate fluctuations is 
  \begin{equation}
\begin{aligned}
    \eta(\theta,\phi) =\sum_{\ell=\ell_{min}}^{\ell_{max}}\sum_{m = -\ell}^{\ell}a_{\ell m}Y_{\ell m}(\theta,\phi)= \sum_{\ell=\ell_{min}}^{\ell_{max}}\bigg[a_{\ell 0}Y_{\ell0}(\theta,\phi)
    \\
    +2\sum_{m=1}^{m=\ell}\bigg(\Re[a_{\ell m}]Y_{\ell m}(\theta,0)\cos(m \phi)
    \\
    -\Im[a_{\ell m}]Y_{\ell m}(\theta,0)\sin(m \phi)\bigg)\bigg]
\end{aligned}
\label{etadfit1}
\end{equation}
where $\ell_{min}$ and $\ell_{max}$ are  the lowest and maximum  multipole  fitted.  The above expression accounts for the fact  that $a_{\ell-m}$ and $a_{\ell m}$ are not independent  coefficients since $\eta$ is a real field and  we only have to choose one of them as a free parameter in the fit.
The $\chi^{2}$ statistic is then minimized with respect to the real and imaginary part of $a_{\ell m}$ (with $m\ge0$). 

We  choose  $\ell_{min}=1$ and run two independent analyses using $\ell_{max}=2$ and $\ell_{max}=3$ respectively.
The resulting best fitting parameters  together with the associated errors,  are reported in tables \ref{Tab_SHfit2} and \ref{Tab_SHfit3}.

\begin{table}[]
\begin{tabular}{|c|c|c|c|c|c|}
\hline
Sample                                                               & $l_d$      & $b_d$     & \begin{tabular}[c]{@{}c@{}}$\hat{C}_1$ \\ ($10^{-4}$)\end{tabular} & \begin{tabular}[c]{@{}c@{}}$\hat{C}_2$\\ ($10^{-4}$)\end{tabular} & $\frac{\chi^{2}_{min}}{dof}$ \\ \hline
\begin{tabular}[c]{@{}c@{}}CF3\\ {[}0.01, 0.05{]}\end{tabular}      & $287\pm5$  & $10\pm4$  & $4.3\pm0.8$                                                  & $1.2\pm0.3$                                                 & 1.26                         \\ \hline
\begin{tabular}[c]{@{}c@{}}CF3g\\ {[}0.01, 0.05{]}\end{tabular}     & $290\pm5$  & $5\pm4$   & $5.3\pm0.9$                                                  & $1.3\pm0.3$                                                 & 1.25                         \\ \hline
\begin{tabular}[c]{@{}c@{}}CF3sn\\ {[}0.01, 0.05{]}\end{tabular}    & $293\pm14$ & $0\pm10$  & $5.5\pm2.6$                                                  & $1.8\pm1.1$                                                 & 1.60                         \\ \hline
\begin{tabular}[c]{@{}c@{}}Pantheon\\ {[}0.01, 0.05{]}\end{tabular} & $311\pm32$ & $35\pm36$ & $3.4\pm2.1$                                                  & $0.5\pm0.9$                                                 & 0.98                         \\ \hline
\begin{tabular}[c]{@{}c@{}}CF3\\ {[}0.01, 0.03{]}\end{tabular}     & $280\pm5$  & $11\pm4$  & $8.0\pm1.4$                                                  & $2.4\pm0.5$                                                 & 1.37                         \\ \hline
\begin{tabular}[c]{@{}c@{}}CF3\\ {[}0.03, 0.05{]}\end{tabular}     & $324\pm17$ & $10\pm12$ & $1.4\pm0.8$                                                  & $1.4\pm0.5$                                                 & 1.16                         \\ \hline
\begin{tabular}[c]{@{}c@{}}CF3sn\\ {[}0.01, 0.03{]}\end{tabular}   & $274\pm14$ & $-5\pm11$ & $8.2\pm4.3$                                                  & $3.1\pm1.9$                                                 & 1.81                         \\ \hline
\begin{tabular}[c]{@{}c@{}}CF3sn\\ {[}0.03, 0.05{]}\end{tabular}   & $325\pm23$ & $6\pm18$  & $5.9\pm5.0$                                                  & $0.9\pm1.1$                                                 & 1.24                         \\ \hline
\end{tabular}
\caption{Best fitting SH parameters obtained by minimizing the $\chi^2$ difference between the expansion field $\hat{\eta}(\bf r)$ and a model given by cutting the SH decomposition at $\ell=2$. A total of eight $a_{\ell m}$ parameters are fitted to the data.}
\label{Tab_SHfit2}
\end{table}

\begin{table}[]
\begin{tabular}{|c|c|c|c|c|c|c|}
\hline
Sample                                                              & $l_d$      & $b_d$     & \begin{tabular}[c]{@{}c@{}}$\hat{C}_1$ \\ ($10^{-4}$)\end{tabular} & \begin{tabular}[c]{@{}c@{}}$\hat{C}_2$\\ ($10^{-4}$)\end{tabular} & \begin{tabular}[c]{@{}c@{}}$\hat{C}_3$\\ ($10^{-4}$)\end{tabular} & $\frac{\chi^{2}_{min}}{dof}$ \\ \hline
\begin{tabular}[c]{@{}c@{}}CF3\\ {[}0.01, 0.05{]}\end{tabular}      & $282\pm5$  & $13\pm4$  & $5.0\pm0.9$                                                  & $1.0\pm0.2$                                                 & $0.6\pm0.2$                                                 & 1.26                         \\ \hline
\begin{tabular}[c]{@{}c@{}}CF3g\\ {[}0.01, 0.05{]}\end{tabular}     & $285\pm5$  & $8\pm4$   & $6.0\pm1.2$                                                  & $1.1\pm0.3$                                                 & $0.7\pm0.2$                                                 & 1.25                         \\ \hline
\begin{tabular}[c]{@{}c@{}}CF3sn\\ {[}0.01, 0.05{]}\end{tabular}    & $285\pm16$ & $2\pm11$  & $5.6\pm3.1$                                                  & $2.4\pm1.4$                                                 & $1.5\pm1.1$                                                 & 1.60                         \\ \hline
\begin{tabular}[c]{@{}c@{}}Pantheon\\ {[}0.01, 0.05{]}\end{tabular} & $247\pm34$ & $20\pm23$ & $7.3\pm8.4$                                                  & $2.0\pm3.2$                                                 & $2.4\pm2.3$                                                 & 1.01                         \\ \hline
\begin{tabular}[c]{@{}c@{}}CF3\\ {[}0.01, 0.03{]}\end{tabular}     & $279\pm6$  & $15\pm5$  & $7.3\pm1.5$                                                  & $2.2\pm0.4$                                                 & $1.1\pm0.3$                                                 & 1.36
\\ \hline
\begin{tabular}[c]{@{}c@{}}CF3\\ {[}0.03, 0.05{]}\end{tabular}     & $323\pm21$ & $14\pm14$ & $1.5\pm1.0$                                                  & $1.6\pm0.7$                                                 & $0.2\pm0.2$                                                 & 1.16                         \\ \hline
\end{tabular}
\caption{Best fitting SH parameters obtained by minimizing the $\chi^2$ difference between the expansion field $\hat{\eta}(\bf r)$ and a model given by cutting the SH decomposition at $\ell=3$. A total of 15 independent parameters are fitted to the data.}
\label{Tab_SHfit3}
\end{table}

The estimates deduced by means of this statistical approach are in excellent agreement with the Fourier determinations (cf. TABLE \ref{Tab_2}. This provides   independent supporting evidence of the robustness of the results obtained in section (\S \ref{sec_results}). Also the minimum value of the normalized $\chi^2$ statistic is of relevance.   
 
Finally, we also quantify the probability that the measured multipolar signals are statistical flukes due to the sparsity and angular anisotropy of the analyzed samples. To this end we compute, using 10000 Monte Carlo simulations,   the probability (p-values) of measuring a power spectrum coefficient greater then a given non-zero value  when the underlying input model does not have any multipole. Results for the CF3, CF3sn samples are shown in FIG. \ref{FIG_fakedetCF3} \ref{FIG_fakedetCF3sn} respectively (see TABLE \ref{Tab_2} for the other samples).  
  
Comparing these results with the values of the power spectrum coefficients $\hat{C}_\ell$  inferred from the CF3g data, we see that the chance of an accidental signal is negligible for $\ell\leq 3$. Similarly, the risk of misinterpreting the CF3sn sample results is insignificant for $\ell\leq 2$.

\vspace{60mm}

\bibliography{biblio_sndipole}

\begin{thebibliography}{89}
\expandafter\ifx\csname natexlab\endcsname\relax\def\natexlab#1{#1}\fi
\expandafter\ifx\csname bibnamefont\endcsname\relax
  \def\bibnamefont#1{#1}\fi
\expandafter\ifx\csname bibfnamefont\endcsname\relax
  \def\bibfnamefont#1{#1}\fi
\expandafter\ifx\csname citenamefont\endcsname\relax
  \def\citenamefont#1{#1}\fi
\expandafter\ifx\csname url\endcsname\relax
  \def\url#1{\texttt{#1}}\fi
\expandafter\ifx\csname urlprefix\endcsname\relax\def\urlprefix{URL }\fi
\providecommand{\bibinfo}[2]{#2}
\providecommand{\eprint}[2][]{\url{#2}}

\bibitem[{\citenamefont{{Peebles}}(1980)}]{Pee1980}
\bibinfo{author}{\bibfnamefont{P.~J.~E.} \bibnamefont{{Peebles}}},
  \emph{\bibinfo{title}{{The large-scale structure of the universe}}}
  (\bibinfo{year}{1980}).

\bibitem[{\citenamefont{Clarkson and
  Maartens}(2010{\natexlab{a}})}]{Clarkson:2010uz}
\bibinfo{author}{\bibfnamefont{C.}~\bibnamefont{Clarkson}} \bibnamefont{and}
  \bibinfo{author}{\bibfnamefont{R.}~\bibnamefont{Maartens}},
  \bibinfo{journal}{Class. Quant. Grav.} \textbf{\bibinfo{volume}{27}},
  \bibinfo{pages}{124008} (\bibinfo{year}{2010}{\natexlab{a}}),
  \eprint{1005.2165}.

\bibitem[{\citenamefont{{Bennett} et~al.}(2013)\citenamefont{{Bennett},
  {Larson}, {Weiland}, {Jarosik}, {Hinshaw}, {Odegard}, {Smith}, {Hill},
  {Gold}, {Halpern} et~al.}}]{Bennet2013}
\bibinfo{author}{\bibfnamefont{C.~L.} \bibnamefont{{Bennett}}},
  \bibinfo{author}{\bibfnamefont{D.}~\bibnamefont{{Larson}}},
  \bibinfo{author}{\bibfnamefont{J.~L.} \bibnamefont{{Weiland}}},
  \bibinfo{author}{\bibfnamefont{N.}~\bibnamefont{{Jarosik}}},
  \bibinfo{author}{\bibfnamefont{G.}~\bibnamefont{{Hinshaw}}},
  \bibinfo{author}{\bibfnamefont{N.}~\bibnamefont{{Odegard}}},
  \bibinfo{author}{\bibfnamefont{K.~M.} \bibnamefont{{Smith}}},
  \bibinfo{author}{\bibfnamefont{R.~S.} \bibnamefont{{Hill}}},
  \bibinfo{author}{\bibfnamefont{B.}~\bibnamefont{{Gold}}},
  \bibinfo{author}{\bibfnamefont{M.}~\bibnamefont{{Halpern}}},
  \bibnamefont{et~al.}, \bibinfo{journal}{\apjs}
  \textbf{\bibinfo{volume}{208}}, \bibinfo{eid}{20} (\bibinfo{year}{2013}),
  \eprint{1212.5225}.

\bibitem[{\citenamefont{Ade et~al.}(2016)}]{Planck:2015igc}
\bibinfo{author}{\bibfnamefont{P.~A.~R.} \bibnamefont{Ade}}
  \bibnamefont{et~al.} (\bibinfo{collaboration}{Planck}),
  \bibinfo{journal}{Astron. Astrophys.} \textbf{\bibinfo{volume}{594}},
  \bibinfo{pages}{A16} (\bibinfo{year}{2016}), \eprint{1506.07135}.

\bibitem[{\citenamefont{{Sarkar} et~al.}(2009)\citenamefont{{Sarkar}, {Yadav},
  {Pandey}, and {Bharadwaj}}}]{Sarkar2009}
\bibinfo{author}{\bibfnamefont{P.}~\bibnamefont{{Sarkar}}},
  \bibinfo{author}{\bibfnamefont{J.}~\bibnamefont{{Yadav}}},
  \bibinfo{author}{\bibfnamefont{B.}~\bibnamefont{{Pandey}}}, \bibnamefont{and}
  \bibinfo{author}{\bibfnamefont{S.}~\bibnamefont{{Bharadwaj}}},
  \bibinfo{journal}{\mnras} \textbf{\bibinfo{volume}{399}},
  \bibinfo{pages}{L128} (\bibinfo{year}{2009}), \eprint{0906.3431}.

\bibitem[{\citenamefont{{Marinoni} et~al.}(2012)\citenamefont{{Marinoni},
  {Bel}, and {Buzzi}}}]{MBB2012}
\bibinfo{author}{\bibfnamefont{C.}~\bibnamefont{{Marinoni}}},
  \bibinfo{author}{\bibfnamefont{J.}~\bibnamefont{{Bel}}}, \bibnamefont{and}
  \bibinfo{author}{\bibfnamefont{A.}~\bibnamefont{{Buzzi}}},
  \bibinfo{journal}{\jcap} \textbf{\bibinfo{volume}{2012}}, \bibinfo{eid}{036}
  (\bibinfo{year}{2012}), \eprint{1205.3309}.

\bibitem[{\citenamefont{{Appleby} and {Shafieloo}}(2014)}]{Appleby2014}
\bibinfo{author}{\bibfnamefont{S.}~\bibnamefont{{Appleby}}} \bibnamefont{and}
  \bibinfo{author}{\bibfnamefont{A.}~\bibnamefont{{Shafieloo}}},
  \bibinfo{journal}{\jcap} \textbf{\bibinfo{volume}{2014}}, \bibinfo{eid}{070}
  (\bibinfo{year}{2014}), \eprint{1405.4595}.

\bibitem[{\citenamefont{{Bengaly} et~al.}(2019)\citenamefont{{Bengaly},
  {Maartens}, {Randriamiarinarivo}, and {Baloyi}}}]{Bengaly2019}
\bibinfo{author}{\bibfnamefont{C.~A.~P.} \bibnamefont{{Bengaly}}},
  \bibinfo{author}{\bibfnamefont{R.}~\bibnamefont{{Maartens}}},
  \bibinfo{author}{\bibfnamefont{N.}~\bibnamefont{{Randriamiarinarivo}}},
  \bibnamefont{and} \bibinfo{author}{\bibfnamefont{A.}~\bibnamefont{{Baloyi}}},
  \bibinfo{journal}{\jcap} \textbf{\bibinfo{volume}{2019}}, \bibinfo{eid}{025}
  (\bibinfo{year}{2019}), \eprint{1905.12378}.

\bibitem[{\citenamefont{{Sarkar} et~al.}(2019)\citenamefont{{Sarkar}, {Pandey},
  and {Khatri}}}]{Sarkar2019}
\bibinfo{author}{\bibfnamefont{S.}~\bibnamefont{{Sarkar}}},
  \bibinfo{author}{\bibfnamefont{B.}~\bibnamefont{{Pandey}}}, \bibnamefont{and}
  \bibinfo{author}{\bibfnamefont{R.}~\bibnamefont{{Khatri}}},
  \bibinfo{journal}{\mnras} \textbf{\bibinfo{volume}{483}},
  \bibinfo{pages}{2453} (\bibinfo{year}{2019}), \eprint{1810.07410}.

\bibitem[{\citenamefont{{Ntelis} et~al.}(2019)\citenamefont{{Ntelis}, {Hawken},
  {Escoffier}, {Ealet}, and {Tilquin}}}]{Ntelis2019}
\bibinfo{author}{\bibfnamefont{P.}~\bibnamefont{{Ntelis}}},
  \bibinfo{author}{\bibfnamefont{A.~J.} \bibnamefont{{Hawken}}},
  \bibinfo{author}{\bibfnamefont{S.}~\bibnamefont{{Escoffier}}},
  \bibinfo{author}{\bibfnamefont{A.}~\bibnamefont{{Ealet}}}, \bibnamefont{and}
  \bibinfo{author}{\bibfnamefont{A.}~\bibnamefont{{Tilquin}}},
  \bibinfo{journal}{arXiv e-prints} \bibinfo{eid}{arXiv:1904.06135}
  (\bibinfo{year}{2019}), \eprint{1904.06135}.

\bibitem[{\citenamefont{Andrade et~al.}(2019)\citenamefont{Andrade, Bengaly,
  Alcaniz, and Capozziello}}]{Andrade:2019kvl}
\bibinfo{author}{\bibfnamefont{U.}~\bibnamefont{Andrade}},
  \bibinfo{author}{\bibfnamefont{C.~A.~P.} \bibnamefont{Bengaly}},
  \bibinfo{author}{\bibfnamefont{J.~S.} \bibnamefont{Alcaniz}},
  \bibnamefont{and}
  \bibinfo{author}{\bibfnamefont{S.}~\bibnamefont{Capozziello}},
  \bibinfo{journal}{Mon. Not. Roy. Astron. Soc.}
  \textbf{\bibinfo{volume}{490}}, \bibinfo{pages}{4481} (\bibinfo{year}{2019}),
  \eprint{1905.08864}.

\bibitem[{\citenamefont{{Payne} et~al.}(2020)\citenamefont{{Payne}, {Banagiri},
  {Lasky}, and {Thrane}}}]{Payne2020}
\bibinfo{author}{\bibfnamefont{E.}~\bibnamefont{{Payne}}},
  \bibinfo{author}{\bibfnamefont{S.}~\bibnamefont{{Banagiri}}},
  \bibinfo{author}{\bibfnamefont{P.~D.} \bibnamefont{{Lasky}}},
  \bibnamefont{and} \bibinfo{author}{\bibfnamefont{E.}~\bibnamefont{{Thrane}}},
  \bibinfo{journal}{\prd} \textbf{\bibinfo{volume}{102}}, \bibinfo{eid}{102004}
  (\bibinfo{year}{2020}), \eprint{2006.11957}.

\bibitem[{\citenamefont{{Pandey} and {Sarkar}}(2021)}]{Pandey2022}
\bibinfo{author}{\bibfnamefont{B.}~\bibnamefont{{Pandey}}} \bibnamefont{and}
  \bibinfo{author}{\bibfnamefont{S.}~\bibnamefont{{Sarkar}}},
  \bibinfo{journal}{\jcap} \textbf{\bibinfo{volume}{2021}}, \bibinfo{eid}{019}
  (\bibinfo{year}{2021}), \eprint{2103.11954}.

\bibitem[{\citenamefont{Gon\c{c}alves et~al.}(2021)\citenamefont{Gon\c{c}alves,
  Carvalho, Andrade, Bengaly, Carvalho, and Alcaniz}}]{Goncalves:2020erb}
\bibinfo{author}{\bibfnamefont{R.~S.} \bibnamefont{Gon\c{c}alves}},
  \bibinfo{author}{\bibfnamefont{G.~C.} \bibnamefont{Carvalho}},
  \bibinfo{author}{\bibfnamefont{U.}~\bibnamefont{Andrade}},
  \bibinfo{author}{\bibfnamefont{C.~A.~P.} \bibnamefont{Bengaly}},
  \bibinfo{author}{\bibfnamefont{J.~C.} \bibnamefont{Carvalho}},
  \bibnamefont{and} \bibinfo{author}{\bibfnamefont{J.}~\bibnamefont{Alcaniz}},
  \bibinfo{journal}{JCAP} \textbf{\bibinfo{volume}{03}}, \bibinfo{pages}{029}
  (\bibinfo{year}{2021}), \eprint{2010.06635}.

\bibitem[{\citenamefont{{Cai} et~al.}(2013)\citenamefont{{Cai}, {Ma}, {Tang},
  and {Tuo}}}]{Cai2013}
\bibinfo{author}{\bibfnamefont{R.-G.} \bibnamefont{{Cai}}},
  \bibinfo{author}{\bibfnamefont{Y.-Z.} \bibnamefont{{Ma}}},
  \bibinfo{author}{\bibfnamefont{B.}~\bibnamefont{{Tang}}}, \bibnamefont{and}
  \bibinfo{author}{\bibfnamefont{Z.-L.} \bibnamefont{{Tuo}}},
  \bibinfo{journal}{\prd} \textbf{\bibinfo{volume}{87}}, \bibinfo{eid}{123522}
  (\bibinfo{year}{2013}), \eprint{1303.0961}.

\bibitem[{\citenamefont{{Chang} et~al.}(2018)\citenamefont{{Chang}, {Lin},
  {Sang}, and {Wang}}}]{Chang2018}
\bibinfo{author}{\bibfnamefont{Z.}~\bibnamefont{{Chang}}},
  \bibinfo{author}{\bibfnamefont{H.-N.} \bibnamefont{{Lin}}},
  \bibinfo{author}{\bibfnamefont{Y.}~\bibnamefont{{Sang}}}, \bibnamefont{and}
  \bibinfo{author}{\bibfnamefont{S.}~\bibnamefont{{Wang}}},
  \bibinfo{journal}{\mnras} \textbf{\bibinfo{volume}{478}},
  \bibinfo{pages}{3633} (\bibinfo{year}{2018}), \eprint{1711.11321}.

\bibitem[{\citenamefont{{Deng} and {Wei}}(2018)}]{DengWei2018}
\bibinfo{author}{\bibfnamefont{H.-K.} \bibnamefont{{Deng}}} \bibnamefont{and}
  \bibinfo{author}{\bibfnamefont{H.}~\bibnamefont{{Wei}}},
  \bibinfo{journal}{European Physical Journal C} \textbf{\bibinfo{volume}{78}},
  \bibinfo{eid}{755} (\bibinfo{year}{2018}), \eprint{1806.02773}.

\bibitem[{\citenamefont{{Sun} and {Wang}}(2018)}]{SunWang2018}
\bibinfo{author}{\bibfnamefont{Z.~Q.} \bibnamefont{{Sun}}} \bibnamefont{and}
  \bibinfo{author}{\bibfnamefont{F.~Y.} \bibnamefont{{Wang}}},
  \bibinfo{journal}{\mnras} \textbf{\bibinfo{volume}{478}},
  \bibinfo{pages}{5153} (\bibinfo{year}{2018}), \eprint{1805.09195}.

\bibitem[{\citenamefont{Andrade et~al.}(2018)\citenamefont{Andrade, Bengaly,
  Santos, and Alcaniz}}]{Andrade:2018eta}
\bibinfo{author}{\bibfnamefont{U.}~\bibnamefont{Andrade}},
  \bibinfo{author}{\bibfnamefont{C.~A.~P.} \bibnamefont{Bengaly}},
  \bibinfo{author}{\bibfnamefont{B.}~\bibnamefont{Santos}}, \bibnamefont{and}
  \bibinfo{author}{\bibfnamefont{J.~S.} \bibnamefont{Alcaniz}},
  \bibinfo{journal}{Astrophys. J.} \textbf{\bibinfo{volume}{865}},
  \bibinfo{pages}{119} (\bibinfo{year}{2018}), \eprint{1806.06990}.

\bibitem[{\citenamefont{{Zhao} et~al.}(2019)\citenamefont{{Zhao}, {Zhou}, and
  {Chang}}}]{Zhao2019}
\bibinfo{author}{\bibfnamefont{D.}~\bibnamefont{{Zhao}}},
  \bibinfo{author}{\bibfnamefont{Y.}~\bibnamefont{{Zhou}}}, \bibnamefont{and}
  \bibinfo{author}{\bibfnamefont{Z.}~\bibnamefont{{Chang}}},
  \bibinfo{journal}{\mnras} \textbf{\bibinfo{volume}{486}},
  \bibinfo{pages}{5679} (\bibinfo{year}{2019}), \eprint{1903.12401}.

\bibitem[{\citenamefont{Hu et~al.}(2020)\citenamefont{Hu, Wang, and
  Wang}}]{Hu:2020mzd}
\bibinfo{author}{\bibfnamefont{J.~P.} \bibnamefont{Hu}},
  \bibinfo{author}{\bibfnamefont{Y.~Y.} \bibnamefont{Wang}}, \bibnamefont{and}
  \bibinfo{author}{\bibfnamefont{F.~Y.} \bibnamefont{Wang}},
  \bibinfo{journal}{Astron. Astrophys.} \textbf{\bibinfo{volume}{643}},
  \bibinfo{pages}{A93} (\bibinfo{year}{2020}), \eprint{2008.12439}.

\bibitem[{\citenamefont{Rahman et~al.}(2022)\citenamefont{Rahman, Trotta,
  Boruah, Hudson, and van Dyk}}]{Rahman:2021mti}
\bibinfo{author}{\bibfnamefont{W.}~\bibnamefont{Rahman}},
  \bibinfo{author}{\bibfnamefont{R.}~\bibnamefont{Trotta}},
  \bibinfo{author}{\bibfnamefont{S.~S.} \bibnamefont{Boruah}},
  \bibinfo{author}{\bibfnamefont{M.~J.} \bibnamefont{Hudson}},
  \bibnamefont{and} \bibinfo{author}{\bibfnamefont{D.~A.} \bibnamefont{van
  Dyk}}, \bibinfo{journal}{Mon. Not. Roy. Astron. Soc.}
  \textbf{\bibinfo{volume}{514}}, \bibinfo{pages}{139} (\bibinfo{year}{2022}),
  \eprint{2108.12497}.

\bibitem[{\citenamefont{Schwarz and Weinhorst}(2007)}]{Schwarz:2007wf}
\bibinfo{author}{\bibfnamefont{D.~J.} \bibnamefont{Schwarz}} \bibnamefont{and}
  \bibinfo{author}{\bibfnamefont{B.}~\bibnamefont{Weinhorst}},
  \bibinfo{journal}{Astron. Astrophys.} \textbf{\bibinfo{volume}{474}},
  \bibinfo{pages}{717} (\bibinfo{year}{2007}), \eprint{0706.0165}.

\bibitem[{\citenamefont{Kashlinsky et~al.}(2009)\citenamefont{Kashlinsky,
  Atrio-Barandela, Kocevski, and Ebeling}}]{Kashlinsky:2008ut}
\bibinfo{author}{\bibfnamefont{A.}~\bibnamefont{Kashlinsky}},
  \bibinfo{author}{\bibfnamefont{F.}~\bibnamefont{Atrio-Barandela}},
  \bibinfo{author}{\bibfnamefont{D.}~\bibnamefont{Kocevski}}, \bibnamefont{and}
  \bibinfo{author}{\bibfnamefont{H.}~\bibnamefont{Ebeling}},
  \bibinfo{journal}{Astrophys. J. Lett.} \textbf{\bibinfo{volume}{686}},
  \bibinfo{pages}{L49} (\bibinfo{year}{2009}), \eprint{0809.3734}.

\bibitem[{\citenamefont{Antoniou and Perivolaropoulos}(2010)}]{Antoniou:2010gw}
\bibinfo{author}{\bibfnamefont{I.}~\bibnamefont{Antoniou}} \bibnamefont{and}
  \bibinfo{author}{\bibfnamefont{L.}~\bibnamefont{Perivolaropoulos}},
  \bibinfo{journal}{JCAP} \textbf{\bibinfo{volume}{12}}, \bibinfo{pages}{012}
  (\bibinfo{year}{2010}), \eprint{1007.4347}.

\bibitem[{\citenamefont{Kalus et~al.}(2013)\citenamefont{Kalus, Schwarz,
  Seikel, and Wiegand}}]{Kalus:2012zu}
\bibinfo{author}{\bibfnamefont{B.}~\bibnamefont{Kalus}},
  \bibinfo{author}{\bibfnamefont{D.~J.} \bibnamefont{Schwarz}},
  \bibinfo{author}{\bibfnamefont{M.}~\bibnamefont{Seikel}}, \bibnamefont{and}
  \bibinfo{author}{\bibfnamefont{A.}~\bibnamefont{Wiegand}},
  \bibinfo{journal}{Astron. Astrophys.} \textbf{\bibinfo{volume}{553}},
  \bibinfo{pages}{A56} (\bibinfo{year}{2013}), \eprint{1212.3691}.

\bibitem[{\citenamefont{Yoon et~al.}(2014)\citenamefont{Yoon, Huterer,
  Gibelyou, Kov\'acs, and Szapudi}}]{Yoon:2014daa}
\bibinfo{author}{\bibfnamefont{M.}~\bibnamefont{Yoon}},
  \bibinfo{author}{\bibfnamefont{D.}~\bibnamefont{Huterer}},
  \bibinfo{author}{\bibfnamefont{C.}~\bibnamefont{Gibelyou}},
  \bibinfo{author}{\bibfnamefont{A.}~\bibnamefont{Kov\'acs}}, \bibnamefont{and}
  \bibinfo{author}{\bibfnamefont{I.}~\bibnamefont{Szapudi}},
  \bibinfo{journal}{Mon. Not. Roy. Astron. Soc.}
  \textbf{\bibinfo{volume}{445}}, \bibinfo{pages}{L60} (\bibinfo{year}{2014}),
  \eprint{1406.1187}.

\bibitem[{\citenamefont{Tiwari and Nusser}(2016)}]{Tiwari:2015tba}
\bibinfo{author}{\bibfnamefont{P.}~\bibnamefont{Tiwari}} \bibnamefont{and}
  \bibinfo{author}{\bibfnamefont{A.}~\bibnamefont{Nusser}},
  \bibinfo{journal}{JCAP} \textbf{\bibinfo{volume}{03}}, \bibinfo{pages}{062}
  (\bibinfo{year}{2016}), \eprint{1509.02532}.

\bibitem[{\citenamefont{{Javanmardi} et~al.}(2015)\citenamefont{{Javanmardi},
  {Porciani}, {Kroupa}, and {Pflamm-Altenburg}}}]{Javanmardi2015}
\bibinfo{author}{\bibfnamefont{B.}~\bibnamefont{{Javanmardi}}},
  \bibinfo{author}{\bibfnamefont{C.}~\bibnamefont{{Porciani}}},
  \bibinfo{author}{\bibfnamefont{P.}~\bibnamefont{{Kroupa}}}, \bibnamefont{and}
  \bibinfo{author}{\bibfnamefont{J.}~\bibnamefont{{Pflamm-Altenburg}}},
  \bibinfo{journal}{\apj} \textbf{\bibinfo{volume}{810}}, \bibinfo{eid}{47}
  (\bibinfo{year}{2015}), \eprint{1507.07560}.

\bibitem[{\citenamefont{Colin et~al.}(2017)\citenamefont{Colin, Mohayaee,
  Rameez, and Sarkar}}]{Colin:2017juj}
\bibinfo{author}{\bibfnamefont{J.}~\bibnamefont{Colin}},
  \bibinfo{author}{\bibfnamefont{R.}~\bibnamefont{Mohayaee}},
  \bibinfo{author}{\bibfnamefont{M.}~\bibnamefont{Rameez}}, \bibnamefont{and}
  \bibinfo{author}{\bibfnamefont{S.}~\bibnamefont{Sarkar}},
  \bibinfo{journal}{Mon. Not. Roy. Astron. Soc.}
  \textbf{\bibinfo{volume}{471}}, \bibinfo{pages}{1045} (\bibinfo{year}{2017}),
  \eprint{1703.09376}.

\bibitem[{\citenamefont{Migkas et~al.}(2020)\citenamefont{Migkas,
  Schellenberger, Reiprich, Pacaud, Ramos-Ceja, and Lovisari}}]{Migkas:2020fza}
\bibinfo{author}{\bibfnamefont{K.}~\bibnamefont{Migkas}},
  \bibinfo{author}{\bibfnamefont{G.}~\bibnamefont{Schellenberger}},
  \bibinfo{author}{\bibfnamefont{T.~H.} \bibnamefont{Reiprich}},
  \bibinfo{author}{\bibfnamefont{F.}~\bibnamefont{Pacaud}},
  \bibinfo{author}{\bibfnamefont{M.~E.} \bibnamefont{Ramos-Ceja}},
  \bibnamefont{and} \bibinfo{author}{\bibfnamefont{L.}~\bibnamefont{Lovisari}},
  \bibinfo{journal}{Astron. Astrophys.} \textbf{\bibinfo{volume}{636}},
  \bibinfo{pages}{A15} (\bibinfo{year}{2020}), \eprint{2004.03305}.

\bibitem[{\citenamefont{Migkas et~al.}(2021)\citenamefont{Migkas, Pacaud,
  Schellenberger, Erler, Nguyen-Dang, Reiprich, Ramos-Ceja, and
  Lovisari}}]{Migkas:2021zdo}
\bibinfo{author}{\bibfnamefont{K.}~\bibnamefont{Migkas}},
  \bibinfo{author}{\bibfnamefont{F.}~\bibnamefont{Pacaud}},
  \bibinfo{author}{\bibfnamefont{G.}~\bibnamefont{Schellenberger}},
  \bibinfo{author}{\bibfnamefont{J.}~\bibnamefont{Erler}},
  \bibinfo{author}{\bibfnamefont{N.~T.} \bibnamefont{Nguyen-Dang}},
  \bibinfo{author}{\bibfnamefont{T.~H.} \bibnamefont{Reiprich}},
  \bibinfo{author}{\bibfnamefont{M.~E.} \bibnamefont{Ramos-Ceja}},
  \bibnamefont{and} \bibinfo{author}{\bibfnamefont{L.}~\bibnamefont{Lovisari}},
  \bibinfo{journal}{Astron. Astrophys.} \textbf{\bibinfo{volume}{649}},
  \bibinfo{pages}{A151} (\bibinfo{year}{2021}), \eprint{2103.13904}.

\bibitem[{\citenamefont{Secrest et~al.}(2021)\citenamefont{Secrest, von
  Hausegger, Rameez, Mohayaee, Sarkar, and Colin}}]{Secrest:2020has}
\bibinfo{author}{\bibfnamefont{N.~J.} \bibnamefont{Secrest}},
  \bibinfo{author}{\bibfnamefont{S.}~\bibnamefont{von Hausegger}},
  \bibinfo{author}{\bibfnamefont{M.}~\bibnamefont{Rameez}},
  \bibinfo{author}{\bibfnamefont{R.}~\bibnamefont{Mohayaee}},
  \bibinfo{author}{\bibfnamefont{S.}~\bibnamefont{Sarkar}}, \bibnamefont{and}
  \bibinfo{author}{\bibfnamefont{J.}~\bibnamefont{Colin}},
  \bibinfo{journal}{Astrophys. J. Lett.} \textbf{\bibinfo{volume}{908}},
  \bibinfo{pages}{L51} (\bibinfo{year}{2021}), \eprint{2009.14826}.

\bibitem[{\citenamefont{Siewert et~al.}(2021)\citenamefont{Siewert,
  Schmidt-Rubart, and Schwarz}}]{Siewert:2020krp}
\bibinfo{author}{\bibfnamefont{T.~M.} \bibnamefont{Siewert}},
  \bibinfo{author}{\bibfnamefont{M.}~\bibnamefont{Schmidt-Rubart}},
  \bibnamefont{and} \bibinfo{author}{\bibfnamefont{D.~J.}
  \bibnamefont{Schwarz}}, \bibinfo{journal}{Astron. Astrophys.}
  \textbf{\bibinfo{volume}{653}}, \bibinfo{pages}{A9} (\bibinfo{year}{2021}),
  \eprint{2010.08366}.

\bibitem[{\citenamefont{Luongo et~al.}(2022)\citenamefont{Luongo, Muccino,
  Colg\'ain, Sheikh-Jabbari, and Yin}}]{Luongo:2021nqh}
\bibinfo{author}{\bibfnamefont{O.}~\bibnamefont{Luongo}},
  \bibinfo{author}{\bibfnamefont{M.}~\bibnamefont{Muccino}},
  \bibinfo{author}{\bibfnamefont{E.~O.} \bibnamefont{Colg\'ain}},
  \bibinfo{author}{\bibfnamefont{M.~M.} \bibnamefont{Sheikh-Jabbari}},
  \bibnamefont{and} \bibinfo{author}{\bibfnamefont{L.}~\bibnamefont{Yin}},
  \bibinfo{journal}{Phys. Rev. D} \textbf{\bibinfo{volume}{105}},
  \bibinfo{pages}{103510} (\bibinfo{year}{2022}), \eprint{2108.13228}.

\bibitem[{\citenamefont{Krishnan et~al.}(2022)\citenamefont{Krishnan, Mohayaee,
  Colg\'ain, Sheikh-Jabbari, and Yin}}]{Krishnan:2021jmh}
\bibinfo{author}{\bibfnamefont{C.}~\bibnamefont{Krishnan}},
  \bibinfo{author}{\bibfnamefont{R.}~\bibnamefont{Mohayaee}},
  \bibinfo{author}{\bibfnamefont{E.~O.} \bibnamefont{Colg\'ain}},
  \bibinfo{author}{\bibfnamefont{M.~M.} \bibnamefont{Sheikh-Jabbari}},
  \bibnamefont{and} \bibinfo{author}{\bibfnamefont{L.}~\bibnamefont{Yin}},
  \bibinfo{journal}{Phys. Rev. D} \textbf{\bibinfo{volume}{105}},
  \bibinfo{pages}{063514} (\bibinfo{year}{2022}), \eprint{2106.02532}.

\bibitem[{\citenamefont{Dainotti et~al.}(2021)\citenamefont{Dainotti,
  De~Simone, Schiavone, Montani, Rinaldi, and Lambiase}}]{Dainotti:2021pqg}
\bibinfo{author}{\bibfnamefont{M.~G.} \bibnamefont{Dainotti}},
  \bibinfo{author}{\bibfnamefont{B.}~\bibnamefont{De~Simone}},
  \bibinfo{author}{\bibfnamefont{T.}~\bibnamefont{Schiavone}},
  \bibinfo{author}{\bibfnamefont{G.}~\bibnamefont{Montani}},
  \bibinfo{author}{\bibfnamefont{E.}~\bibnamefont{Rinaldi}}, \bibnamefont{and}
  \bibinfo{author}{\bibfnamefont{G.}~\bibnamefont{Lambiase}},
  \bibinfo{journal}{Astrophys. J.} \textbf{\bibinfo{volume}{912}},
  \bibinfo{pages}{150} (\bibinfo{year}{2021}), \eprint{2103.02117}.

\bibitem[{\citenamefont{Hoffman et~al.}(2017)\citenamefont{Hoffman, Pomarede,
  Brent~Tully, and Courtois}}]{Hoffman:2017ako}
\bibinfo{author}{\bibfnamefont{Y.}~\bibnamefont{Hoffman}},
  \bibinfo{author}{\bibfnamefont{D.}~\bibnamefont{Pomarede}},
  \bibinfo{author}{\bibfnamefont{R.}~\bibnamefont{Brent~Tully}},
  \bibnamefont{and} \bibinfo{author}{\bibfnamefont{H.}~\bibnamefont{Courtois}}
  (\bibinfo{year}{2017}), \eprint{1702.02483}.

\bibitem[{\citenamefont{Shaya et~al.}(2022)\citenamefont{Shaya, Tully,
  Pomar\`ede, and Peel}}]{Shaya}
\bibinfo{author}{\bibfnamefont{E.~J.} \bibnamefont{Shaya}},
  \bibinfo{author}{\bibfnamefont{R.~B.} \bibnamefont{Tully}},
  \bibinfo{author}{\bibfnamefont{D.}~\bibnamefont{Pomar\`ede}},
  \bibnamefont{and} \bibinfo{author}{\bibfnamefont{A.}~\bibnamefont{Peel}},
  \bibinfo{journal}{Astrophys. J.} \textbf{\bibinfo{volume}{927}},
  \bibinfo{pages}{168} (\bibinfo{year}{2022}), \eprint{2201.12315}.

\bibitem[{\citenamefont{Koda et~al.}(2014)\citenamefont{Koda, Blake, Davis,
  Magoulas, Springob, Scrimgeour, Johnson, Poole, and
  Staveley-Smith}}]{Koda:2013eya}
\bibinfo{author}{\bibfnamefont{J.}~\bibnamefont{Koda}},
  \bibinfo{author}{\bibfnamefont{C.}~\bibnamefont{Blake}},
  \bibinfo{author}{\bibfnamefont{T.}~\bibnamefont{Davis}},
  \bibinfo{author}{\bibfnamefont{C.}~\bibnamefont{Magoulas}},
  \bibinfo{author}{\bibfnamefont{C.~M.} \bibnamefont{Springob}},
  \bibinfo{author}{\bibfnamefont{M.}~\bibnamefont{Scrimgeour}},
  \bibinfo{author}{\bibfnamefont{A.}~\bibnamefont{Johnson}},
  \bibinfo{author}{\bibfnamefont{G.~B.} \bibnamefont{Poole}}, \bibnamefont{and}
  \bibinfo{author}{\bibfnamefont{L.}~\bibnamefont{Staveley-Smith}},
  \bibinfo{journal}{Mon. Not. Roy. Astron. Soc.}
  \textbf{\bibinfo{volume}{445}}, \bibinfo{pages}{4267} (\bibinfo{year}{2014}),
  \eprint{1312.1022}.

\bibitem[{\citenamefont{Ma et~al.}(2011)\citenamefont{Ma, Gordon, and
  Feldman}}]{Ma}
\bibinfo{author}{\bibfnamefont{Y.-Z.} \bibnamefont{Ma}},
  \bibinfo{author}{\bibfnamefont{C.}~\bibnamefont{Gordon}}, \bibnamefont{and}
  \bibinfo{author}{\bibfnamefont{H.~A.} \bibnamefont{Feldman}},
  \bibinfo{journal}{Phys. Rev. D} \textbf{\bibinfo{volume}{83}},
  \bibinfo{pages}{103002} (\bibinfo{year}{2011}), \eprint{1010.4276}.

\bibitem[{\citenamefont{Hellwing et~al.}(2017)\citenamefont{Hellwing, Nusser,
  Feix, and Bilicki}}]{Hellwing}
\bibinfo{author}{\bibfnamefont{W.~A.} \bibnamefont{Hellwing}},
  \bibinfo{author}{\bibfnamefont{A.}~\bibnamefont{Nusser}},
  \bibinfo{author}{\bibfnamefont{M.}~\bibnamefont{Feix}}, \bibnamefont{and}
  \bibinfo{author}{\bibfnamefont{M.}~\bibnamefont{Bilicki}},
  \bibinfo{journal}{Mon. Not. Roy. Astron. Soc.}
  \textbf{\bibinfo{volume}{467}}, \bibinfo{pages}{2787} (\bibinfo{year}{2017}),
  \eprint{1609.07120}.

\bibitem[{\citenamefont{Huterer et~al.}(2017)\citenamefont{Huterer, Shafer,
  Scolnic, and Schmidt}}]{Huterer}
\bibinfo{author}{\bibfnamefont{D.}~\bibnamefont{Huterer}},
  \bibinfo{author}{\bibfnamefont{D.}~\bibnamefont{Shafer}},
  \bibinfo{author}{\bibfnamefont{D.}~\bibnamefont{Scolnic}}, \bibnamefont{and}
  \bibinfo{author}{\bibfnamefont{F.}~\bibnamefont{Schmidt}},
  \bibinfo{journal}{JCAP} \textbf{\bibinfo{volume}{05}}, \bibinfo{pages}{015}
  (\bibinfo{year}{2017}), \eprint{1611.09862}.

\bibitem[{\citenamefont{Bengaly et~al.}(2019)\citenamefont{Bengaly, Larena, and
  Maartens}}]{Larena}
\bibinfo{author}{\bibfnamefont{C.~A.~P.} \bibnamefont{Bengaly}},
  \bibinfo{author}{\bibfnamefont{J.}~\bibnamefont{Larena}}, \bibnamefont{and}
  \bibinfo{author}{\bibfnamefont{R.}~\bibnamefont{Maartens}},
  \bibinfo{journal}{JCAP} \textbf{\bibinfo{volume}{03}}, \bibinfo{pages}{001}
  (\bibinfo{year}{2019}), \eprint{1805.12456}.

\bibitem[{\citenamefont{Strauss and Willick}(1995)}]{Strauss:1995fz}
\bibinfo{author}{\bibfnamefont{M.~A.} \bibnamefont{Strauss}} \bibnamefont{and}
  \bibinfo{author}{\bibfnamefont{J.~A.} \bibnamefont{Willick}},
  \bibinfo{journal}{Phys. Rept.} \textbf{\bibinfo{volume}{261}},
  \bibinfo{pages}{271} (\bibinfo{year}{1995}), \eprint{astro-ph/9502079}.

\bibitem[{\citenamefont{Macpherson and Heinesen}(2021)}]{Macpherson}
\bibinfo{author}{\bibfnamefont{H.~J.} \bibnamefont{Macpherson}}
  \bibnamefont{and} \bibinfo{author}{\bibfnamefont{A.}~\bibnamefont{Heinesen}},
  \bibinfo{journal}{Phys. Rev. D} \textbf{\bibinfo{volume}{104}},
  \bibinfo{pages}{023525} (\bibinfo{year}{2021}),
  \urlprefix\url{https://link.aps.org/doi/10.1103/PhysRevD.104.023525}.

\bibitem[{\citenamefont{Dhawan et~al.}(2022)\citenamefont{Dhawan, Borderies,
  Macpherson, and Heinesen}}]{Dhawan}
\bibinfo{author}{\bibfnamefont{S.}~\bibnamefont{Dhawan}},
  \bibinfo{author}{\bibfnamefont{A.}~\bibnamefont{Borderies}},
  \bibinfo{author}{\bibfnamefont{H.~J.} \bibnamefont{Macpherson}},
  \bibnamefont{and} \bibinfo{author}{\bibfnamefont{A.}~\bibnamefont{Heinesen}}
  (\bibinfo{year}{2022}), \eprint{2205.12692}.

\bibitem[{\citenamefont{{Kristian} and {Sachs}}(1966)}]{Kristian}
\bibinfo{author}{\bibfnamefont{J.}~\bibnamefont{{Kristian}}} \bibnamefont{and}
  \bibinfo{author}{\bibfnamefont{R.~K.} \bibnamefont{{Sachs}}},
  \bibinfo{journal}{\apj} \textbf{\bibinfo{volume}{143}}, \bibinfo{pages}{379}
  (\bibinfo{year}{1966}).

\bibitem[{\citenamefont{{Ellis} et~al.}(1985)\citenamefont{{Ellis}, {Nel},
  {Maartens}, {Stoeger}, and {Whitman}}}]{ellis85}
\bibinfo{author}{\bibfnamefont{G.~F.~R.} \bibnamefont{{Ellis}}},
  \bibinfo{author}{\bibfnamefont{S.~D.} \bibnamefont{{Nel}}},
  \bibinfo{author}{\bibfnamefont{R.}~\bibnamefont{{Maartens}}},
  \bibinfo{author}{\bibfnamefont{W.~R.} \bibnamefont{{Stoeger}}},
  \bibnamefont{and} \bibinfo{author}{\bibfnamefont{A.~P.}
  \bibnamefont{{Whitman}}}, \bibinfo{journal}{\physrep}
  \textbf{\bibinfo{volume}{124}}, \bibinfo{pages}{315} (\bibinfo{year}{1985}).

\bibitem[{\citenamefont{{Lilje} et~al.}(1986)\citenamefont{{Lilje}, {Yahil},
  and {Jones}}}]{Lilje_1986}
\bibinfo{author}{\bibfnamefont{P.~B.} \bibnamefont{{Lilje}}},
  \bibinfo{author}{\bibfnamefont{A.}~\bibnamefont{{Yahil}}}, \bibnamefont{and}
  \bibinfo{author}{\bibfnamefont{B.~J.~T.} \bibnamefont{{Jones}}},
  \bibinfo{journal}{\apj} \textbf{\bibinfo{volume}{307}}, \bibinfo{pages}{91}
  (\bibinfo{year}{1986}).

\bibitem[{\citenamefont{{Courtois} et~al.}(2012)\citenamefont{{Courtois},
  {Hoffman}, {Tully}, and {Gottl{\"o}ber}}}]{Courtois}
\bibinfo{author}{\bibfnamefont{H.~M.} \bibnamefont{{Courtois}}},
  \bibinfo{author}{\bibfnamefont{Y.}~\bibnamefont{{Hoffman}}},
  \bibinfo{author}{\bibfnamefont{R.~B.} \bibnamefont{{Tully}}},
  \bibnamefont{and}
  \bibinfo{author}{\bibfnamefont{S.}~\bibnamefont{{Gottl{\"o}ber}}},
  \bibinfo{journal}{\apj} \textbf{\bibinfo{volume}{744}}, \bibinfo{eid}{43}
  (\bibinfo{year}{2012}), \eprint{1109.3856}.

\bibitem[{\citenamefont{Parnovsky and Parnowski}(2013)}]{Parnovsky}
\bibinfo{author}{\bibfnamefont{S.~L.} \bibnamefont{Parnovsky}}
  \bibnamefont{and} \bibinfo{author}{\bibfnamefont{A.~S.}
  \bibnamefont{Parnowski}}, \bibinfo{journal}{Astrophys. Space Sci.}
  \textbf{\bibinfo{volume}{343}}, \bibinfo{pages}{747} (\bibinfo{year}{2013}),
  \eprint{1210.0895}.

\bibitem[{\citenamefont{Aghanim et~al.}(2020)}]{Planck:2018nkj}
\bibinfo{author}{\bibfnamefont{N.}~\bibnamefont{Aghanim}} \bibnamefont{et~al.}
  (\bibinfo{collaboration}{Planck}), \bibinfo{journal}{Astron. Astrophys.}
  \textbf{\bibinfo{volume}{641}}, \bibinfo{pages}{A1} (\bibinfo{year}{2020}),
  \eprint{1807.06205}.

\bibitem[{\citenamefont{G\'orski et~al.}(2005)\citenamefont{G\'orski, Hivon,
  Banday, Wandelt, Hansen, Reinecke, and Bartelman}}]{Gorski:2004by}
\bibinfo{author}{\bibfnamefont{K.~M.} \bibnamefont{G\'orski}},
  \bibinfo{author}{\bibfnamefont{E.}~\bibnamefont{Hivon}},
  \bibinfo{author}{\bibfnamefont{A.~J.} \bibnamefont{Banday}},
  \bibinfo{author}{\bibfnamefont{B.~D.} \bibnamefont{Wandelt}},
  \bibinfo{author}{\bibfnamefont{F.~K.} \bibnamefont{Hansen}},
  \bibinfo{author}{\bibfnamefont{M.}~\bibnamefont{Reinecke}}, \bibnamefont{and}
  \bibinfo{author}{\bibfnamefont{M.}~\bibnamefont{Bartelman}},
  \bibinfo{journal}{Astrophys. J.} \textbf{\bibinfo{volume}{622}},
  \bibinfo{pages}{759} (\bibinfo{year}{2005}), \eprint{astro-ph/0409513}.

\bibitem[{\citenamefont{{Scolnic} et~al.}(2018)\citenamefont{{Scolnic},
  {Jones}, {Rest}, {Pan}, {Chornock}, {Foley}, {Huber}, {Kessler}, {Narayan},
  {Riess} et~al.}}]{Scolnic2018}
\bibinfo{author}{\bibfnamefont{D.~M.} \bibnamefont{{Scolnic}}},
  \bibinfo{author}{\bibfnamefont{D.~O.} \bibnamefont{{Jones}}},
  \bibinfo{author}{\bibfnamefont{A.}~\bibnamefont{{Rest}}},
  \bibinfo{author}{\bibfnamefont{Y.~C.} \bibnamefont{{Pan}}},
  \bibinfo{author}{\bibfnamefont{R.}~\bibnamefont{{Chornock}}},
  \bibinfo{author}{\bibfnamefont{R.~J.} \bibnamefont{{Foley}}},
  \bibinfo{author}{\bibfnamefont{M.~E.} \bibnamefont{{Huber}}},
  \bibinfo{author}{\bibfnamefont{R.}~\bibnamefont{{Kessler}}},
  \bibinfo{author}{\bibfnamefont{G.}~\bibnamefont{{Narayan}}},
  \bibinfo{author}{\bibfnamefont{A.~G.} \bibnamefont{{Riess}}},
  \bibnamefont{et~al.}, \bibinfo{journal}{\apj} \textbf{\bibinfo{volume}{859}},
  \bibinfo{eid}{101} (\bibinfo{year}{2018}), \eprint{1710.00845}.

\bibitem[{\citenamefont{{Guy} et~al.}(2010)\citenamefont{{Guy}, {Sullivan},
  {Conley}, {Regnault}, {Astier}, {Balland}, {Basa}, {Carlberg}, {Fouchez},
  {Hardin} et~al.}}]{Guy}
\bibinfo{author}{\bibfnamefont{J.}~\bibnamefont{{Guy}}},
  \bibinfo{author}{\bibfnamefont{M.}~\bibnamefont{{Sullivan}}},
  \bibinfo{author}{\bibfnamefont{A.}~\bibnamefont{{Conley}}},
  \bibinfo{author}{\bibfnamefont{N.}~\bibnamefont{{Regnault}}},
  \bibinfo{author}{\bibfnamefont{P.}~\bibnamefont{{Astier}}},
  \bibinfo{author}{\bibfnamefont{C.}~\bibnamefont{{Balland}}},
  \bibinfo{author}{\bibfnamefont{S.}~\bibnamefont{{Basa}}},
  \bibinfo{author}{\bibfnamefont{R.~G.} \bibnamefont{{Carlberg}}},
  \bibinfo{author}{\bibfnamefont{D.}~\bibnamefont{{Fouchez}}},
  \bibinfo{author}{\bibfnamefont{D.}~\bibnamefont{{Hardin}}},
  \bibnamefont{et~al.}, \bibinfo{journal}{\aap} \textbf{\bibinfo{volume}{523}},
  \bibinfo{eid}{A7} (\bibinfo{year}{2010}), \eprint{1010.4743}.

\bibitem[{\citenamefont{{Smith} et~al.}(2012)\citenamefont{{Smith}, {Nichol},
  {Dilday}, {Marriner}, {Kessler}, {Bassett}, {Cinabro}, {Frieman},
  {Garnavich}, {Jha} et~al.}}]{Smith}
\bibinfo{author}{\bibfnamefont{M.}~\bibnamefont{{Smith}}},
  \bibinfo{author}{\bibfnamefont{R.~C.} \bibnamefont{{Nichol}}},
  \bibinfo{author}{\bibfnamefont{B.}~\bibnamefont{{Dilday}}},
  \bibinfo{author}{\bibfnamefont{J.}~\bibnamefont{{Marriner}}},
  \bibinfo{author}{\bibfnamefont{R.}~\bibnamefont{{Kessler}}},
  \bibinfo{author}{\bibfnamefont{B.}~\bibnamefont{{Bassett}}},
  \bibinfo{author}{\bibfnamefont{D.}~\bibnamefont{{Cinabro}}},
  \bibinfo{author}{\bibfnamefont{J.}~\bibnamefont{{Frieman}}},
  \bibinfo{author}{\bibfnamefont{P.}~\bibnamefont{{Garnavich}}},
  \bibinfo{author}{\bibfnamefont{S.~W.} \bibnamefont{{Jha}}},
  \bibnamefont{et~al.}, \bibinfo{journal}{\apj} \textbf{\bibinfo{volume}{755}},
  \bibinfo{eid}{61} (\bibinfo{year}{2012}), \eprint{1108.4923}.

\bibitem[{\citenamefont{Sako et~al.}(2018)}]{Sako}
\bibinfo{author}{\bibfnamefont{M.}~\bibnamefont{Sako}} \bibnamefont{et~al.}
  (\bibinfo{collaboration}{SDSS}), \bibinfo{journal}{Publ. Astron. Soc. Pac.}
  \textbf{\bibinfo{volume}{130}}, \bibinfo{pages}{064002}
  (\bibinfo{year}{2018}), \eprint{1401.3317}.

\bibitem[{\citenamefont{{Riess} et~al.}(1999)\citenamefont{{Riess}, {Kirshner},
  {Schmidt}, {Jha}, {Challis}, {Garnavich}, {Esin}, {Carpenter}, {Grashius},
  {Schild} et~al.}}]{Riess1999}
\bibinfo{author}{\bibfnamefont{A.~G.} \bibnamefont{{Riess}}},
  \bibinfo{author}{\bibfnamefont{R.~P.} \bibnamefont{{Kirshner}}},
  \bibinfo{author}{\bibfnamefont{B.~P.} \bibnamefont{{Schmidt}}},
  \bibinfo{author}{\bibfnamefont{S.}~\bibnamefont{{Jha}}},
  \bibinfo{author}{\bibfnamefont{P.}~\bibnamefont{{Challis}}},
  \bibinfo{author}{\bibfnamefont{P.~M.} \bibnamefont{{Garnavich}}},
  \bibinfo{author}{\bibfnamefont{A.~A.} \bibnamefont{{Esin}}},
  \bibinfo{author}{\bibfnamefont{C.}~\bibnamefont{{Carpenter}}},
  \bibinfo{author}{\bibfnamefont{R.}~\bibnamefont{{Grashius}}},
  \bibinfo{author}{\bibfnamefont{R.~E.} \bibnamefont{{Schild}}},
  \bibnamefont{et~al.}, \bibinfo{journal}{\aj} \textbf{\bibinfo{volume}{117}},
  \bibinfo{pages}{707} (\bibinfo{year}{1999}), \eprint{astro-ph/9810291}.

\bibitem[{\citenamefont{{Jha} et~al.}(2006)\citenamefont{{Jha}, {Kirshner},
  {Challis}, {Garnavich}, {Matheson}, {Soderberg}, {Graves}, {Hicken}, {Alves},
  {Arce} et~al.}}]{Jha}
\bibinfo{author}{\bibfnamefont{S.}~\bibnamefont{{Jha}}},
  \bibinfo{author}{\bibfnamefont{R.~P.} \bibnamefont{{Kirshner}}},
  \bibinfo{author}{\bibfnamefont{P.}~\bibnamefont{{Challis}}},
  \bibinfo{author}{\bibfnamefont{P.~M.} \bibnamefont{{Garnavich}}},
  \bibinfo{author}{\bibfnamefont{T.}~\bibnamefont{{Matheson}}},
  \bibinfo{author}{\bibfnamefont{A.~M.} \bibnamefont{{Soderberg}}},
  \bibinfo{author}{\bibfnamefont{G.~J.~M.} \bibnamefont{{Graves}}},
  \bibinfo{author}{\bibfnamefont{M.}~\bibnamefont{{Hicken}}},
  \bibinfo{author}{\bibfnamefont{J.~F.} \bibnamefont{{Alves}}},
  \bibinfo{author}{\bibfnamefont{H.~G.} \bibnamefont{{Arce}}},
  \bibnamefont{et~al.}, \bibinfo{journal}{\aj} \textbf{\bibinfo{volume}{131}},
  \bibinfo{pages}{527} (\bibinfo{year}{2006}), \eprint{astro-ph/0509234}.

\bibitem[{\citenamefont{{Hicken}
  et~al.}(2009{\natexlab{a}})\citenamefont{{Hicken}, {Wood-Vasey}, {Blondin},
  {Challis}, {Jha}, {Kelly}, {Rest}, and {Kirshner}}}]{Hicken2009a}
\bibinfo{author}{\bibfnamefont{M.}~\bibnamefont{{Hicken}}},
  \bibinfo{author}{\bibfnamefont{W.~M.} \bibnamefont{{Wood-Vasey}}},
  \bibinfo{author}{\bibfnamefont{S.}~\bibnamefont{{Blondin}}},
  \bibinfo{author}{\bibfnamefont{P.}~\bibnamefont{{Challis}}},
  \bibinfo{author}{\bibfnamefont{S.}~\bibnamefont{{Jha}}},
  \bibinfo{author}{\bibfnamefont{P.~L.} \bibnamefont{{Kelly}}},
  \bibinfo{author}{\bibfnamefont{A.}~\bibnamefont{{Rest}}}, \bibnamefont{and}
  \bibinfo{author}{\bibfnamefont{R.~P.} \bibnamefont{{Kirshner}}},
  \bibinfo{journal}{\apj} \textbf{\bibinfo{volume}{700}}, \bibinfo{pages}{1097}
  (\bibinfo{year}{2009}{\natexlab{a}}), \eprint{0901.4804}.

\bibitem[{\citenamefont{{Hicken}
  et~al.}(2009{\natexlab{b}})\citenamefont{{Hicken}, {Challis}, {Jha},
  {Kirshner}, {Matheson}, {Modjaz}, {Rest}, {Wood-Vasey}, {Bakos}, {Barton}
  et~al.}}]{Hicken2009b}
\bibinfo{author}{\bibfnamefont{M.}~\bibnamefont{{Hicken}}},
  \bibinfo{author}{\bibfnamefont{P.}~\bibnamefont{{Challis}}},
  \bibinfo{author}{\bibfnamefont{S.}~\bibnamefont{{Jha}}},
  \bibinfo{author}{\bibfnamefont{R.~P.} \bibnamefont{{Kirshner}}},
  \bibinfo{author}{\bibfnamefont{T.}~\bibnamefont{{Matheson}}},
  \bibinfo{author}{\bibfnamefont{M.}~\bibnamefont{{Modjaz}}},
  \bibinfo{author}{\bibfnamefont{A.}~\bibnamefont{{Rest}}},
  \bibinfo{author}{\bibfnamefont{W.~M.} \bibnamefont{{Wood-Vasey}}},
  \bibinfo{author}{\bibfnamefont{G.}~\bibnamefont{{Bakos}}},
  \bibinfo{author}{\bibfnamefont{E.~J.} \bibnamefont{{Barton}}},
  \bibnamefont{et~al.}, \bibinfo{journal}{\apj} \textbf{\bibinfo{volume}{700}},
  \bibinfo{pages}{331} (\bibinfo{year}{2009}{\natexlab{b}}),
  \eprint{0901.4787}.

\bibitem[{\citenamefont{{Hicken} et~al.}(2012)\citenamefont{{Hicken},
  {Challis}, {Kirshner}, {Rest}, {Cramer}, {Wood-Vasey}, {Bakos}, {Berlind},
  {Brown}, {Caldwell} et~al.}}]{Hicken2012}
\bibinfo{author}{\bibfnamefont{M.}~\bibnamefont{{Hicken}}},
  \bibinfo{author}{\bibfnamefont{P.}~\bibnamefont{{Challis}}},
  \bibinfo{author}{\bibfnamefont{R.~P.} \bibnamefont{{Kirshner}}},
  \bibinfo{author}{\bibfnamefont{A.}~\bibnamefont{{Rest}}},
  \bibinfo{author}{\bibfnamefont{C.~E.} \bibnamefont{{Cramer}}},
  \bibinfo{author}{\bibfnamefont{W.~M.} \bibnamefont{{Wood-Vasey}}},
  \bibinfo{author}{\bibfnamefont{G.}~\bibnamefont{{Bakos}}},
  \bibinfo{author}{\bibfnamefont{P.}~\bibnamefont{{Berlind}}},
  \bibinfo{author}{\bibfnamefont{W.~R.} \bibnamefont{{Brown}}},
  \bibinfo{author}{\bibfnamefont{N.}~\bibnamefont{{Caldwell}}},
  \bibnamefont{et~al.}, \bibinfo{journal}{\apjs}
  \textbf{\bibinfo{volume}{200}}, \bibinfo{eid}{12} (\bibinfo{year}{2012}),
  \eprint{1205.4493}.

\bibitem[{\citenamefont{{Contreras} et~al.}(2010)\citenamefont{{Contreras},
  {Hamuy}, {Phillips}, {Folatelli}, {Suntzeff}, {Persson}, {Stritzinger},
  {Boldt}, {Gonz{\'a}lez}, {Krzeminski} et~al.}}]{Contreras}
\bibinfo{author}{\bibfnamefont{C.}~\bibnamefont{{Contreras}}},
  \bibinfo{author}{\bibfnamefont{M.}~\bibnamefont{{Hamuy}}},
  \bibinfo{author}{\bibfnamefont{M.~M.} \bibnamefont{{Phillips}}},
  \bibinfo{author}{\bibfnamefont{G.}~\bibnamefont{{Folatelli}}},
  \bibinfo{author}{\bibfnamefont{N.~B.} \bibnamefont{{Suntzeff}}},
  \bibinfo{author}{\bibfnamefont{S.~E.} \bibnamefont{{Persson}}},
  \bibinfo{author}{\bibfnamefont{M.}~\bibnamefont{{Stritzinger}}},
  \bibinfo{author}{\bibfnamefont{L.}~\bibnamefont{{Boldt}}},
  \bibinfo{author}{\bibfnamefont{S.}~\bibnamefont{{Gonz{\'a}lez}}},
  \bibinfo{author}{\bibfnamefont{W.}~\bibnamefont{{Krzeminski}}},
  \bibnamefont{et~al.}, \bibinfo{journal}{\aj} \textbf{\bibinfo{volume}{139}},
  \bibinfo{pages}{519} (\bibinfo{year}{2010}), \eprint{0910.3330}.

\bibitem[{\citenamefont{{Rodney} et~al.}(2014)\citenamefont{{Rodney}, {Riess},
  {Strolger}, {Dahlen}, {Graur}, {Casertano}, {Dickinson}, {Ferguson},
  {Garnavich}, {Hayden} et~al.}}]{Rodney}
\bibinfo{author}{\bibfnamefont{S.~A.} \bibnamefont{{Rodney}}},
  \bibinfo{author}{\bibfnamefont{A.~G.} \bibnamefont{{Riess}}},
  \bibinfo{author}{\bibfnamefont{L.-G.} \bibnamefont{{Strolger}}},
  \bibinfo{author}{\bibfnamefont{T.}~\bibnamefont{{Dahlen}}},
  \bibinfo{author}{\bibfnamefont{O.}~\bibnamefont{{Graur}}},
  \bibinfo{author}{\bibfnamefont{S.}~\bibnamefont{{Casertano}}},
  \bibinfo{author}{\bibfnamefont{M.~E.} \bibnamefont{{Dickinson}}},
  \bibinfo{author}{\bibfnamefont{H.~C.} \bibnamefont{{Ferguson}}},
  \bibinfo{author}{\bibfnamefont{P.}~\bibnamefont{{Garnavich}}},
  \bibinfo{author}{\bibfnamefont{B.}~\bibnamefont{{Hayden}}},
  \bibnamefont{et~al.}, \bibinfo{journal}{\aj} \textbf{\bibinfo{volume}{148}},
  \bibinfo{eid}{13} (\bibinfo{year}{2014}), \eprint{1401.7978}.

\bibitem[{\citenamefont{{Graur} et~al.}(2014)\citenamefont{{Graur}, {Rodney},
  {Maoz}, {Riess}, {Jha}, {Postman}, {Dahlen}, {Holoien}, {McCully}, {Patel}
  et~al.}}]{Graur}
\bibinfo{author}{\bibfnamefont{O.}~\bibnamefont{{Graur}}},
  \bibinfo{author}{\bibfnamefont{S.~A.} \bibnamefont{{Rodney}}},
  \bibinfo{author}{\bibfnamefont{D.}~\bibnamefont{{Maoz}}},
  \bibinfo{author}{\bibfnamefont{A.~G.} \bibnamefont{{Riess}}},
  \bibinfo{author}{\bibfnamefont{S.~W.} \bibnamefont{{Jha}}},
  \bibinfo{author}{\bibfnamefont{M.}~\bibnamefont{{Postman}}},
  \bibinfo{author}{\bibfnamefont{T.}~\bibnamefont{{Dahlen}}},
  \bibinfo{author}{\bibfnamefont{T.~W.~S.} \bibnamefont{{Holoien}}},
  \bibinfo{author}{\bibfnamefont{C.}~\bibnamefont{{McCully}}},
  \bibinfo{author}{\bibfnamefont{B.}~\bibnamefont{{Patel}}},
  \bibnamefont{et~al.}, \bibinfo{journal}{\apj} \textbf{\bibinfo{volume}{783}},
  \bibinfo{eid}{28} (\bibinfo{year}{2014}), \eprint{1310.3495}.

\bibitem[{\citenamefont{{Riess} et~al.}(2018)\citenamefont{{Riess}, {Rodney},
  {Scolnic}, {Shafer}, {Strolger}, {Ferguson}, {Postman}, {Graur}, {Maoz},
  {Jha} et~al.}}]{Riess2018}
\bibinfo{author}{\bibfnamefont{A.~G.} \bibnamefont{{Riess}}},
  \bibinfo{author}{\bibfnamefont{S.~A.} \bibnamefont{{Rodney}}},
  \bibinfo{author}{\bibfnamefont{D.~M.} \bibnamefont{{Scolnic}}},
  \bibinfo{author}{\bibfnamefont{D.~L.} \bibnamefont{{Shafer}}},
  \bibinfo{author}{\bibfnamefont{L.-G.} \bibnamefont{{Strolger}}},
  \bibinfo{author}{\bibfnamefont{H.~C.} \bibnamefont{{Ferguson}}},
  \bibinfo{author}{\bibfnamefont{M.}~\bibnamefont{{Postman}}},
  \bibinfo{author}{\bibfnamefont{O.}~\bibnamefont{{Graur}}},
  \bibinfo{author}{\bibfnamefont{D.}~\bibnamefont{{Maoz}}},
  \bibinfo{author}{\bibfnamefont{S.~W.} \bibnamefont{{Jha}}},
  \bibnamefont{et~al.}, \bibinfo{journal}{\apj} \textbf{\bibinfo{volume}{853}},
  \bibinfo{eid}{126} (\bibinfo{year}{2018}), \eprint{1710.00844}.

\bibitem[{\citenamefont{{Riess} et~al.}(2007)\citenamefont{{Riess}, {Strolger},
  {Casertano}, {Ferguson}, {Mobasher}, {Gold}, {Challis}, {Filippenko}, {Jha},
  {Li} et~al.}}]{Riess2007}
\bibinfo{author}{\bibfnamefont{A.~G.} \bibnamefont{{Riess}}},
  \bibinfo{author}{\bibfnamefont{L.-G.} \bibnamefont{{Strolger}}},
  \bibinfo{author}{\bibfnamefont{S.}~\bibnamefont{{Casertano}}},
  \bibinfo{author}{\bibfnamefont{H.~C.} \bibnamefont{{Ferguson}}},
  \bibinfo{author}{\bibfnamefont{B.}~\bibnamefont{{Mobasher}}},
  \bibinfo{author}{\bibfnamefont{B.}~\bibnamefont{{Gold}}},
  \bibinfo{author}{\bibfnamefont{P.~J.} \bibnamefont{{Challis}}},
  \bibinfo{author}{\bibfnamefont{A.~V.} \bibnamefont{{Filippenko}}},
  \bibinfo{author}{\bibfnamefont{S.}~\bibnamefont{{Jha}}},
  \bibinfo{author}{\bibfnamefont{W.}~\bibnamefont{{Li}}}, \bibnamefont{et~al.},
  \bibinfo{journal}{\apj} \textbf{\bibinfo{volume}{659}}, \bibinfo{pages}{98}
  (\bibinfo{year}{2007}), \eprint{astro-ph/0611572}.

\bibitem[{\citenamefont{{Suzuki} et~al.}(2012)\citenamefont{{Suzuki}, {Rubin},
  {Lidman}, {Aldering}, {Amanullah}, {Barbary}, {Barrientos}, {Botyanszki},
  {Brodwin}, {Connolly} et~al.}}]{Suzuki}
\bibinfo{author}{\bibfnamefont{N.}~\bibnamefont{{Suzuki}}},
  \bibinfo{author}{\bibfnamefont{D.}~\bibnamefont{{Rubin}}},
  \bibinfo{author}{\bibfnamefont{C.}~\bibnamefont{{Lidman}}},
  \bibinfo{author}{\bibfnamefont{G.}~\bibnamefont{{Aldering}}},
  \bibinfo{author}{\bibfnamefont{R.}~\bibnamefont{{Amanullah}}},
  \bibinfo{author}{\bibfnamefont{K.}~\bibnamefont{{Barbary}}},
  \bibinfo{author}{\bibfnamefont{L.~F.} \bibnamefont{{Barrientos}}},
  \bibinfo{author}{\bibfnamefont{J.}~\bibnamefont{{Botyanszki}}},
  \bibinfo{author}{\bibfnamefont{M.}~\bibnamefont{{Brodwin}}},
  \bibinfo{author}{\bibfnamefont{N.}~\bibnamefont{{Connolly}}},
  \bibnamefont{et~al.}, \bibinfo{journal}{\apj} \textbf{\bibinfo{volume}{746}},
  \bibinfo{eid}{85} (\bibinfo{year}{2012}), \eprint{1105.3470}.

\bibitem[{\citenamefont{Tully et~al.}(2016)\citenamefont{Tully, Courtois, and
  Sorce}}]{Tully_2016}
\bibinfo{author}{\bibfnamefont{R.~B.} \bibnamefont{Tully}},
  \bibinfo{author}{\bibfnamefont{H.~M.} \bibnamefont{Courtois}},
  \bibnamefont{and} \bibinfo{author}{\bibfnamefont{J.~G.} \bibnamefont{Sorce}},
  \bibinfo{journal}{The Astronomical Journal} \textbf{\bibinfo{volume}{152}},
  \bibinfo{pages}{50} (\bibinfo{year}{2016}),
  \urlprefix\url{https://doi.org/10.3847%2F0004-6256%2F152%2F2%2F50}.

\bibitem[{\citenamefont{Tully et~al.}(2014)\citenamefont{Tully, Courtois,
  Hoffman, and Pomar\`ede}}]{Tully:2014gfa}
\bibinfo{author}{\bibfnamefont{R.~B.} \bibnamefont{Tully}},
  \bibinfo{author}{\bibfnamefont{H.}~\bibnamefont{Courtois}},
  \bibinfo{author}{\bibfnamefont{Y.}~\bibnamefont{Hoffman}}, \bibnamefont{and}
  \bibinfo{author}{\bibfnamefont{D.}~\bibnamefont{Pomar\`ede}},
  \bibinfo{journal}{Nature} \textbf{\bibinfo{volume}{513}}, \bibinfo{pages}{71}
  (\bibinfo{year}{2014}), \eprint{1409.0880}.

\bibitem[{\citenamefont{Fixsen et~al.}(1996)\citenamefont{Fixsen, Cheng, Gales,
  Mather, Shafer, and Wright}}]{Fixsen}
\bibinfo{author}{\bibfnamefont{D.~J.} \bibnamefont{Fixsen}},
  \bibinfo{author}{\bibfnamefont{E.~S.} \bibnamefont{Cheng}},
  \bibinfo{author}{\bibfnamefont{J.~M.} \bibnamefont{Gales}},
  \bibinfo{author}{\bibfnamefont{J.~C.} \bibnamefont{Mather}},
  \bibinfo{author}{\bibfnamefont{R.~A.} \bibnamefont{Shafer}},
  \bibnamefont{and} \bibinfo{author}{\bibfnamefont{E.~L.}
  \bibnamefont{Wright}}, \bibinfo{journal}{Astrophys. J.}
  \textbf{\bibinfo{volume}{473}}, \bibinfo{pages}{576} (\bibinfo{year}{1996}),
  \eprint{astro-ph/9605054}.

\bibitem[{\citenamefont{Tully et~al.}(2008)\citenamefont{Tully, Shaya,
  Karachentsev, Courtois, Kocevski, Rizzi, and Peel}}]{Tully2008}
\bibinfo{author}{\bibfnamefont{R.~B.} \bibnamefont{Tully}},
  \bibinfo{author}{\bibfnamefont{E.~J.} \bibnamefont{Shaya}},
  \bibinfo{author}{\bibfnamefont{I.~D.} \bibnamefont{Karachentsev}},
  \bibinfo{author}{\bibfnamefont{H.~M.} \bibnamefont{Courtois}},
  \bibinfo{author}{\bibfnamefont{D.~D.} \bibnamefont{Kocevski}},
  \bibinfo{author}{\bibfnamefont{L.}~\bibnamefont{Rizzi}}, \bibnamefont{and}
  \bibinfo{author}{\bibfnamefont{A.}~\bibnamefont{Peel}},
  \bibinfo{journal}{Astrophys. J.} \textbf{\bibinfo{volume}{676}},
  \bibinfo{pages}{184} (\bibinfo{year}{2008}), \eprint{0705.4139}.

\bibitem[{\citenamefont{{Marinoni} et~al.}(1998)\citenamefont{{Marinoni},
  {Monaco}, {Giuricin}, and {Costantini}}}]{mari1998}
\bibinfo{author}{\bibfnamefont{C.}~\bibnamefont{{Marinoni}}},
  \bibinfo{author}{\bibfnamefont{P.}~\bibnamefont{{Monaco}}},
  \bibinfo{author}{\bibfnamefont{G.}~\bibnamefont{{Giuricin}}},
  \bibnamefont{and}
  \bibinfo{author}{\bibfnamefont{B.}~\bibnamefont{{Costantini}}},
  \bibinfo{journal}{\apj} \textbf{\bibinfo{volume}{505}}, \bibinfo{pages}{484}
  (\bibinfo{year}{1998}), \eprint{astro-ph/9805081}.

\bibitem[{\citenamefont{Boruah et~al.}(2020)\citenamefont{Boruah, Hudson, and
  Lavaux}}]{Boruah_2020}
\bibinfo{author}{\bibfnamefont{S.~S.} \bibnamefont{Boruah}},
  \bibinfo{author}{\bibfnamefont{M.~J.} \bibnamefont{Hudson}},
  \bibnamefont{and} \bibinfo{author}{\bibfnamefont{G.}~\bibnamefont{Lavaux}},
  \bibinfo{journal}{Monthly Notices of the Royal Astronomical Society}
  \textbf{\bibinfo{volume}{498}}, \bibinfo{pages}{2703} (\bibinfo{year}{2020}),
  \urlprefix\url{https://doi.org/10.1093%2Fmnras%2Fstaa2485}.

\bibitem[{\citenamefont{Hong et~al.}(2014)\citenamefont{Hong, Springob,
  Staveley-Smith, Scrimgeour, Masters, Macri, Koribalski, Jones, and
  Jarrett}}]{Hong_2014}
\bibinfo{author}{\bibfnamefont{T.}~\bibnamefont{Hong}},
  \bibinfo{author}{\bibfnamefont{C.~M.} \bibnamefont{Springob}},
  \bibinfo{author}{\bibfnamefont{L.}~\bibnamefont{Staveley-Smith}},
  \bibinfo{author}{\bibfnamefont{M.~I.} \bibnamefont{Scrimgeour}},
  \bibinfo{author}{\bibfnamefont{K.~L.} \bibnamefont{Masters}},
  \bibinfo{author}{\bibfnamefont{L.~M.} \bibnamefont{Macri}},
  \bibinfo{author}{\bibfnamefont{B.~S.} \bibnamefont{Koribalski}},
  \bibinfo{author}{\bibfnamefont{D.~H.} \bibnamefont{Jones}}, \bibnamefont{and}
  \bibinfo{author}{\bibfnamefont{T.~H.} \bibnamefont{Jarrett}},
  \bibinfo{journal}{Monthly Notices of the Royal Astronomical Society}
  \textbf{\bibinfo{volume}{445}}, \bibinfo{pages}{402} (\bibinfo{year}{2014}),
  \urlprefix\url{https://doi.org/10.1093%2Fmnras%2Fstu1774}.

\bibitem[{\citenamefont{Turnbull et~al.}(2012)\citenamefont{Turnbull, Hudson,
  Feldman, Hicken, Kirshner, and Watkins}}]{turnbull:2011ty}
\bibinfo{author}{\bibfnamefont{S.~J.} \bibnamefont{Turnbull}},
  \bibinfo{author}{\bibfnamefont{M.~J.} \bibnamefont{Hudson}},
  \bibinfo{author}{\bibfnamefont{H.~A.} \bibnamefont{Feldman}},
  \bibinfo{author}{\bibfnamefont{M.}~\bibnamefont{Hicken}},
  \bibinfo{author}{\bibfnamefont{R.~P.} \bibnamefont{Kirshner}},
  \bibnamefont{and} \bibinfo{author}{\bibfnamefont{R.}~\bibnamefont{Watkins}},
  \bibinfo{journal}{Mon. Not. Roy. Astron. Soc.}
  \textbf{\bibinfo{volume}{420}}, \bibinfo{pages}{447} (\bibinfo{year}{2012}),
  \eprint{1111.0631}.

\bibitem[{\citenamefont{Nusser and Davis}(2011)}]{Nusser_2011}
\bibinfo{author}{\bibfnamefont{A.}~\bibnamefont{Nusser}} \bibnamefont{and}
  \bibinfo{author}{\bibfnamefont{M.}~\bibnamefont{Davis}},
  \bibinfo{journal}{The Astrophysical Journal} \textbf{\bibinfo{volume}{736}},
  \bibinfo{pages}{93} (\bibinfo{year}{2011}),
  \urlprefix\url{https://doi.org/10.1088%2F0004-637x%2F736%2F2%2F93}.

\bibitem[{\citenamefont{Scrimgeour et~al.}(2015)\citenamefont{Scrimgeour,
  Davis, Blake, Staveley-Smith, Magoulas, Springob, Beutler, Colless, Johnson,
  Jones et~al.}}]{Scrimgeour_2015}
\bibinfo{author}{\bibfnamefont{M.~I.} \bibnamefont{Scrimgeour}},
  \bibinfo{author}{\bibfnamefont{T.~M.} \bibnamefont{Davis}},
  \bibinfo{author}{\bibfnamefont{C.}~\bibnamefont{Blake}},
  \bibinfo{author}{\bibfnamefont{L.}~\bibnamefont{Staveley-Smith}},
  \bibinfo{author}{\bibfnamefont{C.}~\bibnamefont{Magoulas}},
  \bibinfo{author}{\bibfnamefont{C.~M.} \bibnamefont{Springob}},
  \bibinfo{author}{\bibfnamefont{F.}~\bibnamefont{Beutler}},
  \bibinfo{author}{\bibfnamefont{M.}~\bibnamefont{Colless}},
  \bibinfo{author}{\bibfnamefont{A.}~\bibnamefont{Johnson}},
  \bibinfo{author}{\bibfnamefont{D.~H.} \bibnamefont{Jones}},
  \bibnamefont{et~al.}, \bibinfo{journal}{Monthly Notices of the Royal
  Astronomical Society} \textbf{\bibinfo{volume}{455}}, \bibinfo{pages}{386}
  (\bibinfo{year}{2015}),
  \urlprefix\url{https://doi.org/10.1093%2Fmnras%2Fstv2146}.

\bibitem[{\citenamefont{Carrick et~al.}(2015)\citenamefont{Carrick, Turnbull,
  Lavaux, and Hudson}}]{Carrick_2015}
\bibinfo{author}{\bibfnamefont{J.}~\bibnamefont{Carrick}},
  \bibinfo{author}{\bibfnamefont{S.~J.} \bibnamefont{Turnbull}},
  \bibinfo{author}{\bibfnamefont{G.}~\bibnamefont{Lavaux}}, \bibnamefont{and}
  \bibinfo{author}{\bibfnamefont{M.~J.} \bibnamefont{Hudson}},
  \bibinfo{journal}{Monthly Notices of the Royal Astronomical Society}
  \textbf{\bibinfo{volume}{450}}, \bibinfo{pages}{317} (\bibinfo{year}{2015}),
  \urlprefix\url{https://doi.org/10.1093%2Fmnras%2Fstv547}.

\bibitem[{\citenamefont{Appleby et~al.}(2015)\citenamefont{Appleby, Shafieloo,
  and Johnson}}]{Appleby:2014kea}
\bibinfo{author}{\bibfnamefont{S.}~\bibnamefont{Appleby}},
  \bibinfo{author}{\bibfnamefont{A.}~\bibnamefont{Shafieloo}},
  \bibnamefont{and} \bibinfo{author}{\bibfnamefont{A.}~\bibnamefont{Johnson}},
  \bibinfo{journal}{Astrophys. J.} \textbf{\bibinfo{volume}{801}},
  \bibinfo{pages}{76} (\bibinfo{year}{2015}), \eprint{1410.5562}.

\bibitem[{\citenamefont{{Steinhardt} et~al.}(2020)\citenamefont{{Steinhardt},
  {Sneppen}, and {Sen}}}]{Steinhardt}
\bibinfo{author}{\bibfnamefont{C.~L.} \bibnamefont{{Steinhardt}}},
  \bibinfo{author}{\bibfnamefont{A.}~\bibnamefont{{Sneppen}}},
  \bibnamefont{and} \bibinfo{author}{\bibfnamefont{B.}~\bibnamefont{{Sen}}},
  \bibinfo{journal}{\apj} \textbf{\bibinfo{volume}{902}}, \bibinfo{eid}{14}
  (\bibinfo{year}{2020}), \eprint{2005.07707}.

\bibitem[{\citenamefont{Aluri et~al.}(2022)}]{Aluri:2022hzs}
\bibinfo{author}{\bibfnamefont{P.~K.} \bibnamefont{Aluri}} \bibnamefont{et~al.}
  (\bibinfo{year}{2022}), \eprint{2207.05765}.

\bibitem[{\citenamefont{Zhao et~al.}(2019)\citenamefont{Zhao, Zhou, and
  Chang}}]{Zhao:2019azy}
\bibinfo{author}{\bibfnamefont{D.}~\bibnamefont{Zhao}},
  \bibinfo{author}{\bibfnamefont{Y.}~\bibnamefont{Zhou}}, \bibnamefont{and}
  \bibinfo{author}{\bibfnamefont{Z.}~\bibnamefont{Chang}},
  \bibinfo{journal}{Mon. Not. Roy. Astron. Soc.}
  \textbf{\bibinfo{volume}{486}}, \bibinfo{pages}{5679} (\bibinfo{year}{2019}),
  \eprint{1903.12401}.

\bibitem[{\citenamefont{Kocevski and Ebeling}(2006)}]{Kocevski:2005kr}
\bibinfo{author}{\bibfnamefont{D.~D.} \bibnamefont{Kocevski}} \bibnamefont{and}
  \bibinfo{author}{\bibfnamefont{H.}~\bibnamefont{Ebeling}},
  \bibinfo{journal}{Astrophys. J.} \textbf{\bibinfo{volume}{645}},
  \bibinfo{pages}{1043} (\bibinfo{year}{2006}), \eprint{astro-ph/0510106}.

\bibitem[{\citenamefont{{Kourkchi} et~al.}(2020)\citenamefont{{Kourkchi},
  {Tully}, {Eftekharzadeh}, {Llop}, {Courtois}, {Guinet}, {Dupuy}, {Neill},
  {Seibert}, {Andrews} et~al.}}]{2020ApJ...902..145K}
\bibinfo{author}{\bibfnamefont{E.}~\bibnamefont{{Kourkchi}}},
  \bibinfo{author}{\bibfnamefont{R.~B.} \bibnamefont{{Tully}}},
  \bibinfo{author}{\bibfnamefont{S.}~\bibnamefont{{Eftekharzadeh}}},
  \bibinfo{author}{\bibfnamefont{J.}~\bibnamefont{{Llop}}},
  \bibinfo{author}{\bibfnamefont{H.~M.} \bibnamefont{{Courtois}}},
  \bibinfo{author}{\bibfnamefont{D.}~\bibnamefont{{Guinet}}},
  \bibinfo{author}{\bibfnamefont{A.}~\bibnamefont{{Dupuy}}},
  \bibinfo{author}{\bibfnamefont{J.~D.} \bibnamefont{{Neill}}},
  \bibinfo{author}{\bibfnamefont{M.}~\bibnamefont{{Seibert}}},
  \bibinfo{author}{\bibfnamefont{M.}~\bibnamefont{{Andrews}}},
  \bibnamefont{et~al.}, \bibinfo{journal}{\apj} \textbf{\bibinfo{volume}{902}},
  \bibinfo{eid}{145} (\bibinfo{year}{2020}), \eprint{2009.00733}.

\bibitem[{\citenamefont{Riess et~al.}(2022)}]{Riess:2021jrx}
\bibinfo{author}{\bibfnamefont{A.~G.} \bibnamefont{Riess}}
  \bibnamefont{et~al.}, \bibinfo{journal}{Astrophys. J. Lett.}
  \textbf{\bibinfo{volume}{934}}, \bibinfo{pages}{L7} (\bibinfo{year}{2022}),
  \eprint{2112.04510}.

\bibitem[{\citenamefont{Clarkson and
  Maartens}(2010{\natexlab{b}})}]{Clarkson2010}
\bibinfo{author}{\bibfnamefont{C.}~\bibnamefont{Clarkson}} \bibnamefont{and}
  \bibinfo{author}{\bibfnamefont{R.}~\bibnamefont{Maartens}},
  \bibinfo{journal}{Class. Quant. Grav.} \textbf{\bibinfo{volume}{27}},
  \bibinfo{pages}{124008} (\bibinfo{year}{2010}{\natexlab{b}}),
  \eprint{1005.2165}.

\bibitem[{\citenamefont{Macpherson}(2022)}]{Macpherson2022}
\bibinfo{author}{\bibfnamefont{H.~J.} \bibnamefont{Macpherson}}
  (\bibinfo{year}{2022}), \eprint{2209.06775}.

\end{thebibliography}

\end{document}